\documentclass[a4paper,12pt]{article}

\usepackage{subfigure}
\usepackage{a4}
\usepackage{graphics}
\usepackage{epsfig}
\usepackage{amssymb}
\usepackage{color}
\usepackage{cite}
\usepackage{epsf}
\usepackage{axodraw}
\usepackage{wasysym}
\definecolor{green3}{rgb}{0.,0.7,0.0}
\definecolor{red1}{rgb}{0.9,0,0}

\newcommand{\slashed}[1]{\not\!#1}
\def\lsim{\raise0.3ex\hbox{$\;<$\kern-0.75em\raise-1.1ex\hbox{$\sim\;$}}}
\def\gsim{\raise0.3ex\hbox{$\;>$\kern-0.75em\raise-1.1ex\hbox{$\sim\;$}}}
\DeclareMathAlphabet{\scr}{U}{rsfs}{m}{n}

\begin{document}
\hspace*{\fill} ADP-11-30/T752\\
\hspace*{\fill} DO-TH~11/24\\
\hspace*{\fill} arXiv:1110.6131~[hep-ph]\\
\begin{center}
{\large
 \bf 
CP asymmetries in the supersymmetric trilepton signal at the LHC
}                       
\end{center}
\vspace{0.5cm}
\begin{center}
{\sc S.~Bornhauser}
\end{center}
\begin{center}
{\small \it 
Department of Physics and Astronomy, 
University of New Mexico,
 Albuquerque, NM 87131, USA
}
\end{center}
\begin{center}
\vspace*{.5cm}               
{\sc M.~Drees, H.~Dreiner}
\end{center}
\begin{center}
{\small \it Bethe Center for Theoretical Physics and Physikalisches Institut
  der Universit\"at Bonn, Nussallee 12, D-53115 Bonn, Germany
}
\end{center}
\begin{center}
{\sc O.~J.~P.~\'Eboli}
\end{center}
\begin{center}
{\small \it 
Instituto de F\'{\i}sica,
Universidade de S\~ao Paulo, S\~ao Paulo, SP, Brazil.
}
\end{center}
\begin{center}
{\sc J.~S. Kim}
\end{center}
\begin{center}
{\small \it Fakult\"at f\"ur Physik, TU Dortmund, Otto--Hahn--Str. 4, D44221
  Dortmund, Germany, and\\
ARC Centre of Excellence for
Particle Physics at the Terascale,
School of Chemistry and Physics,
 University of Adelaide,
South Australia 5005,
Australia
}
\end{center}
\begin{center}
{\sc O.~Kittel}
\end{center}
\begin{center}
{\small \it 
Departamento de F\'isica Te\'orica y del Cosmos and CAFPE, \\
Universidad de Granada, E-18071 Granada, Spain
}
\end{center}

\begin{abstract}

In the CP-violating Minimal Supersymmetric Standard Model, we study the
production of a neutralino-chargino pair at the LHC.  For their decays into
three leptons, we analyze CP asymmetries which are sensitive to the CP phases
of the neutralino and chargino sector. We present analytical formulas for the
entire production and decay process, and identify the CP-violating
contributions in the spin correlation terms.  This allows us to define the
optimal CP asymmetries.  We present a detailed numerical analysis of the cross
sections, branching ratios, and the CP observables.  
For light neutralinos, charginos, and squarks, the asymmetries can reach several
$10\%$. We estimate the discovery potential for the LHC to observe CP violation 
in the trilepton channel.

\end{abstract}

\section{Introduction}

The Minimal Supersymmetric extension of the Standard Model
(MSSM)~\cite{mssm} features many potential sources for CP
violation~\cite{Haber:1997if} beyond the Standard Model (SM) of
particle physics. Most of these additional phases are
associated with flavor mixing. In the flavor conserving sector, the
$SU(2)$ gaugino mass parameter $M_2$ is conventionally chosen to be
real and positive.  The CP-violating, complex parameters are then the
higgsino mass parameter $\mu$, the $U(1)$ and $SU(3)$ gaugino mass
parameters $M_1$ and $M_3$, respectively, and the trilinear scalar
coupling parameters $A_f$,
\begin{eqnarray}
\label{eq:phases}
	\mu = |\mu| e^{i \phi_\mu}, \quad  
	M_1 = |M_1| e^{i \phi_{1}}, \quad
	M_3 = |M_3| e^{i \phi_3},   \quad
	A_f = |A_f| e^{i \phi_{A_f}}.
\end{eqnarray}
These phases in general contribute to electric dipole moments
(EDMs), in particular to those of the neutron, and the Thallium and
Mercury atoms.  In fact, scenarios where (at least) one of the phases
appearing in Eq.~(\ref{eq:phases}) is large are often in conflict with
current experimental upper bounds on these EDMs~\cite{EDMs}. However,
large phases are not excluded by these constraints, even if first
generation sleptons are rather light. For example, a certain degree of
fine-tuning allows for cancellations among the different terms
contributing to the EDMs~\cite{EDMcancel,Deppisch:2009nj}.  The EDM
bounds can also be fulfilled by including lepton flavor violating
couplings in the slepton sector~\cite{Bartl:2003ju}. 
See also Refs.~\cite{Abel:1999yz}.

\medskip

It is clear that CP observables outside the low energy EDM sector have to be
measured to independently determine or constrain possible SUSY phases at
colliders.  In that respect, T-odd and CP-odd asymmetries based on triple or
epsilon products have been proposed~\cite{tripleprods}.  For the
ILC~\cite{ILC}, triple product asymmetries have been intensively studied in
the production and decay of neutralinos~\cite{Bartl:2003tr, Bartl:2009pg,
  Bartl:2004ut, NEUT2, Kittel:2011rk, NEUT3, Kittel:2004rp} and
charginos~\cite{Kittel:2004rp, Kittel:2004kd, Bartl:2008fu, CHAR2, CHAR3},
also using transversely polarized beams~\cite{optimal2, Trans}.  At the
LHC~\cite{ATLAS, CMS}, triple product asymmetries have been studied for the
decays of stops~\cite{Deppisch:2009nj, Bartl:2004jr, Ellis:2008hq,
  MoortgatPick:2010wp,Kittel:2011sq}, sbottoms~\cite{Bartl:2006hh},
staus~\cite{Dreiner:2010wj}, and neutralinos which originate from squark
decays~\cite{Bartl:2003ck, MoortgatPick:2009jy}.  The triple-product
asymmetries can be of the order of $60\%$, since they already appear at tree
level due to spin correlations.  In particular, it was shown that the
CP-violating effects in stop decays~\cite{MoortgatPick:2010wp} can be measured
at the $3 \sigma$ confidence level at the LHC for a luminosity of
$\mathcal{L}=300~{\rm fb}^{-1}$, using the technique of momentum
reconstruction for on-shell decay chains~\cite{MoortgatPick:2009jy,
  Buckley:2007th}.

\medskip

We are thus motivated to explore the discovery reach at the LHC for CP 
violation in  the production of a neutralino-chargino pair
\begin{eqnarray}
p + p \to \tilde\chi_i^0 + \tilde\chi_j^\pm;
      \qquad i=2,3,4; \quad j=1,2.
\label{eq:prod}
\end{eqnarray}
The contributing tree-level diagrams involving $\tilde u_L$, $\tilde d_L$
squark and $W$ exchange are shown in Fig.~\ref{Fig:process2}. The phases
$\phi_1$ and $\phi_\mu$ in the neutralino and chargino sector will trigger CP
violation. CP-sensitive contributions to the differential cross section will
originate at tree level from the spin and spin-spin correlations among the
neutralinos and charginos. These can be analyzed with the help of the visible
decay products of the neutralino and chargino. Leptonic decays are especially
useful~\cite{trilepton}. On the one hand, hadronic decays lead to final states
with very large SM backgrounds, and therefore do not lead to viable signals at
the LHC. On the other hand, the construction of T-odd observables requires the
measurement of the charges of final state particles, which is difficult, if
not impossible, for jets. 

\newpage

We therefore use the leptonic two-body decays of
the chargino and neutralino as ``spin analyzers'':
\begin{eqnarray}
\tilde\chi^0_i   &\to& \tilde\ell_R^\mp + \ell_1^\pm \, ,
\label{eq:decayNeut} \nonumber \\
&&\tilde\ell_R^\mp \to \tilde\chi^0_1 + \ell_2^\mp\, ; \\
\tilde\chi^\pm_j   &\to& \tilde\nu_\ell + \ell_3^\pm \, ,
\label{eq:decayChar} \nonumber \\
&&\tilde\nu_\ell \to \tilde\chi^0_1 + \nu_\ell \, .
\end{eqnarray}
See Fig.~\ref{Fig:process} for a schematic picture of the production and decay
process.  The LHC signature is three isolated leptons (two of them with same
flavor and opposite charge) and missing energy, carried away by the two
neutralinos $\tilde\chi^0_1$ and the neutrino $\nu_\ell$.  This process, known
as the trilepton signal~\cite{trilepton,Sullivan:2008ki,Kitano:2008sa}, has
low QCD and SM background, and thus has been studied in detail as a SUSY
discovery channel at the Tevatron~\cite{tev_old, tevatron, Choi:1999mv}, and
also at the LHC~\cite{ATLAS,CMS}.

\medskip

In this paper we first calculate the amplitude squared for the entire
process of neutralino-chargino pair production and decay in the spin
density matrix formalism~\cite{Haber:1994pe}.  
The explicit formulas allow us to identify
the CP-sensitive parts in the spin and spin-spin correlations, see
Section~\ref{CP asymmetries}.  From those we define T-odd asymmetries
of triple- and epsilon products.  In Section~\ref{Numerical analysis},
we numerically analyze these asymmetries, the production cross
sections, and the neutralino and chargino branching ratios in general
MSSM scenarios with complex $\mu$ and $M_1$.  We discuss several
production modes $\tilde\chi_{2,3}^0\tilde\chi_{1,2}^\pm$, and
two-body decays via gauge bosons and three-body decays. The asymmetries
analyzed in Sections \ref{CP asymmetries} and \ref{Numerical analysis}
depend on the 4-momenta of the quarks in the initial state, and of the
chargino and/or neutralino in the final state. These can in general
not be determined exactly. In Section~\ref{construct}, we therefore
discuss approximations for the initial quark and intermediate gaugino
momenta, which work quite efficiently to enhance the asymmetries. 
In Section~\ref{Discovery reach}, we comment on the
minimal luminosity required at the LHC to observe CP-violating effects
in the trilepton mode, before we summarize and conclude in
Section~\ref{Summary and conclusions}. The Appendices contain a review
of chargino and neutralino 
mixing (Appendix~\ref{Chargino and neutralino mixings}), of the relevant parts
of the MSSM Lagrangian with complex couplings~(\ref{Lagrangians and couplings}), 
definitions of momenta and spin vectors as well as details of phase space 
integration~(\ref{Phase space}), analytical expressions for the amplitude 
squared in the spin-density matrix formalism~(\ref{Density matrix formalism}), 
and a discussion of how our asymmetries depend on the lepton 
charges~(\ref{CPasymmetries}).

\setlength{\unitlength}{1pt}
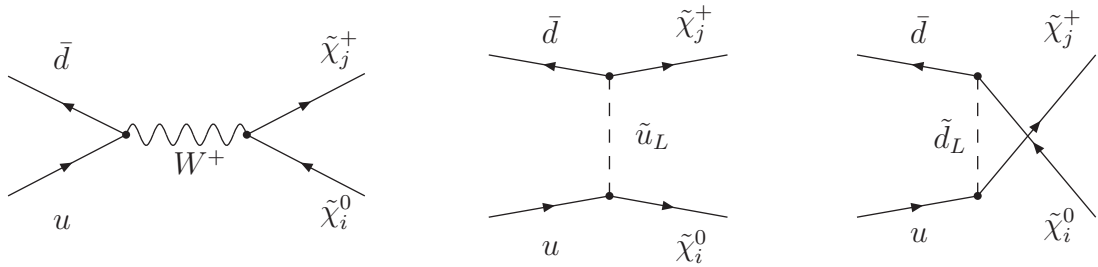
\begin{figure}[t]
\SetScale{1.0}
\begin{picture}(100,100)(20,40)
\Text(50,125)[lb]{{$\bar d$}}
\Text(50,65)[lb]{{$ u$}}
\Text(105,90)[]{{$W^+$}}
\Text(150,125)[lb]{{$\tilde\chi_j^+$}}
\Text(150,65)[lb]{{$\tilde\chi_i^0$}}
\ArrowLine(122,100)(166,123)
\ArrowLine(166,77)(122,100)
\ArrowLine(77,100)(33,122)
\ArrowLine(33,77)(77,100)
\Photon(77,100)(122,100) {4}{4.5}
\Vertex(77,100){1.5}
\Vertex(122,100){1.5}
\end{picture}
\SetScale{1.0}
\begin{picture}(100,100)(-30,40)
\Text(75,135)[lb]{{$\bar d$}}
\Text(75,55)[lb]{{$ u$}}
\Text(110,95)[lb]{{$\tilde u_L$}}
\Text(125,135)[lb]{{$\tilde\chi_j^+$}}
\Text(125,52)[lb]{{$\tilde\chi_i^0$}}
\ArrowLine(100,122)(144,130)
\ArrowLine(100,77)(144,69)
\ArrowLine(100,122)(55,130)
\ArrowLine(55,69)(100,77)
\DashLine(100,77)(100,122){5}  
\Vertex(100,77){1.5}
\Vertex(100,122){1.5}
\end{picture}
\SetScale{1.0}
\begin{picture}(100,100)(-60,40)
\Text(75,135)[lb]{{$\bar d$}}
\Text(75,60)[lb]{{$u$}}
\Text(90,100)[]{{$\tilde d_L$}}
\Text(125,60)[lb]{{$\tilde\chi_i^0$}}
\Text(125,133)[lb]{{$\tilde\chi_j^+$}}
\ArrowLine(144,69)(100,122)
\ArrowLine(100,77)(144,130)
\ArrowLine(100,122)(55,130)
\ArrowLine(55,69)(100,77)
\DashLine(100,77)(100,122){5} 
\Vertex(100,77){1.5}
\Vertex(100,122){1.5}
\end{picture}
\caption{
  Feynman diagrams of neutralino-chargino production, with $s$-channel
  $W$-exchange, $t$-channel $\tilde u_L$-exchange, and $u$-channel $\tilde
  d_L$-exchange.}
\label{Fig:process2}
\end{figure}
\setlength{\unitlength}{1cm}

\begin{figure}[t]
                \scalebox{1}{
\begin{picture}(10,5)(-3.4,0)
	   \ArrowLine(30,90)(0,120)
           \ArrowLine(0, 40)(30,70)
          \ArrowLine(50,90)(80,120)
          \ArrowLine(80, 40)(50,70)
	   \CArc(40,80)(14.5,0,360)
         \Vertex(80,120){1.5}
          \ArrowLine(100,160)(80,120)
           \DashLine(80,120)(130,120){5}
           \Vertex(130,120){1.5}
            \ArrowLine(170,150)(130,120)
            \ArrowLine(130,120)(175,95)
           \Vertex(80,40){1.5}
           \ArrowLine(100,0)(80,40)
           \DashLine(80,40)(130,40){5}
           \Vertex(130,40){1.5}
          \ArrowLine(170,65)(130,40)
          \ArrowLine(130,40)(175,10)
         \put( 0.45,3.85){ $\bar d $}
         \put( 1.60,3.85){ $\tilde\chi_j^+$}
         \put( 0.45,1.6){ $  u $}
         \put( 1.60,1.6){ $\tilde\chi_i^0 $}
         \put( 2.65,0.03){ $\ell_1^\pm$}
         \put(3.55,1.65){ $\tilde\ell^\mp$}
         \put(6.2,2.2){ $\tilde\chi_1^0$}
         \put(6.25,0.03){ $ \ell_2^\mp$}
         \put(3.45,3.64){ $\tilde\nu_\ell$}
         \put( 3.7,5.58){ $\ell_3^+$}
         \put(6.25, 3.2){ $\nu $}
         \put(6.16,5.3){ $\tilde\chi_1^0$}
\end{picture}
}
\caption{Schematic picture of neutralino-chargino production and decay.
         The blob represents the tree level production process, 
         see the Feynman diagrams  in Fig.~\ref{Fig:process2}.
 }
\label{Fig:process}
\end{figure}
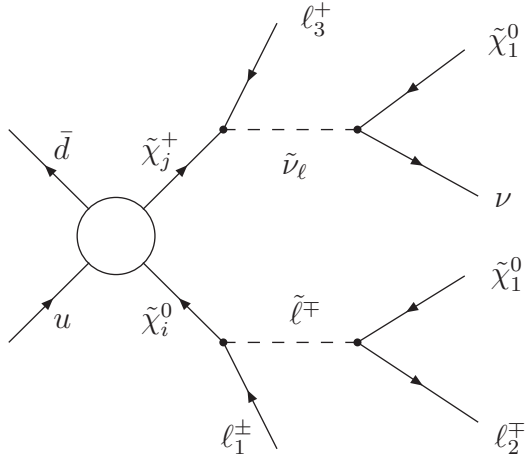

\section{CP asymmetries 
  \label{CP asymmetries}}

In this Section, we first identify the CP-sensitive parts in the amplitude
squared of the entire process of neutralino-chargino pair production and their
subsequent leptonic two-body decays, see
Eqs.~(\ref{eq:prod})-(\ref{eq:decayChar}) and Fig.~\ref{Fig:process}. In order
to probe these parts, we then define several T-odd products of 4-momenta and
the corresponding asymmetries.  Explicit expressions for the squared
amplitude, relevant parts of the MSSM Lagrangian with complex couplings, and
details of phase-space integration, are summarized in Appendices A to D.
In this and the following Section we consider all (products of) 4-momenta to
be observables. Since not all of these momenta can be measured in
actual LHC experiments, we will address solutions to this complication later in 
Section~\ref{construct}. 

\subsection{T-odd observables}
\label{sec:CPterms}

The amplitude squared $|T|^2$ for neutralino-chargino pair production and
decay, see Fig.~\ref{Fig:process}, can be decomposed into contributions from
the neutralino spin correlations, the chargino spin correlations, the
neutralino-chargino spin-spin correlations, and a spin-independent part, see
Eq.~(\ref{eq:matrixelement2}). At tree level, CP-sensitive parts can only
originate from the spin and the spin-spin correlations. They receive
contributions from the exchange of the $W$ and the squarks $\tilde
u_L$,~$\tilde d_L$.

\medskip

The CP-sensitive contributions to the neutralino spin correlations
are \footnote{The squares of the $\tilde u_L$ and $\tilde d_L$ contributions,
  given by $\Sigma_{PD_1}(\tilde u_L \tilde u_L)$ and $\Sigma_{PD_1}(\tilde
  d_L \tilde d_L)$, do not contain sufficiently many different couplings to be
  sensitive to any CP-violating phase. Note that we have assumed vanishing
  $L-R$ mixing for first generation squarks, making their couplings chiral,
  whereas the $W$-$ \tilde \chi_i^0$-$\tilde \chi_j^\pm $ couplings have no fixed
  chirality.} 
\begin{eqnarray}
  \Sigma_{PD_1}(W W) &\propto&
{\rm Im}\{O^L_{ij} O^{R\ast}_{ij} \} \;{\mathcal E}_{\tilde\chi^0_i}, 
\label{eq:neutWW}\\ [2mm]
\Sigma_{PD_1}(W \tilde u_L) &\propto&
{\rm Im}\{f_{u i}^L l_{\tilde u j}^{L\ast} O^{R\ast}_{ij} \}
                      \;{\mathcal E}_{\tilde\chi^0_i},
\label{eq:neutWU}\\ [2mm]
\Sigma_{PD_1}(W \tilde d_L) &\propto&
{\rm Im}\{f_{d i}^{L\ast} l_{\tilde d j}^{L} O^{L\ast}_{ij} \}
                      \;{\mathcal E}_{\tilde\chi^0_i},
\label{eq:neutWD}\\ [2mm]
\Sigma_{PD_1}(\tilde u_L \tilde d_L) &\propto&
{\rm Im}\{f_{u i}^{L\ast} f_{d i}^{L\ast} l_{\tilde u j}^{L} l_{\tilde d j}^{L} \}
	       \;{\mathcal E}_{\tilde\chi^0_i}.
\label{eq:neutUD}
\end{eqnarray} 
These spin correlation terms explicitly depend on the imaginary parts
of the products of the $W$-$\tilde\chi^0_i$-$\tilde\chi^\pm_j$
couplings $O^{L,R}_{ij}$, the $q$-$\tilde q_L$-$\tilde\chi^0_i$
couplings $f_{q i}^{L}$, and the $q^\prime$-$\tilde q_L$-$
\tilde\chi^\pm_j$ couplings $l_{\tilde q j}^{L}$.  The imaginary parts
are manifestly CP-sensitive, \textit{i.e.} sensitive to the phases
$\phi_\mu$, $\phi_{1}$ in the neutralino-chargino sector, and are each
multiplied by the T-odd epsilon product
\begin{equation}
{\mathcal E}_{\tilde\chi^0_i} \,\equiv\, [p_u,p_d,p_{\tilde\chi_i^0},p_{\ell_1}]
\,\equiv \,\varepsilon_{\mu\nu\alpha\beta}~
p_{u}^{\mu}~p_d^{\nu}~p_{\tilde\chi^0_i}^{\alpha}~p_{\ell_1}^{\beta} \, . 
\label{eq:epsilon_neut}                
\end{equation}
We use the convention $\varepsilon_{0123}=1$, and here and in the following
we not put a bar on any of the $u$ or $d$ quark indices, as long as the 
statements made apply for both
$ u \bar d \to \tilde\chi^0_i \tilde\chi^+_j$ or
$ \bar u  d \to \tilde\chi^0_i \tilde\chi^-_j$ production processes.

\medskip

Similarly, the CP-sensitive contributions to the chargino spin correlations are
\begin{eqnarray}
\Sigma_{PD_3}(W W) &\propto&
{\rm Im}\{O^L_{ij} O^{R\ast}_{ij} \} \;{\mathcal E}_{\tilde\chi^\pm_j}, 
\label{eq:charWW}\\ [2mm]
\Sigma_{PD_3}(W \tilde u_L) &\propto&
{\rm Im}\{f_{u i}^L l_{\tilde u j}^{L\ast} O^{R\ast}_{ij} \}
                     \; {\mathcal E}_{\tilde\chi^\pm_j},
\label{eq:charWU}\\ [2mm]
\Sigma_{PD_3}(W \tilde d_L) &\propto&
{\rm Im}\{f_{d i}^{L\ast} l_{\tilde d j}^{L} O^{L\ast}_{ij} \}
                      \;{\mathcal E}_{\tilde\chi^\pm_j},
\label{eq:charWD}\\ [2mm]
\Sigma_{PD_3}(\tilde u_L \tilde d_L) &\propto&
{\rm Im}\{f_{u i}^{L\ast} f_{d i}^{L\ast} l_{\tilde u j}^{L} l_{\tilde d j}^{L} \}
	       \;{\mathcal E}_{\tilde\chi^\pm_j} \, ,
\label{eq:charUD}
\end{eqnarray} 
with the short-hand notation for the epsilon product
\begin{equation}
{\mathcal E}_{\tilde\chi^\pm_j}\, \equiv\, [p_u,p_d,p_{\tilde\chi_j^\pm},p_{\ell_3}]
\,\equiv\,\varepsilon_{\mu\nu\alpha\beta}~
p_{u}^{\mu}~p_d^{\nu}~p_{\tilde\chi^\pm_j}^{\alpha}~p_{\ell_3}^{\beta} \, .
\label{eq:epsilon_char}                
\end{equation}
Note that the chargino spin correlations probe the same coupling combinations
as the neutralino spin correlation terms.

\medskip

Finally, the CP-sensitive contributions to the spin-spin correlations are
\begin{eqnarray}
 \Sigma_{P D_1 D_3}(W W) &\propto&
{\rm Im}\{O^L_{ij} O^{R\ast}_{ij} \} \;f, 
\label{eq:neutcharWW}\\ [2mm]
\Sigma_{P D_1 D_3}(W \tilde u_L) &\propto&
{\rm Im}\{f_{u i}^L l_{\tilde u j}^{L\ast} O^{R\ast}_{ij} \}\; f,
\label{eq:neutcharWU}\\ [2mm]
\Sigma_{P D_1 D_3}(W \tilde d_L) &\propto&
{\rm Im}\{f_{d i}^{L\ast} l_{\tilde d j}^{L} O^{L\ast}_{ij} \}\;f,
\label{eq:neutcharWD}\\ [2mm]
\Sigma_{P D_1 D_3}(\tilde u_L \tilde d_L) &\propto&
{\rm Im}\{f_{u i}^{L\ast} f_{d i}^{L\ast} l_{\tilde u j}^{L} l_{\tilde d j}^{L} \}
	       \;f,
\label{eq:neutcharUD}
\end{eqnarray} 
with the short-hand notation for the kinematical function
\begin{eqnarray}
f &\equiv& \phantom{+} (p_u \cdot p_{\ell_1})
[p_d,p_{\tilde\chi_i^0},p_{\tilde\chi_j^\pm},p_{\ell_3}] 
 + (p_u \cdot p_{\tilde\chi_i^0})[p_d,p_{\tilde\chi_j^\pm},p_{\ell_1},p_{\ell_3} ]
\nonumber \\[1mm]&&
 +(p_d \cdot p_{\ell_3})[p_u,p_{\tilde\chi_i^0},p_{\tilde\chi_j^\pm},p_{\ell_1}]
  +(p_d \cdot p_{\tilde\chi_j^\pm})[p_u,p_{\tilde\chi_i^0},p_{\ell_1},p_{\ell_3}],
\label{eq:xf2} 
\end{eqnarray}
see Eq.~(\ref{eq:f2}) in Appendix~\ref{Squared amplitude of production
and decay}. Again the spin-spin correlations contain the same
dynamics, \textit{i.e.} the same coupling combinations, as the spin
correlations. Note that the neutralino and chargino decays will not
yield additional CP-sensitive contributions, since these are 
two-body decays via scalar particles. The T-odd spin-spin correlation
term~$f$ is analogous to the corresponding term in
neutralino~\cite{Bartl:2009pg} or chargino~\cite{Bartl:2008fu} pair
production at the ILC, since all these processes have the same
kinematical structure.

\medskip

Each of the epsilon products ${\mathcal E}_{\tilde\chi^0_i}$,
Eq.~(\ref{eq:epsilon_neut}), and ${\mathcal E}_{\tilde\chi^\pm_j}$,
Eq.~(\ref{eq:epsilon_char}), or the tensor product~$f$,
Eq.~(\ref{eq:xf2}), is T-odd. That is since each of the spatial
components of the four-momenta changes sign under a naive time
reversal, $t \to -t$.  Due to CPT invariance, these T-odd products 
are multiplied with the CP-sensitive, imaginary parts
of products of couplings,
${\rm Im}\{O^L_{ij} O^{R\ast}_{ij} \} $,
${\rm Im}\{f_{u i}^L l_{\tilde u j}^{L\ast} O^{R\ast}_{ij} \}$,
${\rm Im}\{f_{d i}^{L\ast} l_{\tilde d j}^{L} O^{L\ast}_{ij} \}$, or
${\rm Im}\{f_{u i}^{L\ast} f_{d i}^{L\ast} l_{\tilde u j}^{L} l_{\tilde d j}^{L} \}$.

\subsection{T-odd asymmetries}
\label{sec:TOddProducts}

The task is to define observables that project out the CP-sensitive parts in
the spin and the spin-spin correlation terms in the amplitude squared.  This
can be achieved by defining for the several possible T-odd products ${\mathcal
  T}= {\mathcal E}_{\tilde\chi^0_i}$, ${\mathcal E}_{\tilde\chi^\pm_j}$, or
$f$ the corresponding T-odd asymmetries of the cross section $\sigma$ for
neutralino-chargino production and decay
\begin{equation}
        {\mathcal A}=
        \frac{\sigma({\mathcal T}>0)-\sigma({\mathcal T}<0)}
        {\sigma({\mathcal T}>0)+\sigma({\mathcal T}<0)}
      =
          \frac{\int {\rm Sign}[{\mathcal T}]
                 |T|^2 d{\rm Lips}\; d{\rm PDF}}
           {\int |T|^2 d{\rm Lips}\;d{\rm PDF}},
\label{eq:Toddasym}
\end{equation}
with the amplitude squared $|T|^2$, Eq.~(\ref{eq:matrixelement2}), the Lorentz
invariant phase-space $d{\rm Lips}$~(\ref{eq:phasespace}), and the short-hand
notation $d{\rm PDF}\equiv dx_1 dx_2 f_u(x_1, \mu^2) f_d(x_2, \mu^2)$, see
Eq.~(\ref{eq:sigmahadronic}).

\medskip

We obtain explicit expressions for the asymmetries by inserting the amplitude
squared $|T|^2$~(\ref{eq:matrixelement2}) into Eq.~(\ref{eq:Toddasym}):
\begin{eqnarray}
{\mathcal A}({\mathcal E}_{\tilde\chi^0_i}) &=&
  \frac{\int {\rm Sign}({\mathcal E}_{\tilde\chi^0_i})\,
\Sigma_{PD_1}~D_3 \;  d{\rm Lips} \; d{\rm PDF} }
{\int P~D_1~D_3~d{\rm Lips} \; d{\rm PDF}},
\label{eq:Adependenceneut}\\ [2mm]
{\mathcal A}({\mathcal E}_{\tilde\chi^\pm_j}) &=&
  \frac{\int {\rm Sign}({\mathcal E}_{\tilde\chi^\pm_j})\,
\Sigma_{PD_3}D_1 \;  d{\rm Lips} \; d{\rm PDF} }
{\int P~D_1~D_3~d{\rm Lips} \; d{\rm PDF}},
\label{eq:Adependencechar}\\ [2mm]
{\mathcal A}(f) &=&  \frac{\int {\rm Sign}(f)\, \Sigma_{P D_1 D_3}\;
  d{\rm Lips} \; d{\rm PDF} }
{\int P~D_1~D_3~d{\rm Lips} \; d{\rm PDF}},
\label{eq:Adependenceneutchar}
\end{eqnarray}
using Eqs.~(\ref{eq:dSigmaNeut}) and (\ref{eq:dSigmaChar}), and the
phase space element $d{\rm Lips}$ as given in
Eq.~(\ref{eq:phasespace}), where we have already used the narrow width
approximation of the propagators, see Eq.~(\ref{eq:narrowwidth}). In
the numerators of the asymmetries ${\mathcal A}$,
Eqs.~(\ref{eq:Adependenceneut})-(\ref{eq:Adependenceneutchar}), only
those spin or spin-spin correlations remain, which contain the
corresponding T-odd products ${\mathcal T}={\mathcal E}_{\tilde\chi^0
_i}$, ${\mathcal E}_{\tilde\chi^\pm_j}$, or $f$. The other terms of
the amplitude squared vanish due to the phase space integration over
the sign of the T-odd product, ${\rm Sign}({\mathcal T})$. In the
denominator, all spin and spin-spin correlation terms vanish, and only
the spin-independent parts contribute. This last statement remains
true even after applying acceptance cuts on the final state momenta,
as long as these cuts are CP symmetric.

\medskip

Note that in general the largest asymmetries are obtained by using
T-odd products of 4-momenta that match exactly the kinematical
dependence of the CP-sensitive terms in the amplitude squared. In the
literature, these are therefore sometimes referred to as \emph{optimal
observables}~\cite{optimal}.  Other combinations of momenta lead in
general to smaller asymmetries.\footnote{Note that strictly speaking
the largest observables are the expectation values of the T-odd
products $\langle \mathcal{T} \rangle = \int\mathcal{T}|T|^2 d{\rm
Lips}/ \int |T|^2 d{\rm Lips} $, as used \textit{e.g.} in
Ref.~\cite{Bartl:2009pg,optimal2}. Typically their statistical
significance is increased by $\sim
20\%$~\cite{Bartl:2009pg,Bartl:2008fu}, compared to the asymmetries
given in Eq.~(\ref{eq:Toddasym}). However, we show in
Section~\ref{construct} that these optimal observables are more
difficult to measure experimentally than the asymmetries we analyze.}

\medskip

By construction, the asymmetries ${\mathcal A}$,
Eqs.~(\ref{eq:Adependenceneut})-(\ref{eq:Adependenceneutchar}), are manifestly
Lorentz invariant. Due to the boost between the partonic center-of-mass frame
and the lab frame, the use of triple products, which are not Lorentz
invariant, generally leads to smaller asymmetries.  One famous example is the
triple product of the three outgoing leptons~\cite{Choi:1999mv}
\begin{equation}
 {\mathcal T} \,=\,
	\mathbf{p}_{\ell_1 }\cdot 
        ( \mathbf{p}_{\ell_2 } \times 	\mathbf{p}_{\ell_3})
\,\equiv\, (\mathbf{p}_{\ell_1 },\mathbf{p}_{\ell_2 },\mathbf{p}_{\ell_3} ).
\label{tripleterm}
\end{equation}
That triple product probes the CP-sensitive terms in the spin-spin
correlations and in the neutralino spin correlations, but not in the
chargino spin correlations. Since it does not directly match the
kinematical dependence of the CP-sensitive terms, and since it is not
Lorentz invariant, the corresponding asymmetry will be  reduced by a factor
of about $2$ to $4$ compared to the maximal asymmetries as given in
Eqs.~(\ref{eq:Adependenceneut})-(\ref{eq:Adependenceneutchar}). These
however contain the initial quark and the intermediate
neutralino/chargino momenta, which cannot be measured directly at the
LHC and thus need to be approximated, whereas the lepton momenta
appearing in Eq.~(\ref{tripleterm}) are measurable with high
accuracy. In Section~\ref{construct} we will compare numerically the
sizes of different experimentally measurable CP-asymmetries around a
benchmark scenario.

\medskip

Another complication arises because the initial state at the LHC, which
contains two protons but no antiprotons, is not CP self-conjugate. 
Thus LHC experiments will strictly speaking not be able to
measure true CP-odd asymmetries. However, it is sufficient for our purpose
that our asymmetries are odd under naive time reversal. This ensures that at tree
level they are non-zero only in the presence of non-trivial CP phases. 
Finally, one needs to be careful when summing over events with different 
lepton charges, since some of the asymmetries change sign when the sign 
of a lepton charge is flipped. This is discussed in Appendix~\ref{CPasymmetries}.

\begin{table}[t]
\renewcommand{\arraystretch}{1.7}
\caption{
      MSSM benchmark scenario.
 \label{tab:scenario}}
\begin{center}
\begin{tabular}{cccccccc} \hline
 
  \multicolumn{1}{c}{$M_2$} 
& \multicolumn{1}{c}{$|\mu|$}
& \multicolumn{1}{c}{$\phi_\mu$}  
& \multicolumn{1}{c}{$\phi_{1}$}  
& \multicolumn{1}{c}{$\tan{\beta} $}
& \multicolumn{1}{c}{$ M_{\tilde E}^{\tilde \ell} $}
& \multicolumn{1}{c}{$ M_{\tilde L}^{\tilde \ell} $}
& \multicolumn{1}{c}{$m_{\tilde q}$}
\\\hline
 
  \multicolumn{1}{c}{$240~{\rm GeV}$} 
& \multicolumn{1}{c}{$150~{\rm GeV}$}
& \multicolumn{1}{c}{$0$}  
& \multicolumn{1}{c}{$ \frac{1}{5}\pi$} 
& \multicolumn{1}{c}{$5$}
& \multicolumn{1}{c}{$110~{\rm GeV}$}
& \multicolumn{1}{c}{$150~{\rm GeV}$}
& \multicolumn{1}{c}{$400~{\rm GeV}$}
\\\hline
\end{tabular}
\end{center}
\renewcommand{\arraystretch}{1.0}
\end{table}

\section{Numerical analysis
  \label{Numerical analysis}}

We perform a quantitative study of the CP asymmetries ${\mathcal A}$,
Eqs.~(\ref{eq:Adependenceneut})-(\ref{tripleterm}), 
the $pp\to\tilde\chi_i^0\tilde\chi_j^\pm$ production cross
sections, and branching ratios of the neutralinos and charginos around 
a benchmark scenario defined in Table~\ref{tab:scenario}. 
We fix all relevant parameters directly at the weak scale. 
First we choose rather small values of $|\mu| =150$~GeV and $M_2=240$~GeV,
which results in small chargino and neutralino masses, but still distant 
from the current experimental bounds~\cite{Nakamura:2010zzi}. Larger chargino
and neutralino masses considerably reduce the production cross sections. 
In order to reduce the number of parameters, we use the GUT
inspired\footnote{Note that this choice significantly constrains the
  neutralino sector~\cite{Neutpaper}.}  relation for the modulus of the $U(1)$
gaugino mass parameter $|M_1|=5/3\,\tan^2 \theta_w M_2\approx 0.5M_2$, 
but leave $\mu$ and the phase $\phi_1$ as independent parameters.
In order to enable the neutralino and chargino two-body decays, 
$\tilde\chi^0_2 \to \tilde \ell_R \, \ell_1$  and 
$\tilde\chi^\pm_2 \to\ell_3^\pm\,\tilde\nu_\ell^{(\ast)}$, we fix the
soft-breaking parameters in the slepton sector to $M_{\tilde
E}^{\tilde \ell} = 110$~GeV and $M_{\tilde L}^{\tilde \ell} = 150$~GeV
for $\ell=e,\mu,\tau$. The CP asymmetries, as well as the chargino and
neutralino branching ratios are rather independent of this choice, as
long as the desired neutralino and chargino two-body decays are
kinematically allowed. We fix $\tan\beta=5$ since we observe a mild
dependence of the cross sections and asymmetries on $\tan\beta$. We
take stau mixing into account, and fix the trilinear scalar coupling
parameter $A_\tau=250$~GeV. This choice only has a small impact on the
neutralino and chargino branching ratios. Since its phase does not
contribute to the CP asymmetry, we set $\phi_{A_\tau}=0$. Finally, we
chose a large mass for the charged Higgs boson, \textit{i.e.} we work
in the ``decoupling limit'' of the MSSM Higgs sector. The only Higgs
state of relevance to us is then the lightest neutral Higgs boson,
which has a mass of about 
$115$~GeV.\footnote{ 
As mentioned in the Introduction, the combination of small sparticle
masses and rather large phases tends to give too large electric dipole
moments. This requires some finetuning of parameters not relevant for our
analysis. For example, our benchmark point satisfies the $95\%$~c.l. bound
on the EDM of the electron~\cite{Nakamura:2010zzi} if the phase of $A_e$ lies between
$-1.543$ and $-1.445$ ($-1.0254$ and $-0.9118$) for $|A_e| = 150 \ (300)$~GeV,
indicating a finetuning at the $1$\% level. The EDM of the neutron, and of
atoms, in addition depends on the trilinear soft breaking parameters $A_u$
and $A_d$, as well as on the gluino mass. Since the experimental bound on
the EDM of the neutron is about one order of magnitude weaker than for the
electron, and since squarks are significantly heavier in our benchmark
scenario, somewhat less finetuning is required in the squark sector.}

\medskip

\begin{table}[t]
\renewcommand{\arraystretch}{1.8}
\caption{Superparticle masses and branching ratios for the benchmark scenario
  as given in  Table~\ref{tab:scenario}. The branching ratios are summed over
  $\ell=e,\mu$, and for the neutralino also summed over both slepton charges.
  \label{tab:masses}}
\begin{center}
      \begin{tabular}{|c|c|c|}
\hline
$   m_{\tilde\chi^0_1}  =   89~{\rm GeV}$ &
$ m_{\tilde{e}_R} =   118~{\rm GeV}$ & 
$ m_{\tilde\chi^\pm_1}  =   119~{\rm GeV}$  \\
\hline
$   m_{\tilde\chi^0_2}  =   146~{\rm GeV}$ &
$ m_{\tilde{e}_L} =   157~{\rm GeV}$ & 
$ m_{\tilde\chi^\pm_2}  =   281~{\rm GeV}$   \\
\hline
$   m_{\tilde\chi^0_3}  =   160~{\rm GeV}$ &
$m_{\tilde\tau_1} =   117~{\rm GeV}$ & 
${\rm BR}(\tilde\chi_2^0\to \tilde \ell_R \, \ell) = 66 \%$ \\
\hline
$   m_{\tilde\chi^0_4}  =   281~{\rm GeV}$ &
$ m_{\tilde\nu}   =   137~{\rm GeV}$ &
${\rm BR}(\tilde\chi_2^+\to \tilde\nu_\ell \, \ell^+) = 23\%$ \\
\hline
\end{tabular}
\end{center}
\renewcommand{\arraystretch}{1.0}
\end{table}

Finally, we fix the squark masses $m_{\tilde q_L} = m_{\tilde u_L} =
m_{\tilde d_L} = 400$~GeV to relatively low values, to enhance squark exchange
in the production. For mixed or bino-like gauginos, the squark exchange
channels will give the dominant contributions to the asymmetries and the
$\tilde\chi_i^0\tilde\chi_j^\pm$ production cross sections. 
Note that in the context of the constrained MSSM/mSUGRA, the experimental lower bounds
for squark masses of the first generation  at the $95\%$ confidence level
have recently increased from some  $450$~GeV (in specific $m_{\tilde q}= m_{\tilde g}$
scenarios) for data-sets corresponding to an integrated LHC luminosity of about 
$35~{\rm pb}^{-1}$ for the year 2010~\cite{Aad:2011xm,Caron:2011ge,Chatrchyan:2011qs},
to now up to some $1.1$~TeV, based on data samples corresponding to 
up to $1.14~{\rm fb}^{-1}$, collected in the first half of the year 
2011~\cite{Collaboration:2011iu,Chatrchyan:2011zy}. 
Although we think that the reported special and CP-conserving cMSSM/mSUGRA model 
based bounds can also be transferred to some extend to our general SUSY models considered, 
we still take our benchmark scenario as a starting point for parameter scans,
which we will perform in the following.
In this sense our benchmark scenario has to be seen as the most optimistic,
since it provides the largest CP asymmetries possible.
We discuss and comment on the asymmetries and the cross sections with heavier 
squark masses in detail at the end of the next Subsection.

\medskip

The relevant resulting SUSY masses, branching ratios and production
cross sections for the benchmark scenario are summarized in
Tables~\ref{tab:masses} and~\ref{tab:sigmP}, respectively. Note that
the production cross sections for the charge conjugated gaugino pairs
$pp\to\tilde\chi_i^0\tilde\chi_j^-$ are about half as big as for
$pp\to\tilde\chi_i^0\tilde\chi_j^+$ production, see
Table~\ref{tab:sigmP}. This is due to the PDFs, approximately
reflecting the valence quark ratio $u:d \approx 2:1$, as the partonic
cross sections are identical for the two charge conjugated pairs at
tree level.

\begin{table}[t]
\renewcommand{\arraystretch}{1.8}
\caption{Different production cross sections at $\sqrt{s} = 14$ TeV for the
  scenario of Table~\ref{tab:scenario}. The values in parentheses correspond
  to the production of $\tilde\chi_i^0 \tilde\chi_j^-$.
         \label{tab:sigmP}}
\medskip
      \begin{tabular}{|c|c|c|}
\hline
$  \sigma(p p \to\tilde\chi_2^0\tilde\chi_1^\pm) = 600 \ (332)~{\rm fb} $ &
$  \sigma(p p \to\tilde\chi_3^0\tilde\chi_1^\pm) = 574\ (313)~{\rm fb} $ &
$  \sigma(p p \to\tilde\chi_4^0\tilde\chi_1^\pm) = 2\ (1)~{\rm fb} $ \\
\hline
$  \sigma(p p \to\tilde\chi_2^0\tilde\chi_2^\pm) = 11\ (6)~{\rm fb} $ &
$  \sigma(p p \to\tilde\chi_3^0\tilde\chi_2^\pm) = 27\ (13)~{\rm fb} $ &
$  \sigma(p p \to\tilde\chi_4^0\tilde\chi_2^\pm) = 34\ (16)~{\rm fb} $\\
\hline 
\end{tabular}
\renewcommand{\arraystretch}{1.0}
\end{table}

\subsection{  $\tilde\chi_2^0\tilde\chi_2^\pm$ production
  \label{chi2chi2}}

We are now ready to present some numerical results. We first study the
production of the second lightest neutralino and chargino pair at the LHC,
$pp\to\tilde\chi_2^0\tilde\chi_2^+$, since we find the largest asymmetries
there; we will comment on channels involving the lighter chargino 
and heavier neutralino states in the next Subsection. 
For our reference scenario, the remaining $\tilde
\chi_2^\pm$ branching ratios, not listed in Table~\ref{tab:masses}, are ${\rm
  BR}(\tilde\chi_2^+\to \tilde e_L^+ \,\nu_e) = {\rm BR}(\tilde\chi_2^+\to
\tilde \mu_L^+\, \nu_\mu) \approx {\rm BR}(\tilde\chi_2^+\to \tilde\tau_2^+
\,\nu_\tau) = 12\%$, ${\rm BR}(\tilde\chi_2^+\to \tilde\nu_\tau\, \tau^+) = 12\%$,
${\rm BR}(\tilde\chi_2^+\to \tilde\chi_i^0\, W^+) = 17\%$, summed over
$i=1,2,3$, and ${\rm BR}(\tilde\chi_2^+\to \tilde\chi_1^+\, Z) = 10\%$, ${\rm
  BR}(\tilde\chi_2^+\to \tilde\chi_1^+\, h) = 2\%$~\cite{Kittel:2004rp}.

\begin{figure}[t]
  \centering
  \subfigure[]{
\put(2.2,7.5){ $\mathcal{A}[p_{u},p_{\bar d},p_{\tilde\chi_2^0},p_{\ell_1}]$ in \%}
    \includegraphics[width=0.45\textwidth]{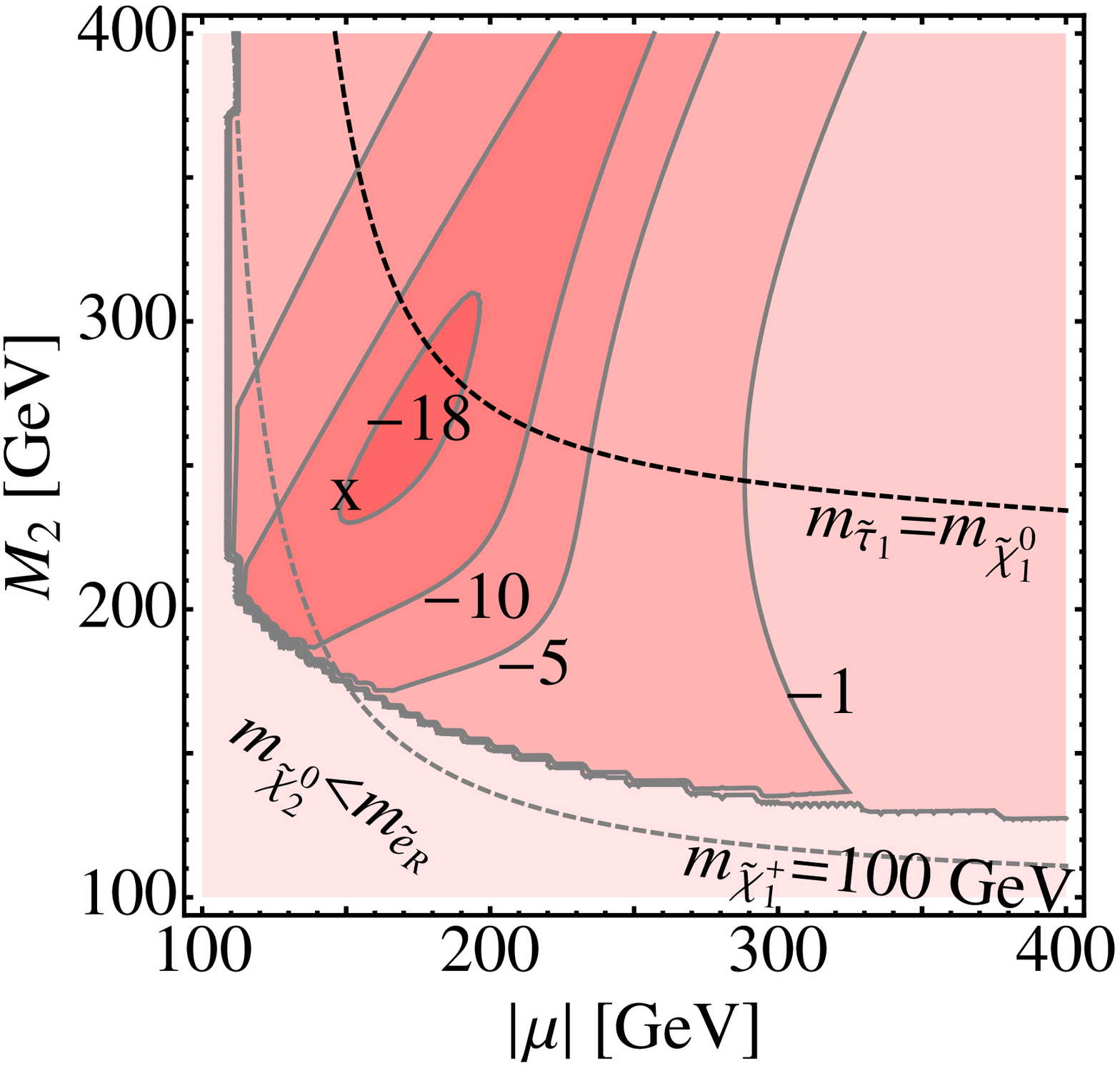}
  }
  \subfigure[]{
\put(2.6,7.5){$ \sigma(pp\to\tilde\chi_2^0\tilde\chi_2^+)$ in~fb}
    \includegraphics[width=0.45\textwidth]{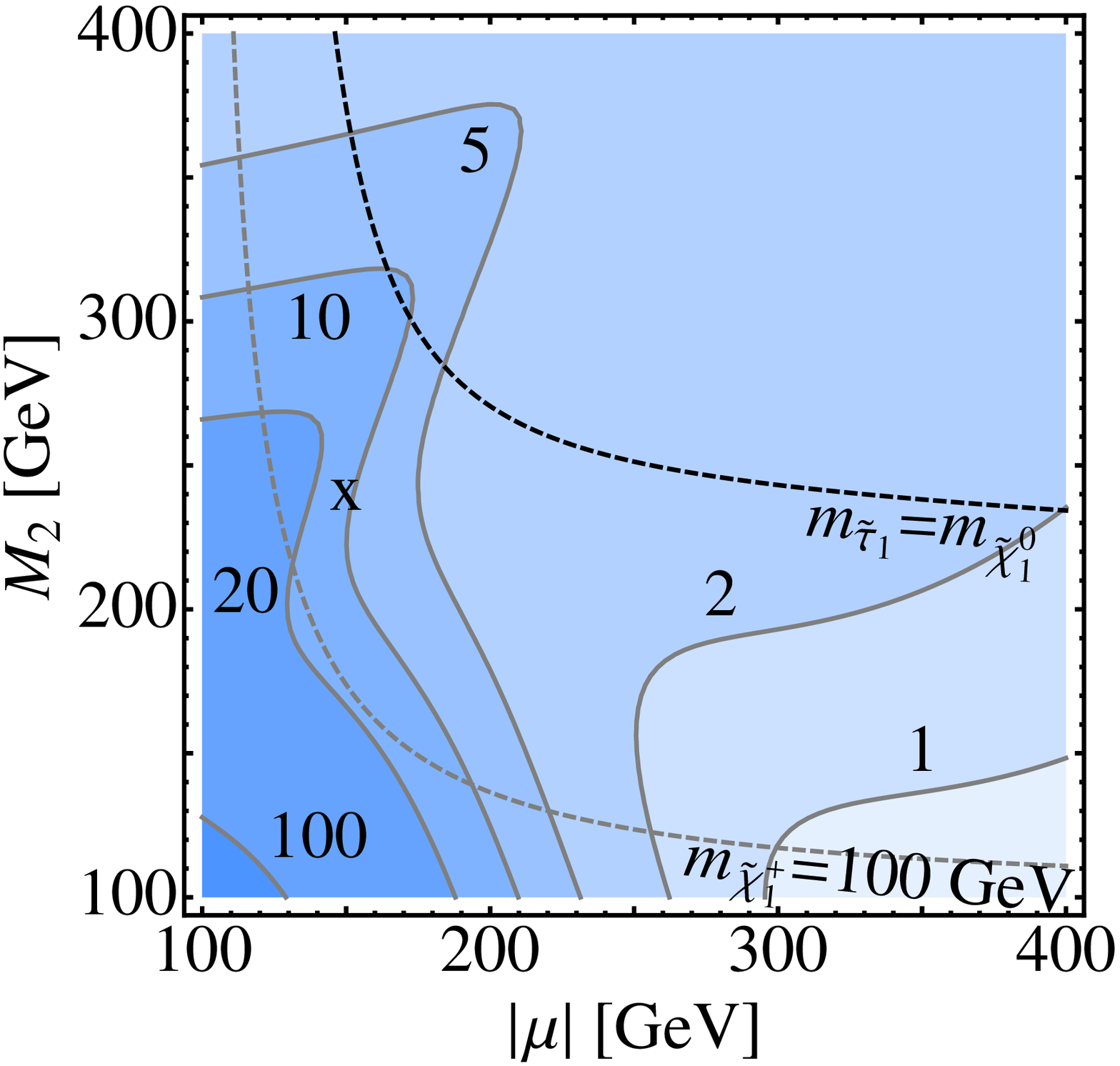}
  }
  \subfigure[]{
\put(2.2,7.5){${\rm BR}(\tilde\chi_2^0\to \tilde e_R^- \,e^+)$ in~$\%$}
    \includegraphics[width=0.45\textwidth]{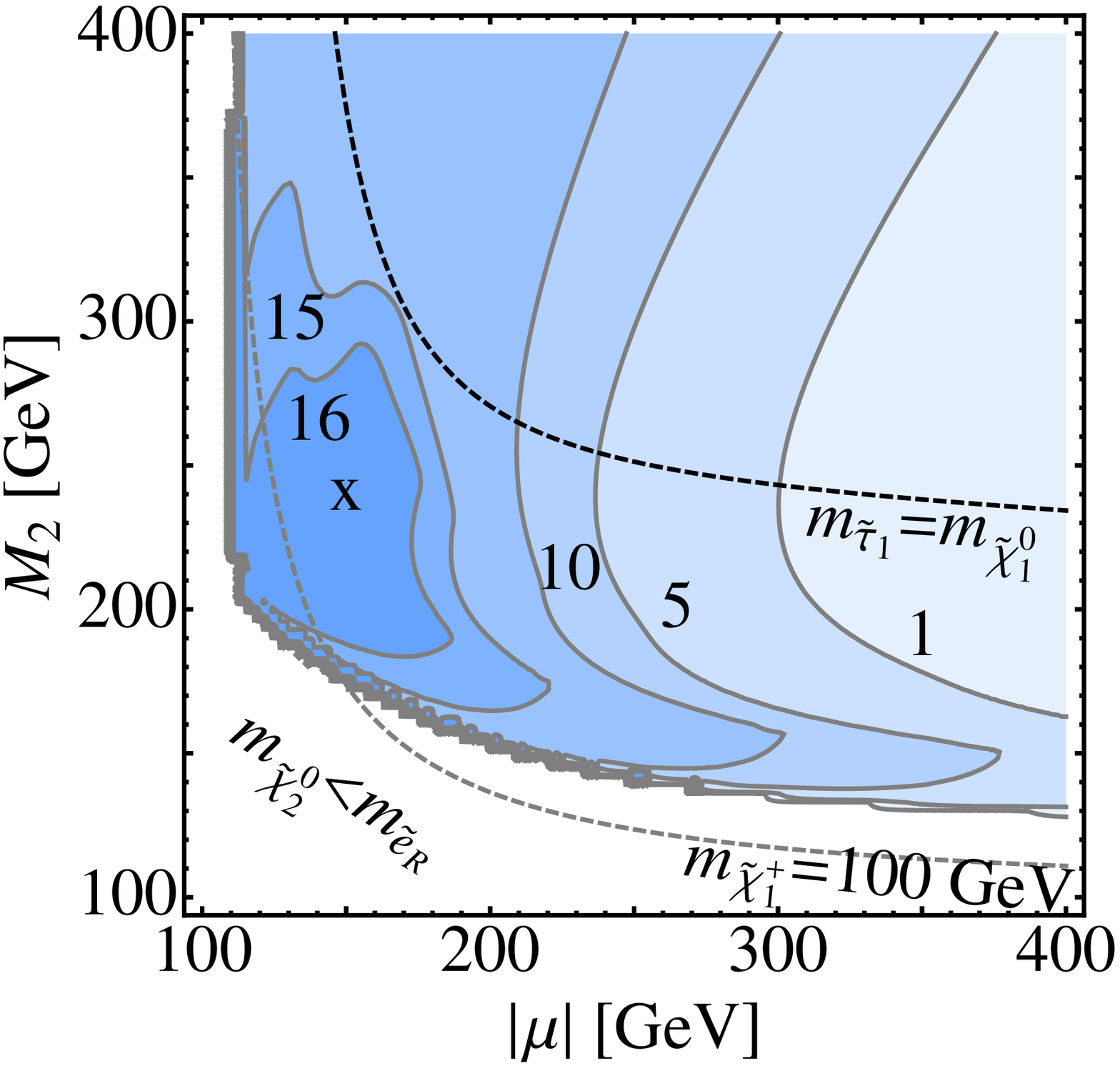}
  }
  \subfigure[]{
\put(2.6,7.5){${\rm BR}(\tilde\chi^+_2 \to  \tilde\nu_e \,e^+)$ in~$\%$}
    \includegraphics[width=0.45\textwidth]{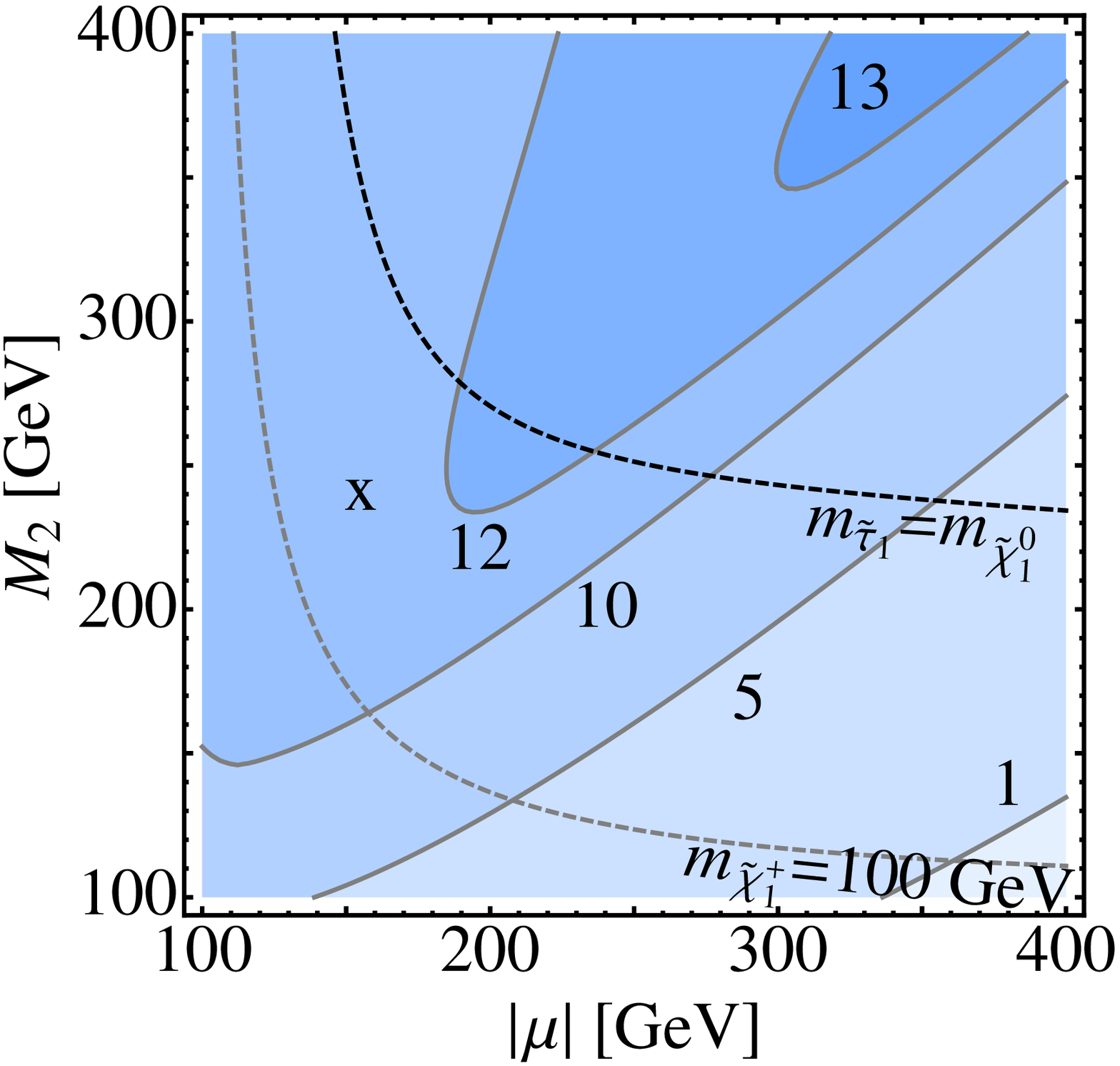}
  }
\caption{
      Contour lines in the $|\mu|$-$M_2$ plane for neutralino-chargino
      pair production $pp\to\tilde\chi_2^0\tilde\chi_2^+$ and subsequent 
      leptonic two-body decays
      $\tilde\chi^0_2 \to \tilde \ell_R^-\,\ell_1^+$;
      $\tilde\ell_R^- \to \tilde\chi^0_1 \,\ell_2^-$ and
      $\tilde\chi^+_2 \to \tilde\nu_\ell \,\ell_3^+ $
      at the LHC at $14$~TeV:
      (a)~CP asymmetry $\mathcal{A}[p_{u},p_{\bar d},p_{\tilde\chi_2^0},p_{\ell_1}]$
      in~$\%$, see Eq.~(\ref{eq:Toddasym}),
      (b)~the production cross section 
      $\sigma(pp\to\tilde\chi_2^0\,\tilde\chi_2^+)$ in~fb,
      (c)~the neutralino branching ratio
      ${\rm BR}(\tilde\chi_2^0\to \tilde e_R^-\, e^+)$ in~$\%$, and
      (d)~the chargino branching ratio
      ${\rm BR}(\tilde\chi^+_2 \to  \tilde\nu_e\, e^+)$ in~$\%$.
      The cross indicates the position of our SUSY benchmark
      scenario, see Table~\ref{tab:scenario}.
      The area above the black dashed line is excluded by
      $ m_{\tilde \tau_1} < m_{\tilde\chi_1^0}$.
      Below the gray dashed line we have $ m_{\tilde\chi^\pm_1} <100$~GeV.
\label{fig:mu_m2_asymm_neut}
      }
\end{figure}

\medskip

In Fig.~\ref{fig:mu_m2_asymm_neut}~(b), we show the $|\mu|$--$M_2$ dependence
of the cross section $\sigma(pp\to\tilde\chi_2^0\,\tilde\chi_2^+)$ for
neutralino-chargino pair production, which reaches several $10$~fb for light
neutralinos and charginos.  Our benchmark scenario is indicated by a cross in
the $|\mu|$--$M_2$ plane.  The neutralino and chargino branching ratios ${\rm
  BR}(\tilde\chi_2^0\to \tilde e_R^-\, e^+)$ and ${\rm BR}(\tilde\chi^+_2 \to
\tilde\nu_e \,e^+)$ are shown in Fig.~\ref{fig:mu_m2_asymm_neut}~(c), (d).
The (Lorentz invariant) asymmetry for the T-odd product ${\mathcal
  E}_{\tilde\chi^0_2} =[p_{u},p_{\bar d},p_{\tilde\chi^0_2},p_{\ell_1^+}]$,
which appears in the neutralino spin correlations, is shown in
Fig.~\ref{fig:mu_m2_asymm_neut}~(a).  For the asymmetry we use the short-hand
notation $\mathcal{A}[p_{u},p_{\bar d},p_{\tilde\chi_2^0},p_{\ell_1^+}] $.
That asymmetry will probe the CP-sensitive parts of the neutralino spin
correlations, as discussed in Section~\ref{CP asymmetries}.
 
\medskip

Looking at Fig.~\ref{fig:mu_m2_asymm_neut}~(a), we see that our
benchmark point lies on a line, approximately reaching from
$(|\mu|,M_2) =(100,150)$~GeV to $(250,400)$~GeV, where the asymmetry
obtains its maximum. In the vicinity of that line there is a
level-crossing of the neutralino states $\tilde\chi_2^0$ and
$\tilde\chi_3^0$ for $\phi_1 = 0$, and between $\tilde\chi_1^0$ and
$\tilde\chi_2^0$ for $\phi_1 = \pi$.  The level crossing of the
neutralinos is also reflected by the abrupt change in the cross
section, see Fig.~\ref{fig:mu_m2_asymm_neut}~(b).  Note that for
non-vanishing phases there is no true level-crossing since the
neutralino masses, \textit{i.e.} the singular values of the neutralino
matrix~\cite{Dreiner:2008tw}, are driven apart by the complex
off-diagonal terms.  

\medskip

%

\begin{figure}[t]
\centering
\includegraphics[clip,width=0.480\textwidth]{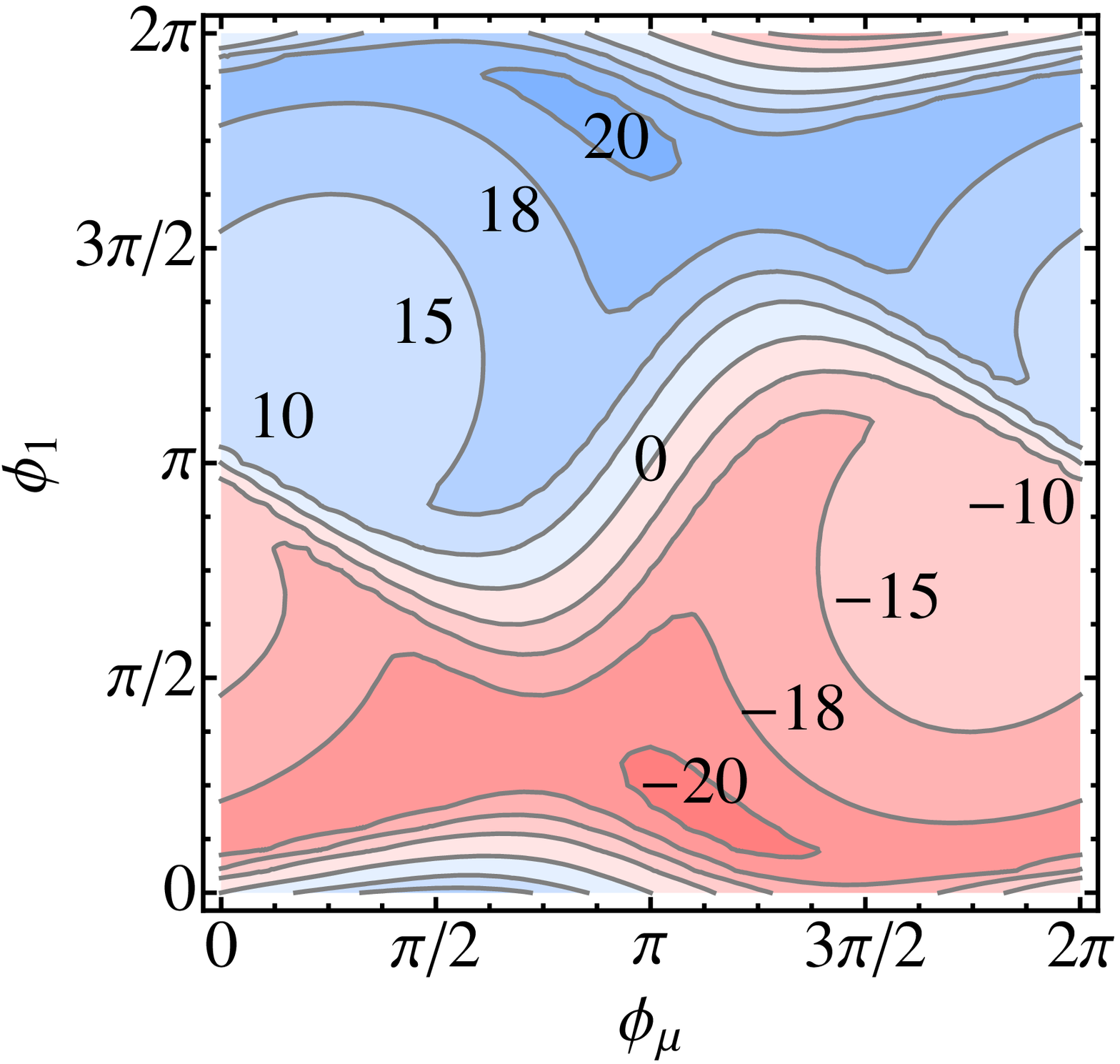}
\put(-5.5,8){ $\mathcal{A}[p_{u},p_{\bar d},p_{\tilde\chi_2^0},p_{\ell_1}]$ in \%}
\includegraphics[clip,width=0.480\textwidth]{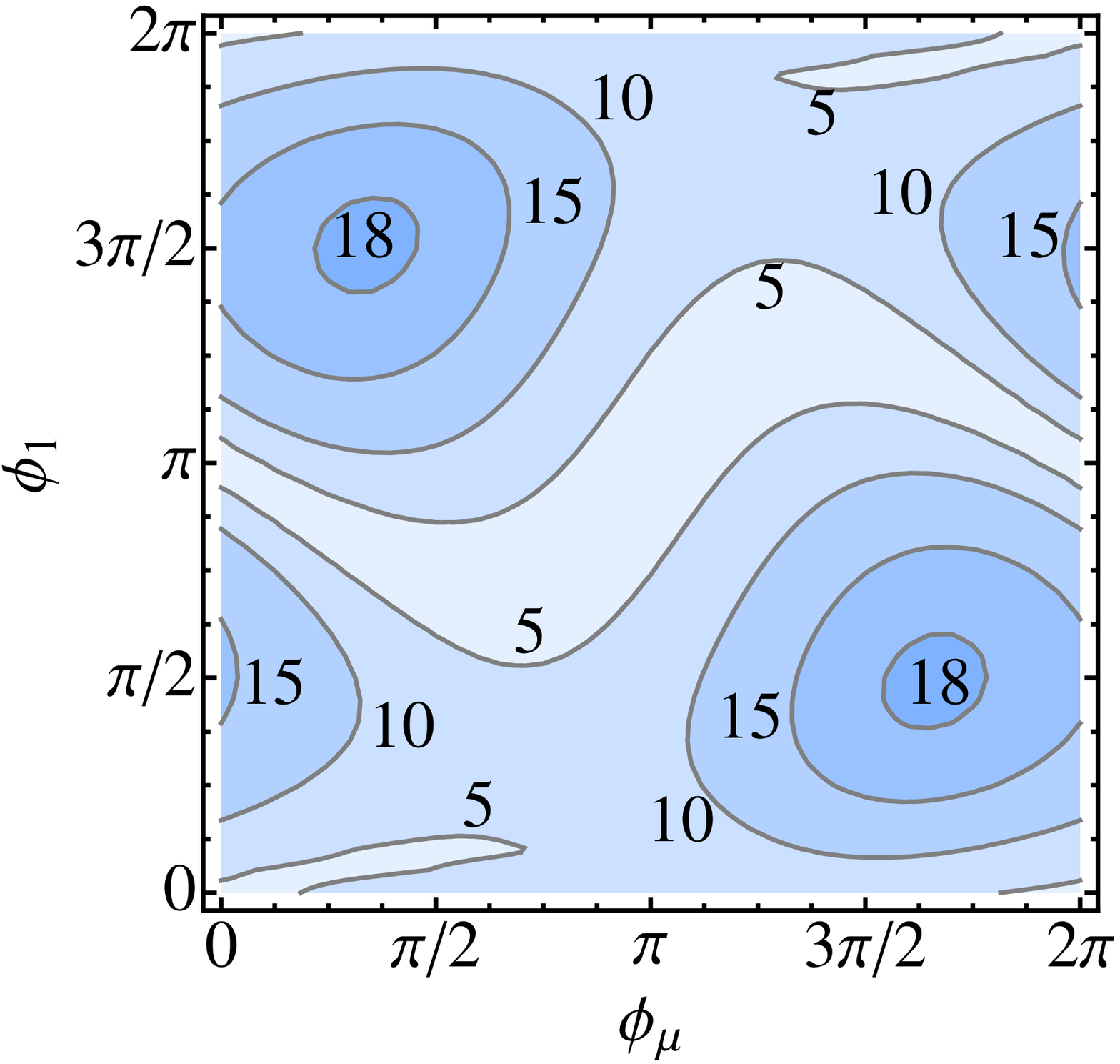}
\put(-5.5,8){$ \sigma(pp\to\tilde\chi_2^0\,\tilde\chi_2^+)$ in~fb}
\caption{
      Phase dependence of (left) the CP asymmetry 
      $\mathcal{A}[p_{ u},p_{\bar d},p_{\tilde\chi_2^0},p_{\ell_1^+}]$
      in~$\%$, see Eq.~(\ref{eq:Toddasym}), 
      and (right)~the production cross section 
      $\sigma(pp\to\tilde\chi_2^0\,\tilde\chi_2^+)$ in~fb,
      for neutralino-chargino
      pair production $pp\to\tilde\chi_2^0\tilde\chi_2^+$ and subsequent 
      leptonic two-body decays
      $\tilde\chi^0_2 \to \tilde \ell_R^-\,\ell_1^+$;
      $\tilde\ell_R^- \to \tilde\chi^0_1 \,\ell_2^-$ and
      $\tilde\chi^+_2 \to \tilde\nu_\ell\, \ell_3^+ $
      at the LHC at $14$~TeV.
      The SUSY parameters are given in Table~\ref{tab:scenario}.
}
\label{fig:phases_asymm_sig}
\end{figure}

\begin{figure}[t]
  \centering
  \subfigure[]{
    \includegraphics[width=0.45\textwidth]{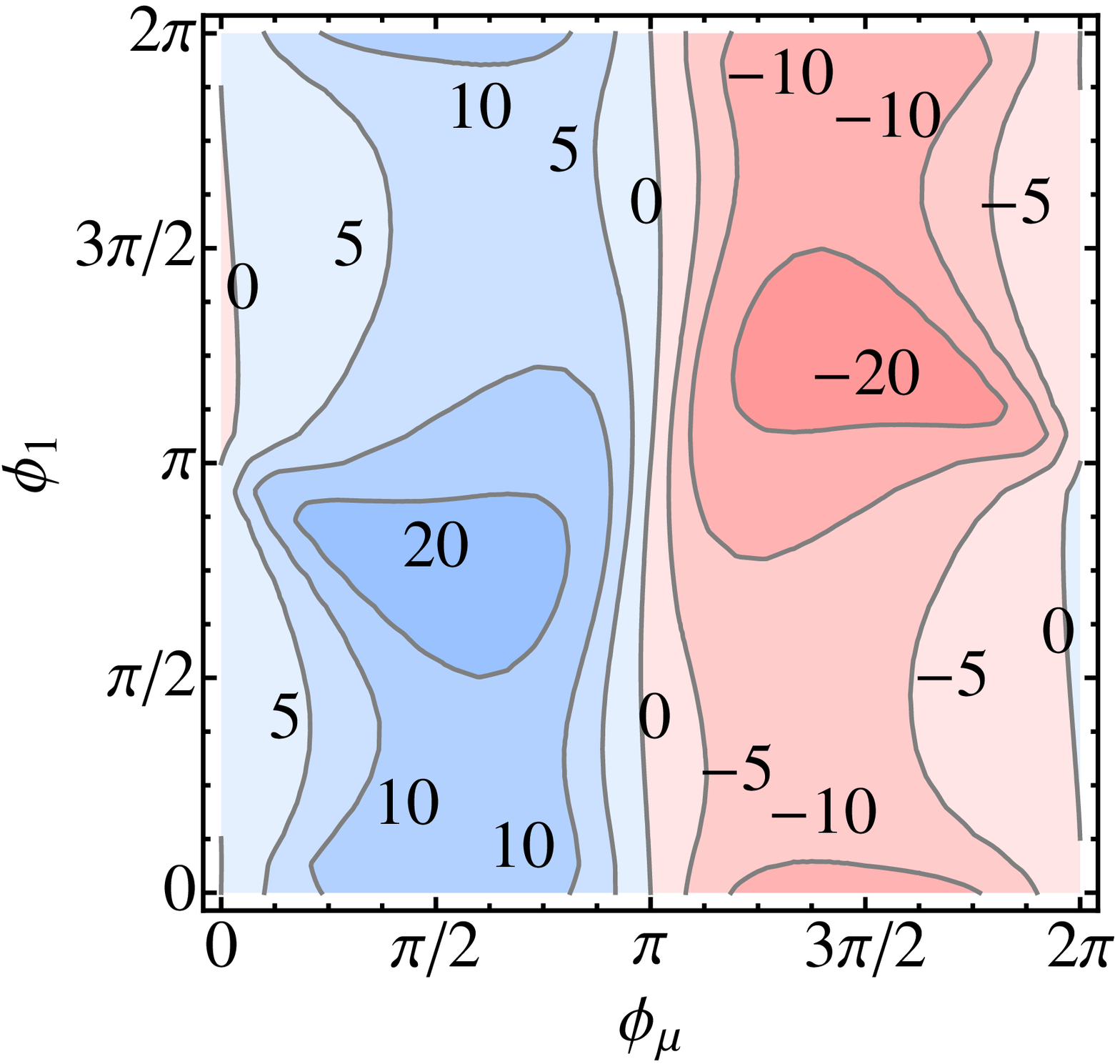}
\put(-4.5,7.5){$WW$ exchange}
  }
  \subfigure[]{
    \includegraphics[width=0.45\textwidth]{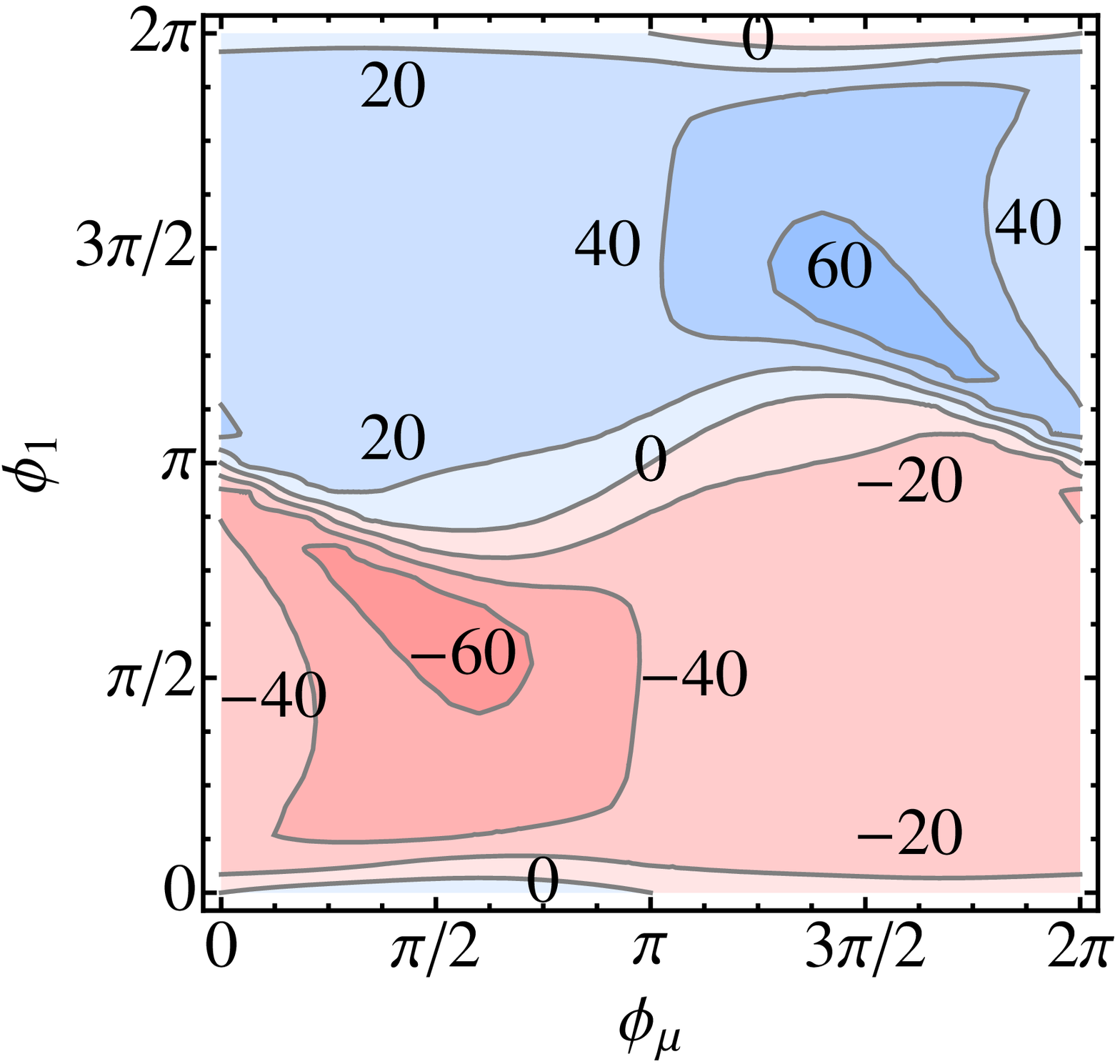}
\put(-4.7,7.5){$W\tilde u_L$ exchange}
  }
  \subfigure[]{
    \includegraphics[width=0.45\textwidth]{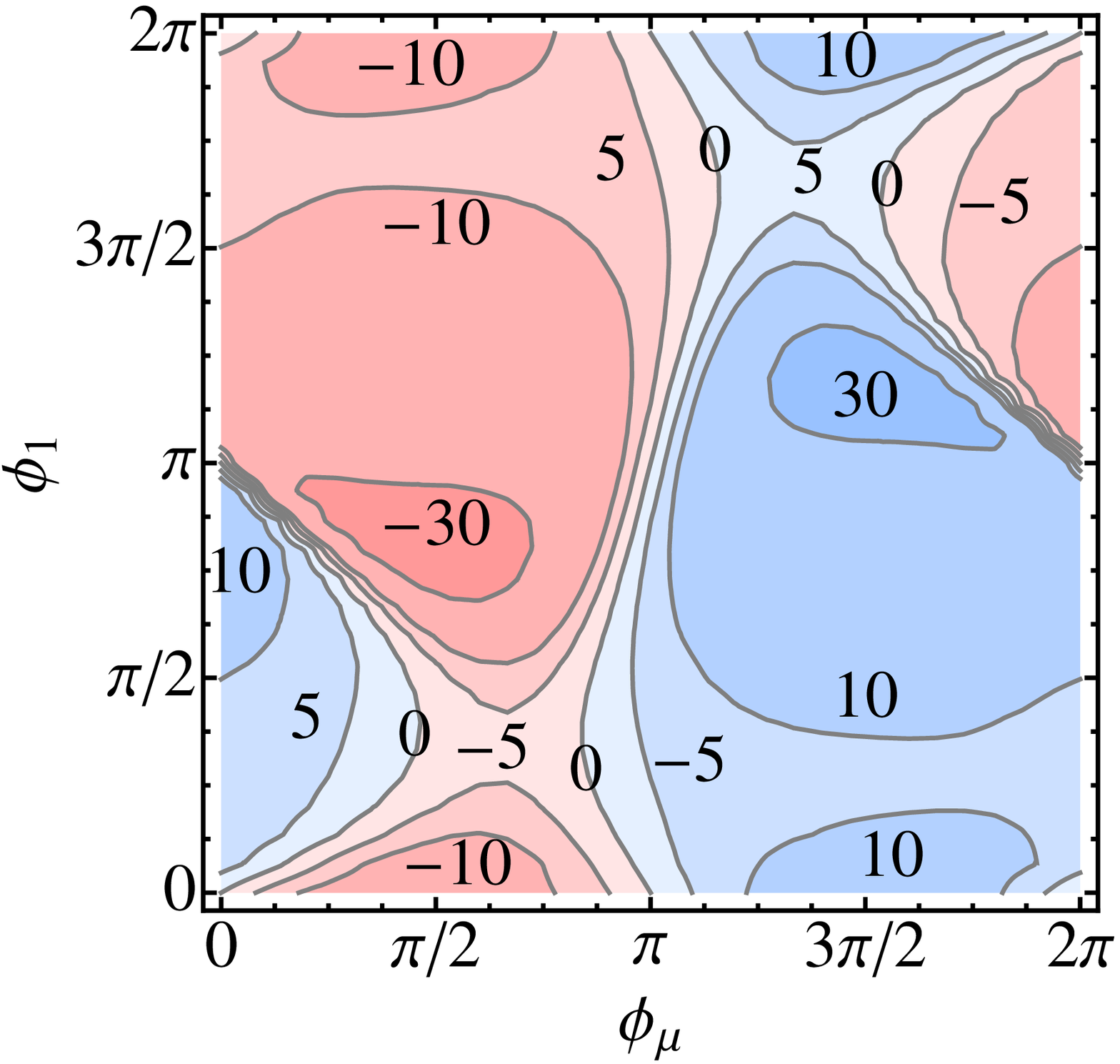}
 \put(-4.5,7.5){$W\tilde d_L$ exchange}
 }
  \subfigure[]{
    \includegraphics[width=0.45\textwidth]{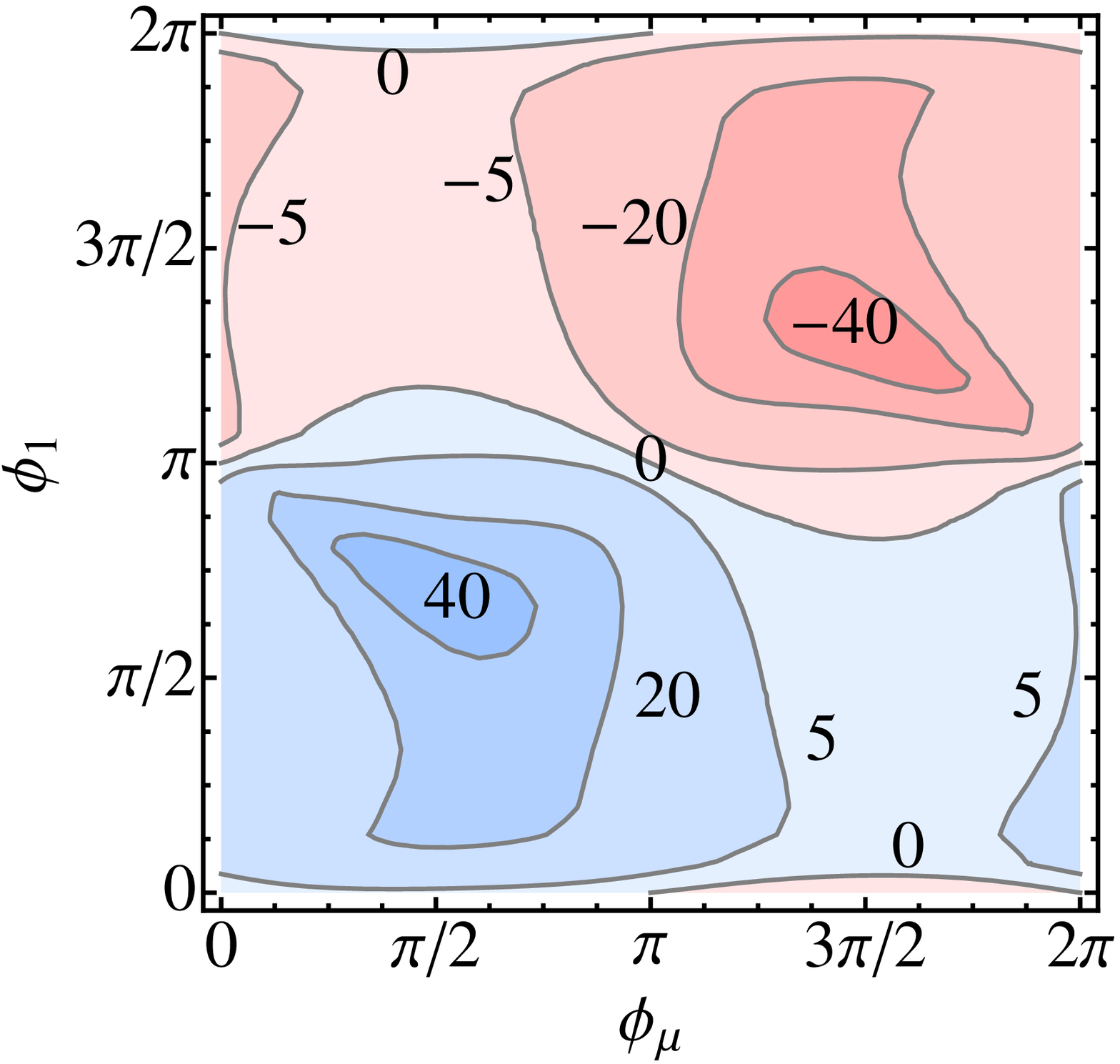}
\put(-4.7,7.5){$\tilde u_L \tilde d_L$ exchange}
  }
\caption{
 Phase dependence of the interference contributions (in~$\%$)
 to the CP asymmetry $\mathcal{A}[p_{u},p_{\bar d},p_{\tilde\chi_2^0},p_{\ell_1^+}]$,
      Eq.~(\ref{eq:Toddasym}), from
      (a)~$WW$ exchange,
      (b)~$W\tilde u_L$ exchange,
      (c)~$W\tilde d_L$ exchange, and
      (d)~$\tilde u_L \tilde d_L$ exchange, 
      for neutralino-chargino
      pair production $pp\to\tilde\chi_2^0\tilde\chi_2^+$ 
      and subsequent leptonic two-body decays
      $\tilde\chi^0_2 \to \tilde \ell_R^-\,\ell_1^+$;
      $\tilde\ell_R^- \to \tilde\chi^0_1 \,\ell_2^-$ and
      $\tilde\chi^+_2 \to \tilde\nu_\ell \,\ell_3^+ $
      at the LHC at $14$~TeV.
      The SUSY parameters are given in Table~\ref{tab:scenario}.
\label{fig:phases_asymm_part}
      }
\end{figure}

In Fig.~\ref{fig:phases_asymm_sig}, we show the phase dependence of
the asymmetry $\mathcal{A}[p_{ u},p_{\bar d},p_{\tilde\chi_2^0},p_{
\ell_1^+}]$ (left), and the corresponding neutralino-chargino cross 
section $\sigma(pp\to\tilde\chi_2^0\,\tilde\chi_2^+)$.  The asymmetry
receives contributions from the different production channels of $W$,
$\tilde u_L$, and $\tilde d_L$ exchange.  The contribution from $WW$
exchange is shown in Fig.~\ref{fig:phases_asymm_part}~(a), that of
$W\tilde u_L$ exchange in Fig.~\ref{fig:phases_asymm_part}~(b), the
$W\tilde d_L$ exchange in Fig.~\ref{fig:phases_asymm_part}~(c), and
finally the $\tilde u_L \tilde d_L$ exchange in
Fig.~\ref{fig:phases_asymm_part}~(d).  We can see that the individual
contributions can be as large as $\pm60\%$, however the different
contributions enter with opposite sign, and thus cancel each other
partly. See the sum of all their contributions to the
asymmetry $\mathcal{A}[p_{u},p_{\bar d},p_{\tilde\chi_2^0},p_{\ell_
1^+}]$ in Fig.~\ref{fig:phases_asymm_sig}~(left).

\medskip

\begin{figure}[t]
\centering
\includegraphics[clip,width=0.480\textwidth]{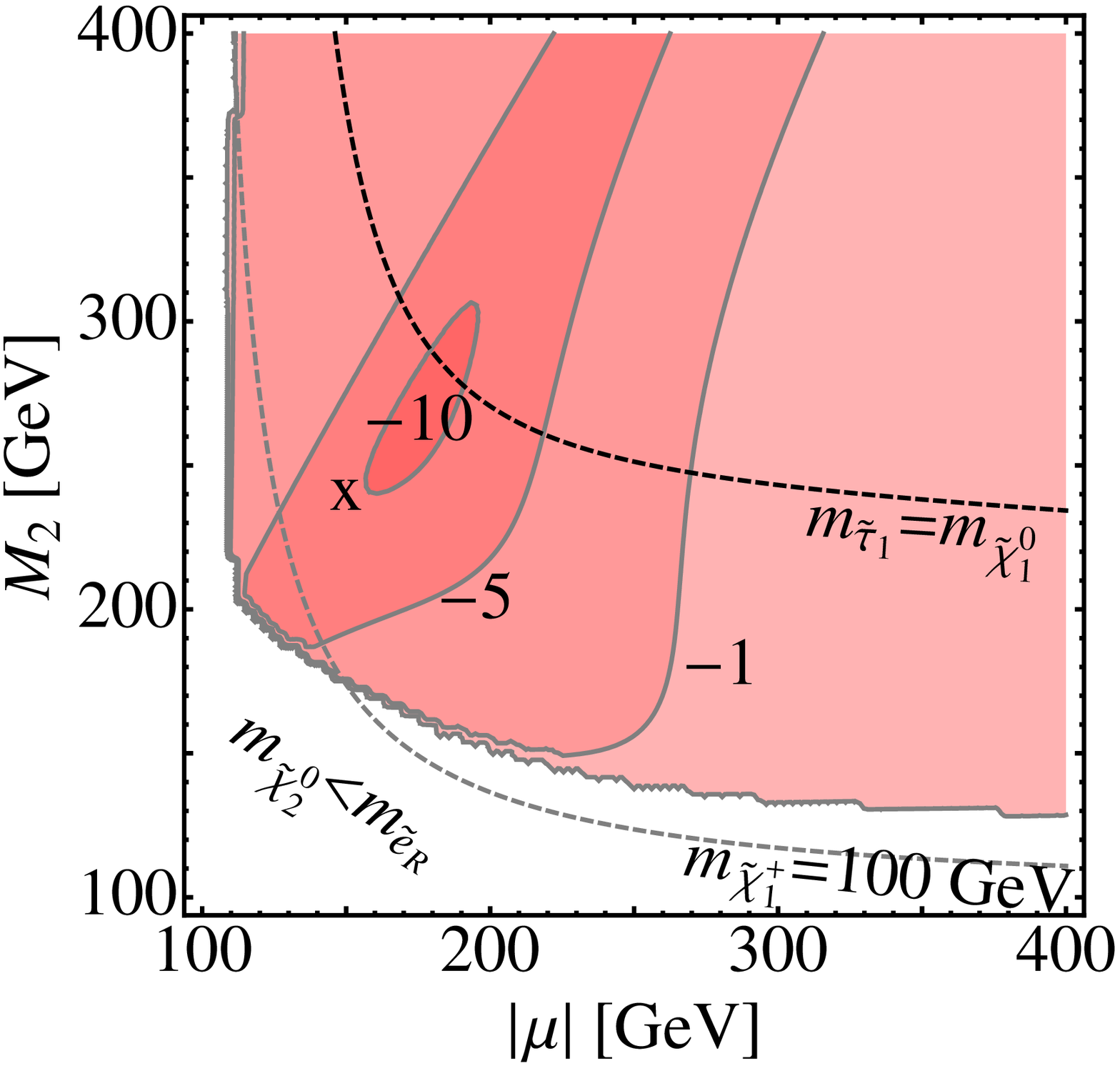}
\put(-5.5,8){ $\mathcal{A}[p_{u},p_{\bar d},p_{\tilde\chi_2^+},p_{\ell_3^+}]$ in \%}
\includegraphics[clip,width=0.480\textwidth]{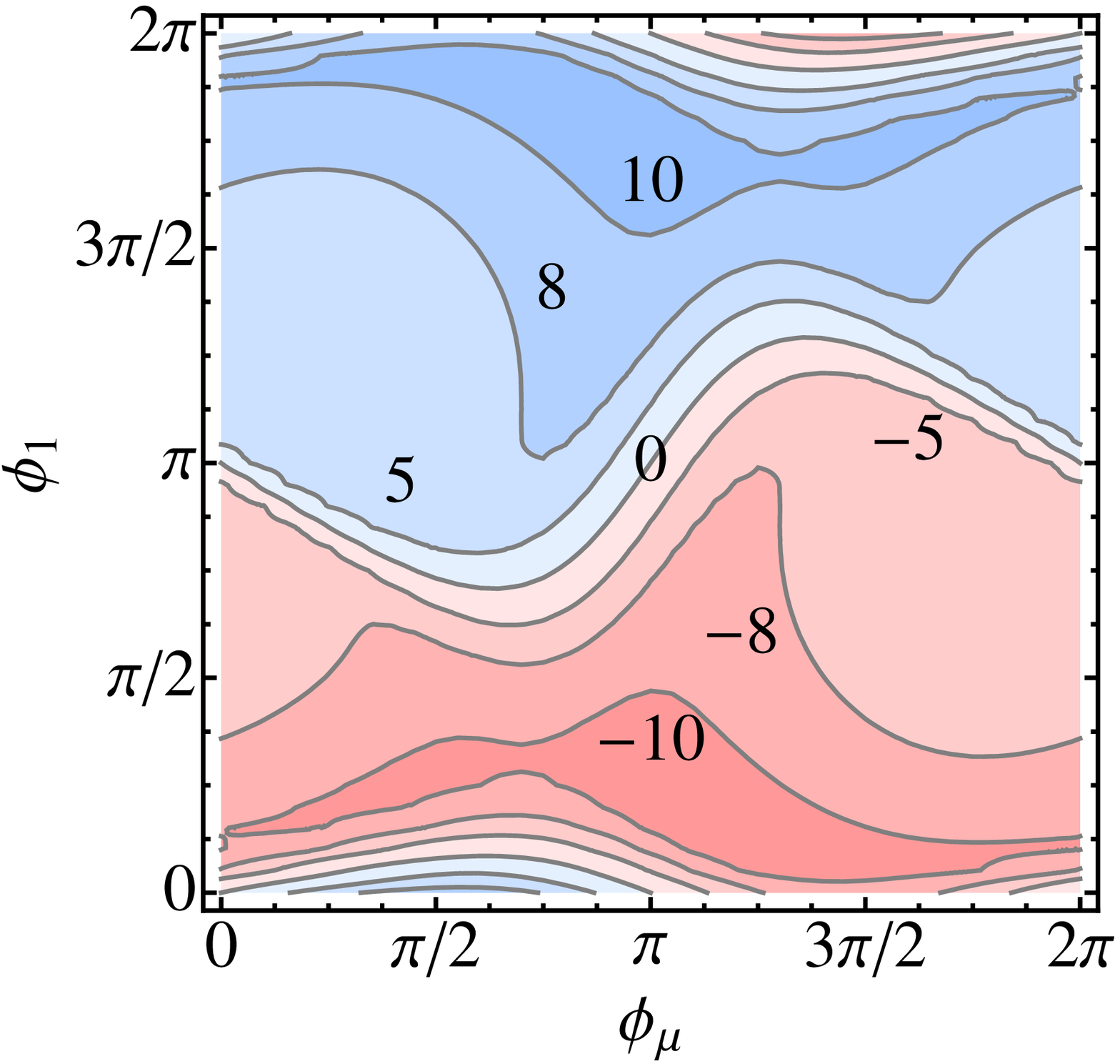}
\put(-5.5,8){$\mathcal{A}[p_{u},p_{\bar d},p_{\tilde\chi_2^+},p_{\ell_3^+}]$ in \%}
\caption{
      Dependence  of the CP asymmetry 
      $\mathcal{A}[p_{u},p_{\bar d},p_{\tilde\chi_2^+},p_{\ell_3^+}]$
      in~$\%$, see Eq.~(\ref{eq:Toddasym}), 
      on $|\mu|$-$M_2$ (left) and on the phases (right)
      for neutralino-chargino
      pair production $pp\to\tilde\chi_2^0\tilde\chi_2^+$ 
      and subsequent  leptonic two-body decays
      $\tilde\chi^0_2 \to \tilde \ell_R^-\,\ell_1^+$;
      $\tilde\ell_R^- \to \tilde\chi^0_1 \,\ell_2^-$ and
      $\tilde\chi^+_2 \to \tilde\nu_\ell \,\ell_3^+ $
      at the LHC at $14$~TeV.
      The SUSY parameters are given in Table~\ref{tab:scenario}.
      Compare with Figs.~\ref{fig:mu_m2_asymm_neut}~(a) and 
      \ref{fig:phases_asymm_sig}~(left), where the asymmetry 
      $\mathcal{A}[p_{u},p_{\bar d},p_{\tilde\chi_2^0},p_{\ell_1^+}]$
      of the neutralino spin correlations is shown.
}
\label{fig:asymm_char}
\end{figure}

In Fig.~\ref{fig:asymm_char}~(left), we show the $|\mu|$--$M_2$ dependence of
the asymmetry of the T-odd product ${\mathcal E}_{\tilde\chi^+_2}
=[p_{u},p_{\bar d},p_{\tilde\chi_2^+},p_{\ell_3^+}]$, which appears in the
chargino spin correlations.  For the asymmetry we use the short-hand notation
$\mathcal{A}[p_{u},p_{\bar d},p_{\tilde\chi_2^+},p_{\ell_3^+}] $, to indicate
the different momenta used.  In Fig.~\ref{fig:asymm_char}~(left), our
benchmark scenario is indicated by a cross in the $|\mu|$--$M_2$ plane.  For
that scenario we show the phase dependence of the asymmetry in the right
panel. By comparing its size with the corresponding asymmetry which probes the
neutralino spin correlations, see Figs.~\ref{fig:mu_m2_asymm_neut} and
\ref{fig:phases_asymm_sig}, for the $|\mu|$--$M_2$ dependence and the phase
dependence, respectively, we see that the asymmetry which probes the chargino
spin-correlations is about half as large. This is only due to kinematics, as
discussed in Section~\ref{CP asymmetries}, since the combinations of products
of imaginary couplings are the same for the neutralino and the chargino spin
correlations.  The asymmetry, $\mathcal{A}[p_{u}, p_{\bar d},
  p_{\tilde\chi_2^+}, p_{\ell_3^+}] $, which probes the chargino spin
correlations, receives contributions from the different production channels of
$W$, $\tilde u_L$ and $\tilde d_L$ exchange. The individual contributions
from $WW$, $W\tilde u_L$, $W\tilde d_L$ and $\tilde u_L \tilde d_L$ exchange
can be as large as $\pm30\%$ (not shown), however they enter with opposite
sign, such that they cancel each other partly in their sum for the asymmetry
$\mathcal{A}[p_{u},p_{\bar d},p_{\tilde\chi_2^0},p_{\ell_1^+}]$.  The
individual contributions have the same dynamical dependence on the CP phases
as those for the neutralino spin correlations, compare
Fig.~\ref{fig:asymm_char}~(right) with Fig.~\ref{fig:phases_asymm_sig}~(left),
and are only about a factor of $2$ smaller.


%

\begin{figure}[t]
\centering
\includegraphics[clip,width=0.480\textwidth]{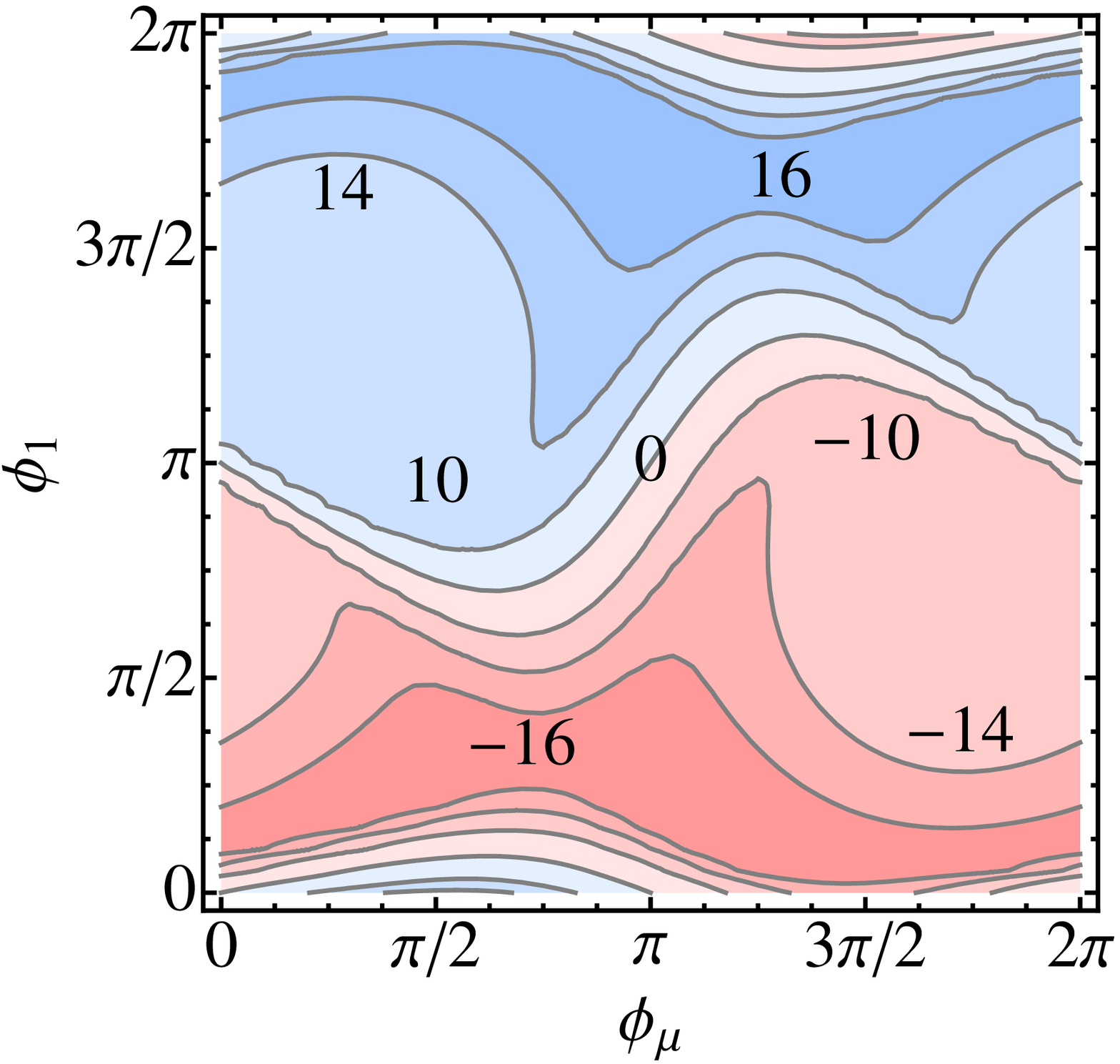}
\put(-4.5,8){$\mathcal{A}[f]$ in $\%$ }
\includegraphics[clip,width=0.480\textwidth]{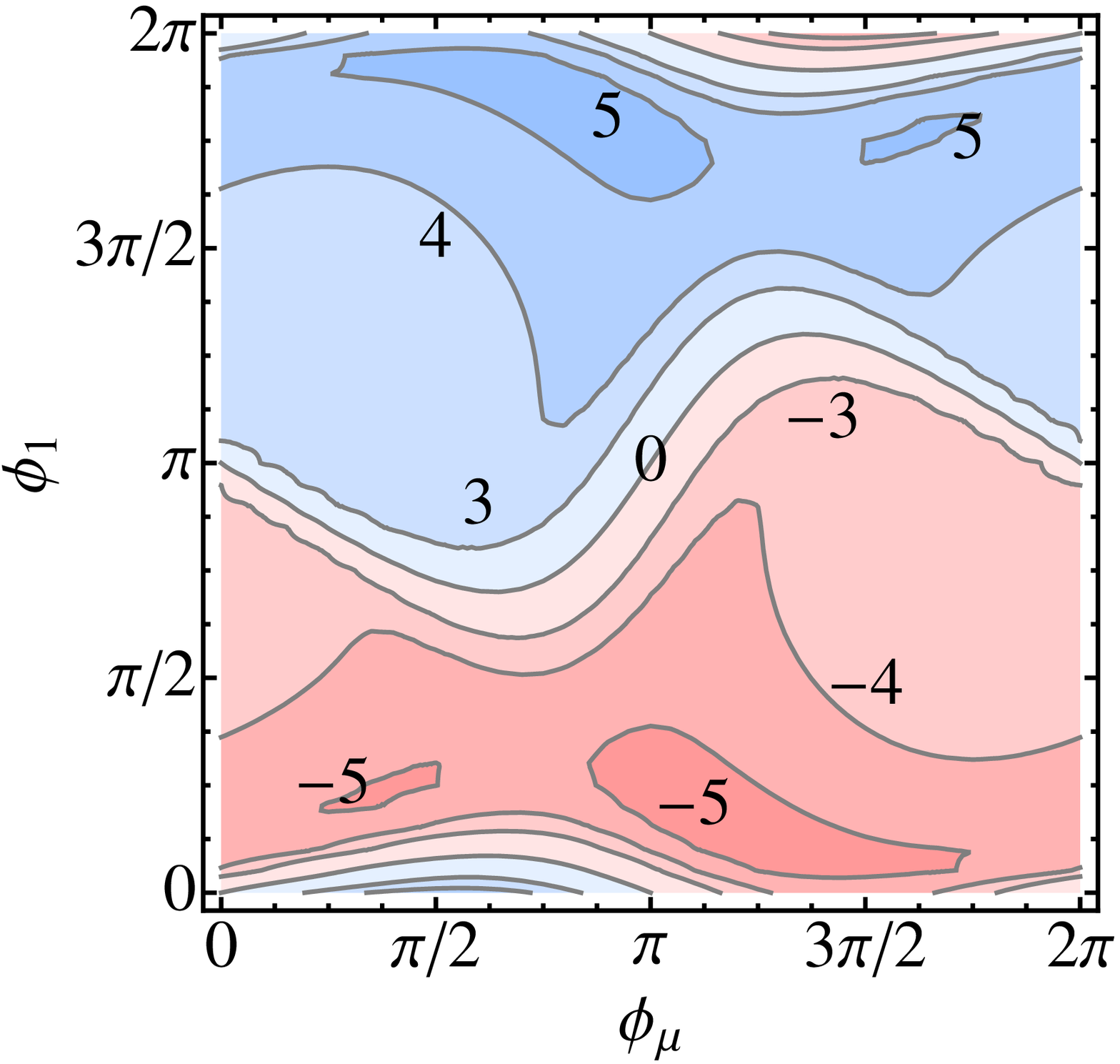}
\put(-5.6,8){$\mathcal{A}(\mathbf{p}_{\ell^+}, \, \mathbf{p}_{\ell^-}, \, \mathbf{p}_{\ell_3^+})$ in~$\%$}
\caption{
      Phase dependence of (left) the CP asymmetry 
      $\mathcal{A}[f]$ in $\%$,
       and (right) the triple product asymmetry
$\mathcal{A}(\mathbf{p}_{\ell^+}, \, \mathbf{p}_{\ell^-}, \, \mathbf{p}_{\ell_3^+})$
      in~$\%$, see Eq.~(\ref{eq:Toddasym}), 
      for neutralino-chargino
      pair production $pp\to\tilde\chi_2^0 \, \tilde\chi_2^+$ and subsequent 
      leptonic two-body decays
      $\tilde\chi^0_2 \to \tilde \ell_R^\mp \, \ell^\pm$;
      $\tilde\ell_R^\mp \to \tilde\chi^0_1 \, \ell^\mp$ and
      $\tilde\chi^+_2 \to \tilde\nu_\ell \, \ell_3^+ $
      at the LHC at $14$~TeV.
      The SUSY parameters are given in Table~\ref{tab:scenario}.
      Compare their sizes with the other asymmetries in
      Figs.~\ref{fig:phases_asymm_sig}~(left) and 
      \ref{fig:asymm_char}~(right).
}
\label{fig:phases_asymm_spinspin}
\end{figure}

\medskip

In Fig.~\ref{fig:phases_asymm_spinspin}~(left), we show the phase dependence
of the CP asymmetry $\mathcal{A}[f]$, which probes the CP-sensitive parts $f$,
see Eq.~(\ref{eq:xf2}), of the spin-spin correlations in neutralino-chargino
production and decay.  That asymmetry is Lorentz-invariant, in contrast to the
triple product asymmetry $\mathcal{A}(\mathbf{p}_{\ell_1},\mathbf{p}_{\ell_2
},\mathbf{p}_{\ell_3})$, see Eq.~(\ref{tripleterm}), which we show in
Fig.~\ref{fig:phases_asymm_spinspin}~(right).  Since the triple product
asymmetry is not Lorentz invariant, and since it is not an optimal observable
in the sense that the triple product does not match the kinematical dependence
of the CP-sensitive terms in the spin or spin-spin correlations, it is greatly
reduced compared to the other asymmetries. In Section~\ref{construct}, we will 
discuss alternatives to the triple product asymmetry.

\medskip

\begin{figure}[t]
\centering
\begin{picture}(16,8)
\put(-4,-20){\includegraphics{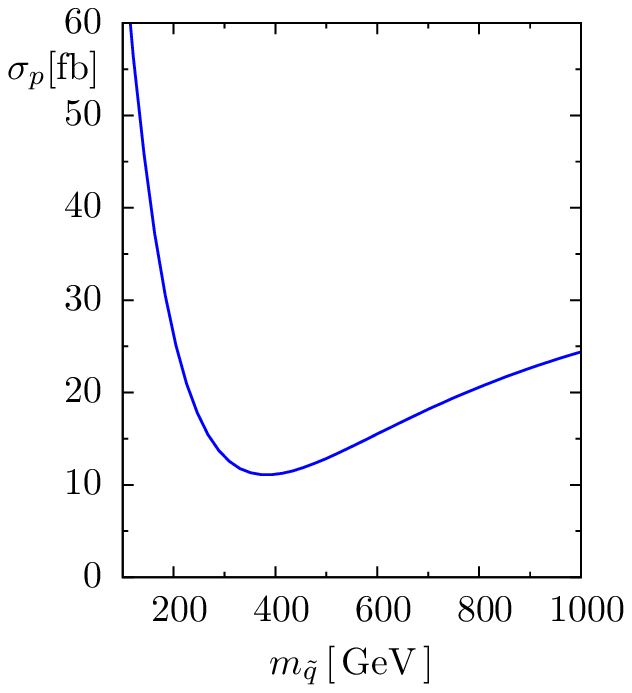}}
\put(5.0,-20){\includegraphics{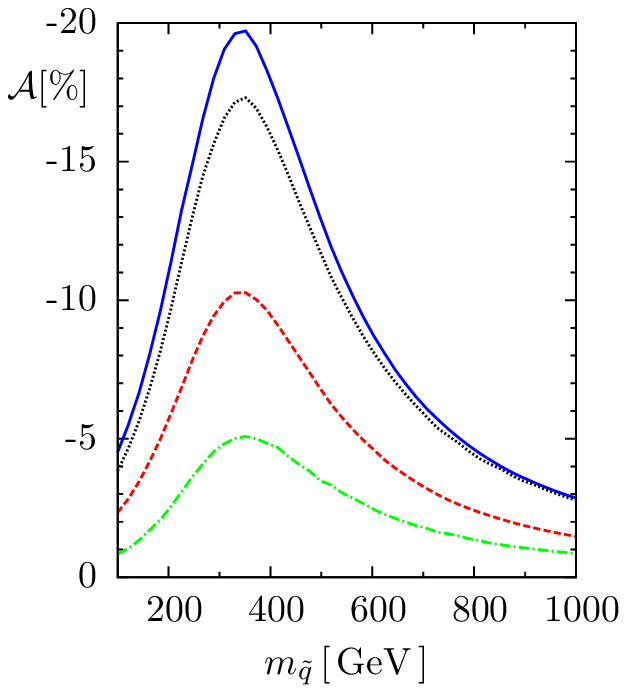}}
\end{picture}
\vspace{.2cm}
\caption{
     Squark mass dependence $m_{\tilde q} = m_{\tilde u_L}=m_{\tilde d_L}$ of 
      (left) the cross section 
      $\sigma_p(pp\to\tilde\chi_2^0\,\tilde\chi_2^+)$ 
      for neutralino-chargino pair production at the LHC at $14$~TeV;
      (right) of the different CP asymmetries, Eq.~(\ref{eq:Toddasym}),
      for the subsequent leptonic two-body decays
      $\tilde\chi^0_2 \to \tilde \ell_R^-\,\ell_1^+$;
      $\tilde\ell_R^- \to \tilde\chi^0_1 \,\ell_2^-$ and
      $\tilde\chi^+_2 \to \tilde\nu_\ell \,\ell_3^+ $,
      $\mathcal{A}[p_{u},p_{\bar d},p_{\tilde\chi^0_2},p_{\ell_1^+}]$
      (solid, blue),
      $\mathcal{A}[f]$
      (dotted, black),
      $\mathcal{A}[p_{u},p_{\bar d},p_{\tilde\chi_2^+},p_{\ell_3^+}]$
      (dashed, red),
$\mathcal{A}(\mathbf{p}_{\ell^+}, \, \mathbf{p}_{\ell^-}, \, \mathbf{p}_{\ell_3^+})$
      (dot-dashed, green).
       See the other SUSY parameters in Table~\ref{tab:scenario}.
}
\label{fig:asym_sigma_mq}
\end{figure}

In Fig.~\ref{fig:asym_sigma_mq}, we show the dependence of the cross section
$\sigma_p(pp\to\tilde\chi_2^0\,\tilde\chi_2^+)$ (left), and of the four
different asymmetries (right) on the squark masses, $m_{\tilde q} = m_{\tilde
  u_L}=m_{\tilde d_L}$.  Since the different interference contribution depend
sensitively on the squark masses, also the cross section and the asymmetries
are very sensitive. For increasing squark masses, the $W$ boson contribution
dominates, which results in the asymptotic values
$\sigma_p(pp\to\tilde\chi_2^0\,\tilde\chi_2^+)\to 35$~fb,
whereas the asymmetries almost vanish in this limit.

\clearpage
\newpage

\subsection{Remarks on  $\tilde\chi_2^0\,\tilde\chi_1^\pm$ and  
                     $\tilde\chi_3^0\,\tilde\chi_1^\pm$ production
  \label{chi1chi2}}
%

As we can see from Table~\ref{tab:masses}, the cross sections for
$pp\to\tilde\chi_{2,3}^0\,\tilde\chi_1^\pm$ production are much larger than the
$pp\to\tilde\chi_{2,3}^0\,\tilde\chi_2^\pm$ cross sections.\footnote{Note that
  the $pp\to\tilde\chi_{2,3}^0\,\tilde\chi_1^\pm$ channels can wash out the
  strong CP signal from $\tilde\chi_2^0\,\tilde\chi_2^\pm$ production, if the
  different $\tilde\chi_j^0\,\tilde\chi_i^\pm$ modes are not separated.
  However, for the present benchmark scenario, the pollution from the
  $\tilde\chi_1^\pm$ three-body decays can probably be made quite small, in
  particular if final states are excluded where all three leptons have the
  same flavor. Since $\tilde\chi_1^\pm$ decays produce rather soft leptons,
  they can be suppressed by putting a cut on the energy or $p_{\rm T}$ of the
  unpaired lepton $\ell_3$. Similarly, one may be able to suppress
  contributions from $\tilde\chi_{3,4}^0\,\tilde\chi_2^\pm$ contributions
  through cuts on the paired leptons $\ell_1\ell_2$. In particular, the
  neutralino two-body decay chain has a well-defined upper edge of the $\ell_1
  \ell_2$ invariant mass distributions, with most events not too far from this
  edge. A cut on the $\ell_1 \ell_2$ invariant mass should therefore suppress
  unwanted contributions. Depending on the masses, this could also suppress
  $\tilde\chi_2^0\to \tilde \ell_L \,\ell$, if open.}
However, in scenarios with $ M_{\tilde E}^{\tilde \ell} <  M_{\tilde
  L}^{\tilde \ell} $, it is difficult to simultaneously satisfy the
inequalities
\begin{equation}
 m_{\tilde{\ell}_R} < m_{\tilde\chi^0_2} < m_{\tilde{\ell}_L} 
  \qquad  {\rm and} \qquad  m_{\tilde\nu}< m_{\tilde\chi^\pm_1} \, ,\qquad (\ell=e,\mu);
\end{equation}
these are needed to enable the neutralino and chargino two-body decays,
Eqs.~(\ref{eq:decayNeut}), (\ref{eq:decayChar}), respectively, while
suppressing undesirable $\tilde \chi_2^0 \rightarrow \tilde \ell_L \,\ell$ decays.
These decays, if allowed, reduce the branching ratio ${\rm
  BR}(\tilde\chi_2^0\to \tilde \ell_R\, \ell)$ we are interested in. Moreover,
the CP asymmetry for the neutralino ${\mathcal A}[p_u, p_d,
  p_{\tilde\chi_i^0}, p_{\ell_1}]$ changes sign with an intermediate $\tilde
\ell_L$ compared to an intermediate $\tilde \ell_R$ in the neutralino decay
chain~\cite{Bartl:2003tr,Bartl:2009pg}. Thus the two contributions, from
intermediate $\tilde \ell_R$ and $\tilde \ell_L$, will tend to cancel, if they
cannot be distinguished experimentally on an event-by-event
basis.\footnote{Clearly the size of the branching ratios, and hence the degree
  to which these two contributions cancel, also depend also on the wino/bino
  admixtures of~$\tilde\chi^0_{i}$.} Note that in the MSSM one generally has
$m_{\tilde \chi_2^0} \gsim m_{\tilde \chi_1^\pm}$, while $SU(2)$ gauge
invariance implies $m_{\tilde \ell_L} \sim m_{\tilde \nu}$. Thus some fine-tuning
of parameters is needed to achieve $m_{\tilde\chi^\pm_1} > m_{\tilde\nu}$
without getting also automatically $m_{\tilde\chi^0_2} > m_{\tilde{\ell}_L} $.

\medskip

As an example of a somewhat fine-tuned scenario, we keep the hierarchy
$M_{\tilde L}^{\tilde \ell} > M_{\tilde E}^{\tilde \ell}$ as in our benchmark
scenario, Table~\ref{tab:scenario}, but choose a slightly smaller $M_{\tilde
  L}^{\tilde \ell} = 120$~GeV (and close to $M_{\tilde R}^{\tilde \ell} =
110$~GeV), which enables the two body decay $\tilde\chi^\pm_1 \to \tilde\nu_\ell
\, \ell^\pm$.  We also adopt the phase $\phi_1 = 0.9 \pi$, which gives maximal
asymmetries for this scenario. We then scan the parameter space around that
modified benchmark point. Although the production cross section $pp \to
\tilde\chi_2^0 \, \tilde\chi_1^+$ reaches up to $2$~pb for small $M_2, \,
|\mu|\lsim 200$~GeV, the asymmetry $\mathcal{A}[p_u, p_{\bar d},
  p_{\tilde\chi_2^0}, p_{\ell_1^+}]$ does not exceed $0.5\%$. The neutralino
branching ratio ${\rm BR}(\tilde\chi_2^0\to \tilde e_R^-\, e^+)$ gets
slightly reduced since now the $\tilde \nu$ are lighter than $\tilde \ell_R$,
and reaches not more than $10\%$. The chargino branching ratio can reach ${\rm
  BR}(\tilde\chi^+_1 \to \tilde\nu_e\, e^+) = 33\%$, since only the chargino
decay channels into leptons are open. Similarly, the $pp \to \tilde\chi_3^0
\,\tilde\chi_1^+$ production cross section reaches $1$~pb, with ${\rm
  BR}(\tilde\chi_3^0\to \tilde e_R^-\, e^+) < 10\%$, but the asymmetries are
again small, $\mathcal{A}[p_u,p_{\bar d},p_{\tilde\chi_2^0},p_{\ell_1^+}]\lsim
1 \%$.

\medskip

In a second approach, we invert the hierarchy between the left and right
slepton states, taking $M_{\tilde L}^{\tilde \ell} < M_{\tilde E}^{\tilde
  \ell}$. Then we choose $M_{\tilde E}^{\tilde \ell}$ sufficiently heavy to
close the channel $\tilde\chi_2^0\to \tilde \ell_R \, \ell$, but still allow for
$\tilde\chi_1^\pm\to \tilde\nu_\ell\, \ell^\pm$ and $\tilde\chi_2^0\to \tilde
\ell_L \,\ell$, such that there is only the contribution from the left slepton
to the neutralino asymmetry, and no cancellations appear. However we cannot
find asymmetries larger than 
$\mathcal{A}[p_u,p_{\bar d}, p_{\tilde\chi_2^0},  p_{\ell_1^+}] > 0.5\%$.

\subsection{Note on other chargino and neutralino decays
  \label{threebody}}
%

So far we have assumed rather large CP phases and small sfermion masses. This
will generate too large electric dipole moments of SM fermions, unless the
first and second generation $A$-parameters are fine-tuned accurately in both
size and phase. This could be ameliorated by choosing sufficiently large slepton
masses. The phase sensitivity of spin and spin-spin correlations could then be
studied via chargino and neutralino decays into real $Z^0$ and $W^\pm$ gauge
bosons; background reduction and charge determination would again force one to
focus on purely leptonic decays of the gauge bosons.

\medskip

However, the asymmetries for the neutralino decay $\tilde\chi_i^0 \to Z
\,\tilde\chi_1^0$ with the subsequent leptonic decay $Z \to \ell\,\bar\ell$
are reduced by a factor $f_Z=( |L|^2 -|R|^2)/( |L|^2 +|R|^2)\approx0.1
5$, where $L \ (R)$ is the left (right) SM coupling of the $Z$ to
charged leptons. Thus one loses almost an order of magnitude in all
neutralino spin asymmetries.  In addition, the overall statistical
significance is reduced, since $\rm{BR}(Z\to \ell\,\bar\ell)\approx 6.7
\%$ summed over $\ell=e,\mu$. This reduces the total cross section for
the purely leptonic final state, see the discussion at the end of
Section~$3$ in Ref.~\cite{Bartl:2004ut}. Due to the Majorana
properties of the neutralinos, the neutralino spin asymmetries
identically vanish for decays into a Higgs bosons, $\tilde\chi_i^0 \to
H \,\tilde\chi_1^0$, since the couplings obey $|c_L|^2 =|c_R|^2$.

\medskip

A similar argument holds for the chargino decay into a $W$ boson,
$\tilde\chi_i^\pm \to \tilde\chi_1^0 \,W^\pm$. If the $W$ momentum is
reconstructed, which is in principle possible for hadronic decays, the
reduction factor $f_W = ( |l|^2 -|r|^2)/( |l|^2 +|r|^2)$ to the chargino spin
asymmetries due to the left $(l)$ and right $(r)$
$\tilde\chi_i^\pm$-$\tilde\chi_1^0$-$W$ couplings is typically of the order of
$0.2$ to $0.4$~\cite{Kittel:2005ma}. For leptonic decays the overall
statistical significance is reduced due to the branching ratio $\rm{BR}(W \to
\ell\nu_\ell)\approx 21\%$, for $\ell=e,\mu$.  Lastly, for the chargino decay
via a $W$, and the neutralino decay via an on-shell $Z$ boson, one would have
to fight the large SM background from $pp\to ZW^\pm$ production, of order
$16$~pb~\cite{ATLAS}. Thus maximally P-violating chargino and neutralino decays, 
which can be realized only if sleptons are light, are ideal for analyzing the 
CP-violating effects in chargino-neutralino production.

\medskip

There are also regions of parameter space where some of the charginos and
neutralinos only have three-body decays. This happens if the sleptons (and
squarks) are heavier than the charginos and neutralinos in question; in
addition, the mass difference between these states and the lightest neutralino
has to be smaller than the masses of the $W$ and $Z$ boson.  Three-body
decays, which proceed via the exchange of virtual sfermions, gauge and Higgs
bosons, could then provide additional CP-violating contributions to the 
asymmetries~\cite{NEUT3,CHAR3}. (This also happens for two-body decays into 
spin-$1$ $Z$~\cite{Bartl:2004ut} and $W$~\cite{Kittel:2004kd} bosons.) Although that
could be interesting, as those CP-violating contributions can be of the order
of $10\%$~\cite{Bartl:2004ut,NEUT3,CHAR3,Kittel:2004kd,Bartl:2003ck, MoortgatPick:2009jy}, 
it could be more difficult to disentangle the two
different CP-violating contributions from production and decay. Since the
 three-body decay scenario would require quite different calculations
and phenomenology, we defer its analysis to another work.

\section{Constructing accessible asymmetries at the LHC
  \label{construct}}

So far we have discussed those CP asymmetries, based on the
optimal epsilon products, which exactly match the kinematical
dependence of the CP-sensitive terms in the amplitude squared. For
example, in the neutralino spin correlations this is the epsilon
product ${\mathcal E}_{\tilde\chi^0_i} =[p_u,p_d,p_{\tilde\chi_i^0},
p_{\ell_1}]$, see Eq.~(\ref{eq:epsilon_neut}).  This epsilon product
contains the quark and neutralino momenta, which are not directly
accessible at the LHC.
Even if we knew all particle masses, we would not be able to completely
reconstruct the momenta in the event. There are $8$ unknowns: 
the $3$-momenta of the $\tilde\chi_1^0$ from the neutralino decay, and the 
invisible sneutrino ($\tilde\nu_\ell \to \nu_\ell \, \tilde\chi_1^0$) from 
the chargino decay, as well as the $z$-components of the $u$ and $\bar d$ momenta, 
but only $7$ kinematical constraints: $4$ from energy-momentum conservation, 
and $3$ invariant mass constraints for the decays of the chargino, the neutralino,
and  the slepton. A complete reconstruction is thus impossible, 
since the decay chain on the chargino side is too short due to the invisible sneutrino.

\medskip

In this Section we thus discuss different methods to approximate the 
intermediate neutralino and the initial partonic momenta.
We have checked that all our results for the 
asymmetries and cross sections are in agreement with the 
public code \texttt{MadGraph}~\cite{Alwall:2011uj}.
%
%
Since the estimates we present in the following cannot be perfect, 
we quantify the dilution of the
asymmetries ${\mathcal A}$ due to the momenta replacements, and the
dilution of the cross section $\sigma$ due to selection cuts. The
figure of merit will be the statistical significance, which roughly
scales like
\begin{eqnarray}
S &=& |{\mathcal A}| \; \sqrt{{\mathcal L} \; \sigma },
\label{significance}
\end{eqnarray}
for a given luminosity  ${\mathcal L}$, assuming $100\%$ acceptance.
For our benchmark point, with $pp\to \tilde\chi_2^0\,\tilde\chi_2^+$
production and decay, see Table~\ref{tab:scenario}, we have $\sigma(p
p \to\tilde\chi_2^0\,\tilde\chi_2^\pm) = 11~{\rm fb}$, ${\rm BR}(\tilde
\chi_2^0\to \tilde \ell_R\, \ell) = 66\%$, ${\rm BR}(\tilde\chi_2^+\to 
\tilde\nu_\ell\, \ell^+) = 23\%$.  The task is to obtain larger 
significances than that of the triple product asymmetry $\mathcal{A}
( \mathbf{p}_{\ell^+}, \mathbf{p}_{\ell^-},\mathbf{p}_{\ell_3^+}) = 
-4.7 \%$, for which we have $S = 0.6$ for ${\mathcal L}=100~{\rm fb}
^{-1}$.

\subsection{Lepton preselection
  \label{preselection}}

As a preselection, we apply cuts on the
transverse momentum and rapidity of each lepton
\begin{eqnarray}
p_{\rm T} > 10~{\rm GeV}; \qquad
|\eta| < 2.5; \qquad
\Delta R = \sqrt{ \Delta\phi^2 + \Delta\eta^2} > 0.4,
\end{eqnarray}
where $\Delta\phi$ is the difference of the azimuthal angles of a lepton
pair in radiant, and $\Delta\eta$ their rapidity difference.
The preselection are standard cuts to isolate leptons, which are for example
included as basic cuts in \texttt{MadGraph}~\cite{Alwall:2011uj}.
About $50\%$ of our signal events pass these cuts.

\subsection{Replacing the neutralino momentum
  \label{neutmom}}

We replace the neutralino momentum $p_{\tilde\chi_2^0}$ by the lepton momenta 
$\ell_{1,2} $ from its decay, such that the epsilon product  becomes
$[p_u,p_d,p_{\tilde\chi_2^0},p_{\ell_1}] \to
[p_u,p_d,p_{\ell_1}+ p_{\ell_2},p_{\ell_1}]=
[p_u,p_d,p_{\ell_2},p_{\ell_1}]$.
The corresponding asymmetry is reduced from $21\%$ to $14\%$. 
%
%
We can improve the approximation of the neutralino momentum by
focusing on events where the lepton pair $\ell_1 \ell_2$ from the
neutralino decay has a large invariant mass, $m_{\ell_1 \ell_2}> b 
\; \mathrm{max}(m_{\ell_1\ell_2})$, which optimizes the 
significance for $b\approx 0.5$. This cut ensures that the sum
of the 3-momenta of the leptons is small in the $\tilde\chi_2^0$
rest-frame.  Note that a cut on this quantity might be needed
to reduce SM backgrounds where the lepton pair originates from a
virtual photon.  In addition we only keep those lepton pairs with a
large transverse momentum sum, $(p_{{\rm T}_{\ell_1}} + p_{{\rm T}_{
\ell_2}}) > c \; m_{\tilde\chi_2^0}$, which optimizes the significance
for $c\approx 0.2$. About two thirds of the preselected events pass
both these cuts, but the asymmetry $\mathcal{A}[p_u,p_d,p_{\ell_2},p_
{\ell_1}]$ is enhanced from $14\%$ to $18\%$.

\medskip

Note that the sign of the asymmetry depends on the charge of the
near $\ell_1$ and far lepton $\ell_2$ from the neutralino decay in the
following way, see also the discussion in Appendix~\ref{CPasymmetries},
\begin{eqnarray}
 \mathcal{A} [p_u,p_{d},p_{\ell_1^+} ,p_{\ell_2^-}] &=&
-\mathcal{A} [p_u,p_{d},p_{\ell_1^-} ,p_{\ell_2^+}] \\&=&
+\mathcal{A} [p_u,p_{d},p_{\ell_2^+} ,p_{\ell_1^-}] \\&=&
-\mathcal{A} [p_u,p_{d},p_{\ell_2^-} ,p_{\ell_1^+}].
\label{charge} 
\end{eqnarray}
The sign change in the first step originates from
Eq.~(\ref{eq:SigmaaD1}), and the sign change in the second and third
steps is due to the interchange of the momenta of $\ell_1$ and $\ell_2$
in the antisymmetric epsilon product. These relations 
are important, since we need not determine from which vertex, near or
far, the leptons $\ell_1$ and $\ell_2$ originate.  Instead in the
epsilon product one just groups them according to their
charge, and uses for example the asymmetry $\mathcal{A}[p_u,
p_{d}, p_{\ell^+}, p_{\ell^-}]$, see also the discussion in Appendix~G
of Ref.~\cite{Deppisch:2009nj}.  One can ensure that
the two leptons stem from the neutralino decay, if one requires that
the lepton $\ell_3$ from the chargino decay has different flavor than
the opposite-sign same-flavor lepton pair $\ell_1^\pm \ell_2^\mp$ from
the neutralino decay.

\subsection{Approximating the quark momenta
  \label{quarkmom}}

The sign of ${\mathcal E}_{\tilde\chi^0_i} = [p_u,p_{\bar
d},p_{\tilde\chi_i^0},p_{\ell_1}]$, Eq.~(\ref{eq:epsilon_neut}),
depends on the direction of the incoming
$u$-quark. (The $\bar d$ antiquark then obviously comes from the opposite
direction.) In most events $pp \to \tilde\chi^0_i\,\tilde\chi^+_j$ the
$u$-quark will be more energetic than the antiquark $\bar d$, due to
the characteristic momentum distributions of valence $u$ and
sea-quarks $\bar d$ in the PDFs. So the entire event will be mostly
boosted in the direction of the incoming $u$-quark. The z-component of
the sum of all three lepton momenta, $\mathbf{p}_\ell^z=\mathbf{p}_
{\ell_1}^z+\mathbf{p}_{\ell_2}^z+\mathbf{p}_{\ell_3}^z$, coincides in
$75\%$ of the cases with the direction of the incoming $u$-quark
momentum.  However, guessing wrong will immediately reduce the
asymmetry, since then the event is included with the wrong sign. The
efficiency of the guess can be considerably enhanced to over $90\%$,
if we instead require that only the lepton pair from the neutralino
decay has a large component in the direction of the beam, $|
\mathbf{p}_{\ell_1}^z + \mathbf{p}_{\ell_2}^z|>d\;m_{\tilde\chi_2^0}$.
About half of the preselected events pass the cut for $d \approx 0.6$,
which optimizes the significance.  With these approximated quark
momenta,
\begin{eqnarray}
   p_u^{\rm aprx} &=& (1, 0, 0, \eta), 
\label{eq:quarkapproxu} \\
  p_{\bar d}^{\rm aprx} &=& (1, 0, 0,-\eta), 
\quad {\rm with}~\eta = 
     {\rm Sign}[ \mathbf{p}_{\ell_1}^z + \mathbf{p}_{\ell_2}^z ] = \pm 1,
\label{eq:quarkapproxd}
\end{eqnarray}
we obtain for the asymmetry
$\mathcal{A}[p_u^{\rm aprx}, p_{\bar d}^{\rm aprx}, p_{\tilde\chi_2^0}, p_{\ell_1^+}]
 =-20 \%$.

\medskip

If we now also replace the neutralino momentum and include the cuts on the invariant mass 
and the transverse momentum of the lepton pair, we obtain an asymmetry of 
$\mathcal{A}[p_u^{\rm aprx}, p_{\bar d}^{\rm aprx}, p_{\ell_1^+}, p_{\ell_2^-}] = 17 \%$, 
with a cut efficiency of $35\%$ after preselection cuts, 
such that the corresponding significance is
 $S = 0.95$, for ${\mathcal L}=100~{\rm fb}^{-1}$.
This value has to be compared with the significance  
$S = 0.45$ for the triple product asymmetry 
$\mathcal{A}( \mathbf{p}_{\ell^+}, \mathbf{p}_{\ell^-}, \mathbf{p}_{\ell_3^+}) 
= -4.7 \%$,
with the lepton preselection cuts only.
Although our cuts and momenta approximations can more than double the significance,
it will be difficult to measure the asymmetries at the LHC  due to the low production 
cross section. We will comment on this issue in the next Section.


\section{Discovery reach at the LHC
  \label{Discovery reach}}

The trilepton signal for CP-conserving neutralino-chargino pair production 
at the LHC has been studied by the ATLAS~\cite{ATLAS} and CMS~\cite{CMS} 
collaborations. 
They have focused on the benchmark point SU2, given by
$m_0 = 3.55$~TeV, $m_{1/2}= 350$~GeV, $A_0=0$, $\tan\beta=10$, $\mu>0$,
which lies within the focus point region of the mSUGRA parameter space.
Thus it is characterized by heavy squarks and sleptons
 of order $3$~TeV, but relatively light charginos and neutralinos
 of order $100$~GeV to $300$~GeV.
While the heavy squarks and sleptons only provide rather small 
production cross sections,
the light gauginos provide good SUSY discovery potential in multi-lepton events. 

\medskip

In general, the most important SM backgrounds to the trilepton signal
are from $ t\bar t$, $bZ$, and $WZ$ production~\cite{ATLAS,CMS}.
The fully leptonic  $Z W$ events can be reduced by rejecting 
lepton pairs (with opposite signs and same flavor) which have
an invariant mass $\approx\pm 10$~GeV around $m_Z$.
The leptonic decays of $t\bar t$, $Z b$, can generate a third, but rather soft, 
lepton from the leptonic $b$ decay. Stringent isolation cuts on the lepton 
tracks can reduce those backgrounds, and also soft leptons from bremsstrahlung, 
other hadron decays, and photon conversions, which otherwise  have large
contributions.
The dominant source of low momentum trileptons are heavy flavor SM processes, 
like $pp\to qZ(\gamma^\ast)$, $q\bar q Z(\gamma^\ast)$, for $q = b,c$,
as pointed out in Ref~\cite{Sullivan:2008ki}, and a cut on  
 $E_{\rm T}^{\rm miss}$ around $30$~GeV is proposed.
\medskip

For the CP-conserving SU2 scenario, the  ATLAS collaboration~\cite{ATLAS}
has shown that with appropriate cuts, the SUSY trilepton signal gets reduced 
from about $33$~fb to $3$~fb, with a remaining SM background
of $21$~fb~\cite{ATLAS}.
Thus ATLAS expects a $5\sigma$ discovery over background in the CP-conserving
 SU2 scenario for a luminosity of $80$~$\rm{fb}^{-1}$.
If we assume a similar loss in the signal events, due to the cuts, 
of about an order of magnitude, 
we can estimate the discovery potential for CP violation in the
trilepton signal.  For our scenario for $\tilde\chi_2^0\tilde\chi_2^+$ production as 
given in Table~\ref{tab:scenario}, 
we have shown that the asymmetry 
$\mathcal{A}[p_u^{\rm aprx}, p_{\bar d}^{\rm aprx}, p_{\ell_1^+}, p_{\ell_2^-}]$ 
can reach up to $17 \%$.  The necessary cuts for the asymmetry will  be in 
some sense equivalent to those which isolate the signal.
We would then need at least a luminosity of
 $\mathcal{L}=n^2/(\mathcal{A}^2\sigma {\rm BR})  =n^2 \times 200~\rm{fb}^{-1}$ 
for $n$ standard deviations. Clearly the exact answer can only be given after 
performing a detailed experimental study.
%
%
However  the trilepton signal is probably not best suited to study SUSY CP violation at 
the LHC, and thus we also defer a detailed Monte Carlo study taking into 
account the above mentioned backgrounds and cuts.


\section{Summary and conclusions 
  \label{Summary and conclusions}}

In the complex MSSM, we have analyzed the potential to observe CP violation
from the gaugino and higgsino phases  $\phi_1$ and $\phi_\mu$
in neutralino-chargino pair production at the LHC, 
$p  p \to \tilde\chi_i^0 \,\tilde\chi_j^\pm$. Their subsequent leptonic
two-body decays  give rise to a trilepton signal, with low QCD and SM backgrounds.
The trilepton signal is well known as a clean SUSY discovery channel at the Tevatron, 
and also at the LHC.

\medskip
In order to find the optimal CP asymmetries in the trilepton signal at the LHC,
we have calculated the amplitude squared in the spin density matrix formalism.
From the explicit formulas we have identified the CP-sensitive contributions, 
which already appear at tree level in the neutralino and chargino spin and spin-spin 
correlations.
We have then defined optimal CP asymmetries, which base on epsilon products
that exactly match the kinematic dependence of the CP-violating spin correlations.
After performing a systematic scan in the MSSM parameter space, we have found
that these asymmetries can reach up to  $20\%$ 
for scenarios with light squarks, neutralinos and charginos, in particular for 
parameter  points near level crossings where the neutralinos strongly mix.
Only the cancellations between the different exchange contributions from squark 
$\tilde u_L, \tilde d_L$ and $W$ boson exchange in the production, prevent the 
asymmetries from attaining larger values.

\medskip

These optimal asymmetries would however require a reconstruction of
the initial quark and intermediate neutralino/chargino momenta, which
is only possible to a certain degree at the LHC. We thus have
discussed different replacement and approximation strategies, which
are best realized by the analysis of triple products of the outgoing
three lepton momenta, and by efficient methods to estimate the initial
partonic systems.  For example, we have shown that, by using
appropriate cuts on the final lepton momenta, the direction of the
initial quark momenta can be guessed right in over $90\%$ of the
events, such that the resulting asymmetries can still reach values up
to $17\%$.  We have checked that all our results are in agreement with the 
public code \texttt{MadGraph}~\cite{Alwall:2011uj}.

\medskip

These washout effects, compared to the optimal asymmetries, are 
caused by the strong partonic boosts at the LHC, and by cancellations between the 
different contributions from the spin and spin-spin correlations.
Due to these effects, we conclude that SUSY CP violation in 
the trilepton signal will be difficult to observe, in particular when
squarks and gauginos are much heavier than $400$~GeV.  
We have estimated that one would need at least a luminosity of
 $\mathcal{L}  =n^2 \times 200~\rm{fb}^{-1}$, 
for $n$ standard deviations to observe CP-violating effects at the LHC.

\section{Acknowledgments}
We thank Uli Nierste for inspiring discussions and motivations to analyze 
triple products in the trilepton signal. We thank
K.~Rolbiecki for helpful comments and discussions.
SB and OK thank the Instituto de F\'{\i}sica,
Universidade de S\~ao Paulo, for kind hospitality. 
This work was supported in part by a grant funded jointly by
the DFG (Germany) and FAPESP (S\~ao Paulo State, Brazil),
and in part by the BMBF, and by CNPq-Brazil.
OK was supported by MICINN project FPA.2006-05294, and by a CPAN (Spain) fellowship.
JK was supported in part by the ARC Centre of Excellence for Particle Physics at the
Terascale and in part by the Initiative and Networking Fund of the Helmholtz Association, 
contract HA-101 (``Physics at the Terascale'').

\bigskip

\begin{appendix}
\noindent{\Large\bf Appendix}
\setcounter{equation}{0}
\renewcommand{\thesubsection}{\Alph{section}.\arabic{subsection}}
\renewcommand{\theequation}{\Alph{section}.\arabic{equation}}

\setcounter{equation}{0}

\section{Chargino and neutralino mixings}
  \label{Chargino and neutralino mixings}

The masses and mixing angles of the charginos follow from their mass
matrix~\cite{mssm}
\begin{equation}
{\mathcal M}_{\tilde\chi^\pm} = \left( \begin{array}{cc}
M_2 & m_W\sqrt{2}\sin\beta\\
m_W\sqrt{2}\cos\beta & \mu
\end{array}\right),
\end{equation}
with the $SU(2)$ gaugino mass parameter $M_2$, the higgsino mass parameter
$\mu$, the ratio $\tan\beta = v_2/v_1$ of the vacuum expectation values of the
two neutral Higgs fields, and the mass $m_W$ of the $W$ boson. The chargino
mass matrix can be diagonalized by two complex unitary $2\times 2$
matrices~\cite{mssm},
\begin{equation}
U^*~{\mathcal M}_{\tilde\chi^\pm}~V^{-1} = 
{\rm diag}(m_{\tilde\chi^\pm_1},m_{\tilde\chi^\pm_2}),
\end{equation}
such that the chargino masses satisfy $m_{\tilde\chi^\pm_j} \ge 0$.

\medskip

The complex symmetric mass matrix of the neutralinos in the photino, zino,
higgsino basis ($\tilde{\gamma},\tilde{Z}, \tilde{H}^0_a, \tilde{H}^0_b$), is
given by~\cite{Bartl:1986hp}
\begin{equation}
{\mathcal M}_{\tilde\chi^0} =
\left(\begin{array}{cccc}
M_2 \, s^2_w + M_1 \, c^2_w & 
(M_2-M_1) \, s_w c_w & 0 & 0 \\
(M_2-M_1) \, s_w c_w & 
M_2 \, c^2_w + M_1 \, s^2_w & m_Z  & 0 \\
0 & m_Z &  \mu \, s_{2\beta} & -\mu \, c_{2\beta} \\
0 &  0  & -\mu \, c_{2\beta} & -\mu \, s_{2\beta} 
\end{array}\right) \,.
\label{eq:neutmass}
\end{equation}
Here $s_w = \sin\theta_w$, $c_w = \cos\theta_w$, $\theta_w$ being the weak
mixing angle, $s_{2\beta} = \sin(2\beta)$, $c_{2\beta} = \cos(2\beta)$, $M_1$
is the ${\rm U(1)}$ gaugino mass parameter, and $m_Z$ is the mass of the $Z$
boson. We diagonalize the neutralino mass matrix by a complex, unitary
$4\times 4$ matrix~\cite{mssm}
\begin{eqnarray}
N^* \, {\mathcal M}_{\tilde\chi^0} \, N^{\dagger}&=&
{\rm diag}(m_{\tilde\chi^0_1},\dots,m_{\tilde\chi^0_4}),
\label{eq:neutn}
\end{eqnarray}
such that the neutralino masses satisfy $ m_{\tilde\chi^0_i} \ge 0$.
The $SU(2)$ gaugino mass parameter $M_2$
has been chosen real and positive by absorbing its possible phase via field
redefinitions.\footnote{The mass can be made real and positive by a phase
  transformation of the $SU(2)$ gaugino fields. We then also need to redefine
  the fermion and/or sfermion fields in order to keep the $SU(2)$ gaugino
  fermion sfermion couplings real and positive. Keeping the fermion sfermion
  couplings to higgsinos as well as $U(1)_Y$ and $SU(3)$ gauginos real then
  requires phase transformations of these fields as well. This illustrates
  that only the {\em relative} phases between $M_2$ and the other gaugino
  masses, and between $M_2$ and $\mu$, are physical.} Thus we parametrize
CP-violation in the neutralino and chargino sector by the phases of the
complex parameters $M_1=|M_1|e^{i\phi_1}$ and $\mu=|\mu|e^{i\phi_\mu }$.


\section{Lagrangians and couplings}
\label{Lagrangians and couplings}

The interaction Lagrangians for $W$ boson exchange in the
production are~\cite{mssm} 
\begin{eqnarray}
{\scr L}_{W^{-}ud}&=&
	 -\frac{g}{\sqrt 2}\, W_{\mu}^{-}\bar{d}\, \gamma^{\mu}\, P_L\, u
	 + \mbox{h.c.},\\
{\scr L}_{W^{-}\tilde\chi^+\tilde\chi^0}&=&
	 g \, W_{\mu}^{-}\bar{\tilde\chi}^0_i\gamma^{\mu}
	\left[O_{ij}^L P_L+O_{ij}^R P_R \right]\tilde\chi^+_j +\mbox{h.c.},
\end{eqnarray}
with the weak coupling constant $g=e/\sin\theta_w$, $e>0$, $P_{L, R}=(1\mp \gamma_5)/2$.
In the photino, zino, higgsino basis
($\tilde{\gamma},\tilde{Z}, \tilde{H}^0_a, \tilde{H}^0_b$),
the couplings are
\begin{eqnarray}
O_{ij}^L &=& 
        -1/\sqrt{2}\Big( \cos\beta N_{i4}-\sin\beta N_{i3}
        \Big)V_{j2}^{*}
        +\Big( \sin\theta_w N_{i1}+\cos\theta_w N_{i2} \Big) V_{j1}^{*},
	 \qquad \\
O_{ij}^R &= &
         +1/\sqrt{2}\Big( \sin\beta N^{*}_{i4}+\cos\beta
         N^{*}_{i3}\Big) U_{j2}
         +\Big( \sin\theta_w N^{*}_{i1}+\cos\theta_w N^{*}_{i2} \Big)U_{j1}.
	   \quad
\end{eqnarray}
The interaction Lagrangians for $u$-squark exchange are 
\begin{eqnarray}
{\scr L}_{u \tilde u\tilde\chi^0}
&=& g\,\bar u\, f_{ui}^{L}\,P_R 
            \,\tilde\chi_i^0\,\tilde u_L
+ {\rm h.c.} ,\\[2mm]
{\scr L}_{d \tilde u\tilde\chi^+}
&=& g\,\bar d\,l_{\tilde u j}^{L}\,P_R 
            \,\tilde\chi_j^{+C}\,\tilde u_L
   + {\rm h.c.} ,
\end{eqnarray}
and those for $d$-squark exchange are
\begin{eqnarray}
{\scr L}_{d \tilde d\tilde\chi^0}
&=& g\,\bar d\,f_{di}^{L}\,P_R 
            \,\tilde\chi_i^0\,\tilde d_L
+ {\rm h.c.} ,\\[2mm]
{\scr L}_{u \tilde d\tilde\chi^+}
&=& g\,\bar u\,l_{\tilde d j}^{L}\,P_R 
            \,\tilde\chi_j^{+}\,\tilde d_L
   + {\rm h.c.} .
\end{eqnarray}
In the photino, zino, higgsino basis, the couplings are
\begin{eqnarray}
f_{u i}^L &=& -\sqrt{2}\bigg[\frac{1}{\cos
        \theta_w}\left(\frac{1}{2}-\frac{2}{3}\sin^2\theta_w\right)N_{i2}+
        \frac{2}{3}\sin \theta_w N_{i1}\bigg],
\label{eq:flu}\\[2mm]
l_{\tilde u j}^{L}  &=& -V_{j1},
\label{eq:luj}
\end{eqnarray}
and
\begin{eqnarray}
f_{d i}^L &=& -\sqrt{2}\bigg[\frac{1}{\cos
        \theta_w}\left(-\frac{1}{2}+\frac{1}{3}\sin^2\theta_w\right)N_{i2}-
        \frac{1}{3}\sin \theta_w N_{i1}\bigg],
\label{eq:fld}\\[2mm]
l_{\tilde d j}^{L}  &=& -U_{j1}.
\label{eq:ldj}
\end{eqnarray}
We neglect mixings in the quark and squark generations.
The interaction Lagrangians for neu\-tra\-lino decay
$\tilde\chi_i^0 \to \tilde\ell_{R,L}^\pm \,\ell^\mp$, followed by
$\tilde\ell_{R,L}^\pm \to \tilde\chi_1^0 \,\ell^\pm$ with 
$\ell = e,\mu$ are~\cite{Bartl:1986hp}
\begin{eqnarray}
      {\scr L}_{\ell \tilde \ell \tilde\chi^0} & = & 
             g \bar\ell f_{\ell i}^L  P_R \tilde\chi_i^0  \tilde \ell_L 
     +       g \bar\ell f_{\ell i}^R  P_L \tilde\chi_i^0  \tilde \ell_R
     +       \mbox{h.c.},
\label{eq:slechie}
\end{eqnarray}
with the couplings~\cite{Bartl:1986hp} 
\begin{eqnarray}
f_{\ell i}^L &=& \sqrt{2}\bigg[\frac{1}{\cos
        \theta_w}\left(\frac{1}{2}-\sin^2\theta_w\right)N_{i2}+
         \sin \theta_w N_{i1}\bigg],
\label{eq:fl}\\[2mm]
f_{\ell i}^R &=& \sqrt{2} \sin \theta_w
        \left(\tan\theta_w N_{i2}^*-N_{i1}^*\right).
\label{eq:fr}
\end{eqnarray}
The interaction Lagrangians for chargino decay
$\tilde\chi_j^\pm \to \ell^\pm \,\tilde\nu_\ell$,
followed by
$\tilde\nu_\ell \to \tilde\chi_1^0 \,\nu_\ell$  
are~\cite{Bartl:1986hp}
\begin{eqnarray}
{\scr L}_{\ell \tilde\nu\tilde\chi^+} &=&
     - g U_{j1}^{*} \bar{\tilde\chi}^+_j P_{L} \nu \tilde\ell_L^*
     - g V_{j1}^{*} \bar{\tilde\chi}_j^{+C} P_L \ell 
       \tilde{\nu}^{*}+\mbox{h.c.},\quad \ell=e,\mu,\\[2mm]
{\cal L}_{\nu \tilde\nu\tilde\chi^0} &=&
	 g f_{\nu k}^{L} \bar{\nu} P_R \tilde{\chi}^0_k \tilde{\nu}_L+
	\mbox{h.c.},
\end{eqnarray}
with 
\begin{eqnarray}
f_{\nu k}^L &=& -\frac{\sqrt{2}}{2}\frac{1}{\cos\theta_w}N_{k2}.
\end{eqnarray}

For completeness, we also give the couplings in the
bino, wino, higgsino $\tilde H_{1,2}^0$ basis~\cite{mssm}.
In that basis the neutralino mass matrix is diagonalized by the complex, 
unitary matrix $Z$, similar to Eq.~(\ref{eq:neutn}). 
The quark-squark gaugino couplings are
\begin{eqnarray}
f_{u i}^L \, = \, \frac{1}{\sqrt{2}} \Big(  \frac{1}{3}t_w Z_{i1} + Z_{i2}  \Big),
\qquad
f_{d i}^L  \,= \, \frac{1}{\sqrt{2}} \Big(  \frac{1}{3}t_w Z_{i1} - Z_{i2}  \Big),
\end{eqnarray}
with $t_w = \tan\theta_w$, and $l_{\tilde u j}^{L}$ and $l_{\tilde d j}^{L}$ as given in 
Eqs.~(\ref{eq:luj}),  (\ref{eq:ldj}).
With these definitions the CP-sensitive imaginary parts of product of couplings 
become
\begin{eqnarray}
  {\rm Im}\{O^L_{ij} O^{R\ast}_{ij} \} &=&
 {\rm Im}\Big\{
-  \frac{1}{2}        Z_{i4}Z_{i3} U_{j2}^\ast V_{j2}^\ast
-  \frac{1}{\sqrt{2}} Z_{i4}Z_{i2} U_{j1}^\ast V_{j2}^\ast  
      \nonumber\\[2mm]&&
    \phantom{a}
+  \frac{1}{\sqrt{2}} Z_{i2}Z_{i3} U_{j2}^\ast V_{j1}^\ast 
+                     Z_{i2}^2     U_{j1}^\ast V_{j1}^\ast 
 \Big\}, 
\\[4mm]
{\rm Im}\{f_{u i}^L l_{\tilde u j}^{L\ast} O^{R\ast}_{ij} \} &=&
 \frac{1}{\sqrt{2}} {\rm Im}\Big\{ V_{j1}^\ast
  \Big(
  \frac{t_w}{3} Z_{i1}Z_{i3}  U_{j2}^\ast
 +\frac{t_w}{3} Z_{i1}Z_{i2}  U_{j1}^\ast
      \nonumber\\[2mm]&&
\phantom{lalalalal}
 +\frac{1}{\sqrt{2}} Z_{i2}Z_{i3}  U_{j2}^\ast
 +                   Z_{i2}^2  U_{j1}^\ast
  \Big)
\Big\}, 
\\[4mm]
{\rm Im}\{f_{d i}^{L\ast} l_{\tilde d j}^{L} O^{L\ast}_{ij} \} &=&
 \frac{1}{\sqrt{2}} {\rm Im}\Big\{ U_{j1}^\ast
  \Big(
  \frac{t_w}{3} Z_{i1}Z_{i4}  V_{j2}^\ast
 -\frac{t_w}{3} Z_{i1}Z_{i2}  V_{j1}^\ast
      \nonumber\\[2mm]&&
\phantom{lalalalal}
 -\frac{1}{\sqrt{2}} Z_{i2}Z_{i4}  V_{j2}^\ast
 +                   Z_{i2}^2  V_{j1}^\ast
  \Big)
\Big\}, 
\\[4mm]
{\rm Im}\{f_{u i}^{L\ast} f_{d i}^{L\ast} l_{\tilde u j}^{L} l_{\tilde d j}^{L} \} &=&
\frac{1}{2} {\rm Im}\Big\{ 
  U_{j1}^\ast V_{j1}^\ast
 \Big(  Z_{i2}^2 - \frac{t_w^2}{3^2} Z_{i1}^2
 \Big)
\Big\}.
\end{eqnarray}

\section{Kinematics and phase space}
\label{Phase space}

\bigskip
In the center-of-mass~(cms) system of the incoming quarks,
we parametrize the momenta with the scattering angle  
$\hat \theta \varangle (\hat{\mathbf p}_{u},\hat{\mathbf p}_{\tilde\chi_i^0})$,
and the azimuth $\hat\phi$ to be chosen zero,
   \begin{eqnarray}
	&&\hat p_u^{\mu} = \hat E_b(1,0,0, 1),\quad
         \hat p_d^{\mu} = \hat E_b(1,0,0,-1),\\[2mm]
&& \hat p_{\tilde\chi_i^0}^{\mu} = (\hat E_i,-\hat q\sin\hat\theta,0, \hat q\cos\hat\theta),\quad
   \hat p_{\tilde\chi_j^\pm}^{\mu} = (\hat E_j,\hat q\sin\hat\theta,0,-\hat q\cos\hat\theta),\qquad
   \end{eqnarray}
with the cms energy of the partons $\hat E_b=\sqrt{\hat s}/2$, and 
\begin{eqnarray}
 &&  \hat E_i = \frac{\hat s+m_i^2-m_j^2}{2 \sqrt{\hat s}},\;
     \hat E_j = \frac{\hat s+m_j^2-m_i^2}{2 \sqrt{\hat s}},\;
      \hat q = \frac{\lambda^{\frac{1}{2}}
             (\hat s,m_i^2,m_j^2)}{2 \sqrt{\hat s}}, \qquad
\end{eqnarray}
with $\lambda(x,y,z) = x^2+y^2+z^2-2(xy+xz+yz)$.
We label variables in the cms system by a hat in our notation.
The energies and the $z$-components of the momenta of the
outgoing neutralino $\tilde\chi_i^0$  and chargino $\tilde\chi^\pm_j$ 
in the laboratory~(lab) frame
\begin{eqnarray}
 E = \gamma(\hat E +\beta \hat p_z),\quad 
 p_z =\gamma(\hat p_z + \beta \hat E),
\end{eqnarray}
 are obtained by a Lorentz boost with~\cite{Byckling} 
\begin{eqnarray}
\beta = \frac{x_1 -x_2}{x_1 +x_2}, \quad
\gamma = \frac{1}{\sqrt{1-\beta^2 }}= 
\frac{x_1 +x_2}{2\sqrt{x_1 x_2}}.
\label{eq:boost}
\end{eqnarray}
The partons have energy fractions
$E_u = x_1 E_p$,  $E_d = x_2 E_p$,
of the proton energy $E_p=\sqrt{s}/2$
in the laboratory~(lab) frame, such that 
$\hat s = x_1 x_2 s$, and
\begin{eqnarray}
	&& p_u^{\mu} = E_u(1,0,0, 1),\quad
           p_d^{\mu} = E_d(1,0,0,-1).
\end{eqnarray}
In the laboratory frame,
the momenta for the subsequent decays of the neutralino
$\tilde\chi_i^0  \to \tilde\ell_R^\mp \,\ell_1^\pm$; 
$\tilde\ell_R^\mp\to \tilde\chi_1^0 \,\ell_2^\mp$
and of the chargino
$\tilde\chi^\pm_j \to \tilde\nu_\ell  \,\ell_3^\pm$;
$\tilde\nu_\ell \to \tilde\chi^0_1 \, \nu_\ell$,
are given by
\begin{eqnarray}
  p_{\ell_1}^{\mu} &=& E_{\ell_1}(1,\, \sin \theta_1 \cos \phi_1,
                      \, \sin \theta_1 \sin \phi_1,
                      \, \cos \theta_1), \\[2mm]
  p_{\ell_2}^{\mu} &=& E_{\ell_2}(1,\, \sin \theta_2 \cos \phi_2,
                      \, \sin \theta_2 \sin \phi_2,
                      \, \cos \theta_2),\\[2mm]
  p_{\ell_3}^{\mu} &=& E_{\ell_3}(1,\, \sin \theta_3 \cos \phi_3,
                      \, \sin \theta_3 \sin \phi_3,
                      \, \cos \theta_3),
\label{eq:momenta2}
\end{eqnarray}
with the energies
\begin{eqnarray}
&&E_{\ell_1} = \frac{ m^2_i-m^2_{\tilde\ell} }
            {2(E_i-|{\mathbf p}_i|\cos\theta_{D_1}  )},\quad
E_{\ell_2} = \frac{ m^2_{\tilde\ell}- m^2_{\tilde\chi_1^0} }
            {2(E_{\tilde\ell}-|{\mathbf p}_{\tilde\ell}|\cos\theta_{D_2}  )},\\[2mm]
&&
E_{\ell_3} = \frac{m^2_j-m^2_{\tilde\nu_\ell} }
             {2(E_j-|{\mathbf p}_j|\cos\theta_{D_3}  )}, 
\label{eq:energies}
\end{eqnarray}
and the decay angles,
$\theta_{D_1} \varangle ({\mathbf p}_i,{\mathbf p}_{\ell_1})$,
$\theta_{D_2} \varangle ({\mathbf p}_{\tilde\ell},{\mathbf p}_{\ell_2})$, and
$\theta_{D_3} \varangle ({\mathbf p}_j,{\mathbf p}_{\ell_3})$.

For the description of the polarization of the neutralino 
$\tilde\chi^0_i$ and chargino  $\tilde\chi^\pm_j$
we choose three spin vectors in the laboratory frame
\begin{eqnarray}
s^{1,\,\mu}_i = \left(0,\frac{{\bf s}^2_i\times{\bf s}^3_i}
	{|{\bf s}^2_i\times{\bf s}^3_i|}\right),\;
s^{2,\,\mu}_i = \left(0, \frac{{\bf p}_u\times{\bf p}_i}
	{|{\bf p}_u\times{\bf p}_i|}\right),\;
s^{3,\,\mu}_i = \frac{1}{m_i} \left(|{\bf p}_i|, 
	\frac{E_i}{|{\bf p}_i|}{\bf p}_i \right),\qquad 
\label{eq:spinvecNeut}\\[3mm]
s^{1,\,\mu}_j=\left(0,\frac{{\bf s}^2_j\times{\bf s}^3_j}
	{|{\bf s}^2_j\times{\bf s}^3_j|}\right),
s^{2,\,\mu}_j=\left(0, \frac{{\bf p}_u\times{\bf p}_j}
	{|{\bf p}_u\times{\bf p}_j|}\right),
s^{3,\,\mu}_j=\frac{1}{m_j} \left(|{\bf p}_j|, 
	\frac{E_j}{|{\bf p}_j|}{\bf p}_j \right). \qquad
\label{eq:spinvecChar}
\end{eqnarray}
They form an orthonormal set
\begin{eqnarray}
&&s^a_i\cdot s^b_i=-\delta^{ab}, \quad
s^a_i\cdot  e_i=0, \qquad
s^a_j\cdot s^b_j=-\delta^{ab}, \quad
s^a_j\cdot  e_j=0,
\end{eqnarray}
with the unit momentum vectors 
$ e^{\mu} = p^{\mu}/m$.

The Lorentz invariant phase-space element
for neutralino-chargino production and their subsequent two-body decay
chain, see Eqs.~(\ref{eq:prod})-(\ref{eq:decayChar}),
can be decomposed into two-body  phase-space elements~\cite{Kittel:2004rp,Byckling}
\begin{eqnarray}
 d{\rm Lips}(\hat s; p_{\tilde\nu_\ell},p_{\tilde\chi_1^0},
                p_{\ell_1},p_{\ell_2},p_{\ell_3})=
\frac{1}{(2\pi)^3 }
d{\rm Lips}(\hat s;p_i,p_j)~
d s_i~d{\rm Lips}(s_i;p_{\ell_1},p_{\tilde\ell_{R}})
\nonumber \\  \times
d s_{\tilde\ell_{R}}~d{\rm Lips}(s_{\tilde\ell_{R}},p_{\tilde\chi_1^0},p_{\ell_2})~
d s_j~d{\rm Lips}(s_j;p_{\ell_3},p_{\tilde\nu_\ell}). \quad
\label{eq:phasespace}
 \end{eqnarray}
The several parts of the phase space elements are given by
 \begin{eqnarray}
d{\rm Lips}(\hat s;p_i,p_j)&=&
\frac{\hat q}{8\pi \sqrt{\hat s}}~
\sin\hat\theta~ d\hat\theta,\\
d{\rm Lips}(s_i;p_{\ell_1},p_{\tilde\ell_{R}})&=&
	\frac{1}{2(2\pi)^2}~
	\frac{|{\mathbf p}_{\ell_1}|^2}{m_i^2-m_{\tilde\ell_{R}}^2}
	~d\Omega_1,\\
	d{\rm Lips}(s_{\tilde\ell_{R}} ;p_{\tilde\chi_1^0},p_{\ell_2})&=&
\frac{1}{2(2\pi)^2}~
	\frac{|{\mathbf p}_{\ell_2}|^2}{m_{\tilde\ell_{R}}^2-m_{\tilde\chi_1^0}^2}
	~d\Omega_2,\\
	d{\rm Lips}(s_j;p_{\ell_3},p_{\tilde\nu_\ell})&=&
\frac{1}{2(2\pi)^2}~
	\frac{|{\mathbf p}_{\ell_3}|^2}{m_j^2-m_{\tilde\nu_\ell}^2}
	~d\Omega_3,
\end{eqnarray}
with $\hat s=x_1 x_2 s$, $s_k=p^2_k$, and 
$ d\Omega_k=\sin\theta_k~ d\theta_k~ d\phi_k$, $k=1,2,3$.


\section{Density matrix formalism}
  \label{Density matrix formalism}

For the calculation of the amplitude squared of neutralino-chargino
production, Eq.~(\ref{eq:prod}), and their two-body decay chains,
Eqs.~(\ref{eq:decayNeut})-(\ref{eq:decayChar}), we use the
spin-density matrix formalism of Ref.~\cite{Haber:1994pe}. The
amplitudes squared in the helicity formalism are given,
\textit{e.g.} in Ref.~\cite{Choi:1999mv}.

\subsection{Production matrices}
     \label{Production matrices}

For the production of the neutralino-chargino pair,
\begin{eqnarray} 
u(p_u) + \bar d(p_d) \to \tilde\chi_i^0(p_i,\lambda_i) + 
                         \tilde\chi_j^+(p_j,\lambda_j), 
\label{eq:productionB}
\end{eqnarray}
with momentum $p$ and helicity $\lambda$, the un-normalized spin-density matrix is 
\begin{eqnarray} 
      \rho_P(\tilde\chi^0_i\tilde\chi^\pm_j)
        ^{\lambda_i \lambda_i^\prime \lambda_j \lambda_j^\prime}&=&
      T_P^{\lambda_i \lambda_j}(T_P^{\lambda_i' \lambda_j^\prime})^\ast.
\label{eq:rhoPdef}
\end{eqnarray}
The helicity amplitudes are  
\begin{eqnarray}
T_P^{\lambda_i\lambda_j}(s,W) &=&\frac{g^2}{\sqrt 2}
		\Delta_s(W)\left[ \bar{v}(p_d)\gamma^\mu P_L u(p_u)\right]
       \left[ \bar u(p_j,\lambda_j) \gamma_\mu 
       (O^{L\ast}_{ij} P_L+O^{R\ast}_{ij} P_R) v(p_i,\lambda_i) \right], 
\nonumber \\ \label{eq:TW}\\
T_P^{\lambda_i\lambda_j}(t,\tilde u) &=& 
           g^2 \Delta_t(\tilde u_L)
	   \left[ \bar u(p_i, \lambda_i)f_{u i}^{L\ast} P_L u(p_u) \right]
	   \left[ \bar v(p_d) l_{\tilde u j}^{L} P_R v(p_j,\lambda_j) \right],
\label{eq:Tu}\\
T_P^{\lambda_i\lambda_j}(u, \tilde d) &=& 
           -g^2 \Delta_u(\tilde d_L)
        \left[ \bar v(p_d)f_{d i}^L P_R v(p_i,\lambda_i) \right]
        \left[ \bar u(p_j, \lambda_j) l_{\tilde d j}^{L\ast} P_L u(p_u) \right],
\label{eq:Td}
\end{eqnarray}
with the propagators  
\begin{equation}
   \Delta_s(W)          =  \frac{i}{s-m_W^2},\quad
   \Delta_t(\tilde u_L) =  \frac{i}{t-m^2_{\tilde u_L }},\quad
   \Delta_u(\tilde d_L) =  \frac{i}{u-m^2_{\tilde d_L }},
\label{eq:propagators2}
\end{equation}
and 
$s=(p_u+p_d)^2$, $t=(p_u-p_i)^2$, $u=(p_u-p_j)^2$.
The Feynman diagrams are shown in  Fig.~\ref{Fig:process2}.


Having introduced a set of spin four-vectors $s^a_i$ with $a=1,2,3,$
for the neutralino $\tilde\chi_i^0$, see Eq.~(\ref{eq:spinvecNeut}),
and $s^b_j$ with $b=1,2,3,$ for the chargino $\tilde\chi_j^+$, see
Eq.~(\ref{eq:spinvecChar}), the production matrix~(\ref{eq:rhoPdef})
can be expanded in terms of the Pauli matrices
\begin{equation} 
\rho_{P  }(\tilde\chi^0_i\tilde\chi^\pm_j)^
	     {\lambda_i\lambda^\prime_i\lambda_j\lambda^\prime_j}
 =
       \delta_{\lambda_i\lambda^\prime_i}~\delta_{\lambda_j\lambda^\prime_j}~P
      +\delta_{\lambda_j\lambda^\prime_j}~
        \sigma_{\lambda_i\lambda^\prime_i}^a~\Sigma_P^a
     +\delta_{\lambda_i\lambda^\prime_i}~
         \sigma_{\lambda_j\lambda^\prime_j}^b~\Sigma_P^b
      +\sigma_{\lambda_i\lambda^\prime_i}^a~
       \sigma_{\lambda_j\lambda^\prime_j}^b~
     \Sigma_P^{ab},
\label{eq:rhoP}
\end{equation}
using the Bouchiat-Michel formulas
for massive spin $1/2$ particles~\cite{Bouchiat-Michel}
\begin{eqnarray}
u(p,\lambda^\prime)~\bar u(p,\lambda)&=&\frac{1}{2}~[\delta_{\lambda\lambda^\prime}+
                \gamma_5\slashed s^a \sigma^a_{\lambda\lambda^\prime}](\slashed p+m),\\
v(p,\lambda^\prime)~\bar v(p,\lambda)&=&\frac{1}{2}~[\delta_{\lambda^\prime\lambda}+
                \gamma_5\slashed s^a \sigma^a_{\lambda'\lambda}](\slashed p-m).
\end{eqnarray}


The expansion coefficient $P$ of the production density matrix~(\ref{eq:rhoP})
is independent of the chargino and neutralino polarizations.
It can be composed into contributions from the 
 different production channels
\begin{equation}
	P=P(W W)+P(\tilde u_L \tilde u_L)+P(\tilde d_L \tilde d_L)+
          P(W \tilde u_L)+P(W \tilde d_L)+P(\tilde u_L \tilde d_L),
	\label{eq:exP}
\end{equation}
with
\begin{eqnarray}
P(W W) &=& \frac{g^4}{2}|\Delta_s(W)|^2 \Big[
        |O^L_{ij}|^2 (p_u \cdot p_i)(p_d\cdot p_j )
       +|O^R_{ij}|^2 (p_u \cdot p_j)(p_d\cdot p_i )
                \nonumber \\&& 
      +{\rm Re}\{O^L_{ij} O^{R\ast}_{ij} \} m_i m_j (p_u \cdot p_d)
      \Big], 
\label{eq:PWW} \\[2mm]
P(\tilde u_L \tilde u_L) &=& 
              \frac{g^4}{4}|\Delta_t(\tilde u_L)|^2 |f_{u i}^L|^2
	      |l_{\tilde u j}^{L}|^2 (p_u \cdot p_i)(p_d\cdot p_j ),
\label{eq:Puu} \\ [2mm]
P(\tilde d_L \tilde d_L) &=& 
              \frac{g^4}{4}|\Delta_u(\tilde d_L)|^2 |f_{d i}^L|^2
	      |l_{\tilde d j}^{L}|^2 (p_u \cdot p_j)(p_d\cdot p_i ),
\label{eq:Pdd} \\ [2mm]
P(W \tilde u_L) &=& \frac{\sqrt 2}{4}g^4\Delta_s(W) \Delta_t^\ast(\tilde u_L) \Big[
       2 {\rm Re}\{f_{u i}^L l_{\tilde u j}^{L\ast} O^{L\ast}_{ij} \}
	           (p_u \cdot p_i)(p_d\cdot p_j )
      \nonumber \\&&
       + {\rm Re}\{f_{u i}^L l_{\tilde u j}^{L\ast} O^{R\ast}_{ij} \}
                   m_i m_j(p_u \cdot p_d)
        \Big], 
\label{eq:PWu} \\[2mm]
P(W \tilde d_L) &=& -\frac{\sqrt 2}{4}g^4\Delta_s(W) \Delta_u^\ast(\tilde d_L) \Big[
       2 {\rm Re}\{f_{d i}^{L\ast} l_{\tilde d j}^{L} O^{R\ast}_{ij} \}
	           (p_u \cdot p_j)(p_d\cdot p_i )
      \nonumber \\&&
       + {\rm Re}\{f_{d i}^{L\ast} l_{\tilde d j}^{L} O^{L\ast}_{ij} \}
                   m_i m_j(p_u \cdot p_d)
        \Big], 
\label{eq:PWd} \\[2mm]%
P(\tilde u_L \tilde d_L) &=& 
              -\frac{g^4}{4} \Delta_t(\tilde u_L)\Delta_u^\ast(\tilde d_L)  
{\rm Re}\{f_{u i}^{L\ast} f_{d i}^{L\ast} l_{\tilde u j}^{L} l_{\tilde d j}^{L} \}
	      m_i m_j (p_u \cdot p_d),
\label{eq:Pud} 
\end{eqnarray}
with the couplings as defined in Appendix~\ref{Lagrangians and couplings}.
The terms for $P$ are the same for the charge conjugated process,
$\bar u  d \to \tilde\chi_i^0  \tilde\chi_j^-$.

\newpage

The coefficients $\Sigma^a_P$, which describe the polarization of the 
neutralino $\tilde\chi^0_i$, decompose into
\begin{eqnarray}
\Sigma_P^a &=&
  \Sigma_P^a(W W)+\Sigma_P^a(\tilde u_L \tilde u_L)+\Sigma_P^a(\tilde d_L \tilde d_L)+
 \nonumber\\[2mm]&&
 \Sigma_P^a(W \tilde u_L)+\Sigma_P^a(W \tilde d_L)+\Sigma_P^a(\tilde u_L \tilde d_L),
\label{eq:SigmaNeut}
\end{eqnarray}
with
\begin{eqnarray}
\Sigma_P^a(W W) &=&  \frac{g^4}{2}|\Delta_s(W)|^2 \Big\{
        |O^L_{ij}|^2 m_i (p_d \cdot p_j )(p_u \cdot s_i^a)
       -|O^R_{ij}|^2 m_i (p_u \cdot p_j)(p_d \cdot s_i^a )
                \nonumber \\&& 
      +{\rm Re}\{O^L_{ij} O^{R\ast}_{ij} \} m_j \left[  
      (p_d \cdot p_i)(p_u \cdot s_i^a )- (p_u \cdot p_i )(p_d \cdot s_i^a)
      \right]
               \nonumber \\&&
      -{\rm Im}\{O^L_{ij} O^{R\ast}_{ij} \} m_j[p_u,p_d,p_i,s_i^a]
\Big\},
\label{eq:SigmaPaWW} \\ [2mm]
\Sigma_P^a(\tilde u_L \tilde u_L) &=& 
              \frac{g^4}{4}|\Delta_t(\tilde u_L)|^2 |f_{u i}^L|^2
	      |l_{\tilde u j}^{L}|^2 m_i(p_d \cdot p_j)(p_u\cdot s_i^a ),
\label{eq:SigmaPauu} \\ [2mm]
\Sigma_P^a(\tilde d_L \tilde d_L) &=& 
              -\frac{g^4}{4}|\Delta_u(\tilde d_L)|^2 |f_{d i}^L|^2
	      |l_{\tilde d j}^{L}|^2 m_i(p_u \cdot p_j)(p_d\cdot s_i^a ),
\label{eq:SigmaPadd} \\ [2mm]
\Sigma_P^a(W \tilde u_L) &=& 
      \frac{\sqrt 2}{4}g^4 \Delta_s(W) \Delta_t^\ast(\tilde u_L) \Big\{
       2 {\rm Re}\{f_{u i}^L l_{\tilde u j}^{L\ast} O^{L\ast}_{ij} \}
	           m_i(p_d \cdot p_j)(p_u\cdot s_i^a )
      \nonumber \\&&
       + {\rm Re}\{f_{u i}^L l_{\tilde u j}^{L\ast} O^{R\ast}_{ij} \}
 m_j \left[(p_d \cdot p_i)(p_u \cdot s_i^a)-(p_u \cdot p_i)(p_d \cdot s_i^a)  \right]
      \nonumber \\&&
      -{\rm Im}\{f_{u i}^L l_{\tilde u j}^{L\ast} O^{R\ast}_{ij} \}
                 m_j[p_u,p_d,p_i,s_i^a]
        \Big\}, 
\label{eq:SigmaPaWu} \\[2mm]
\Sigma_P^a(W \tilde d_L) &=& 
      \frac{\sqrt 2}{4}g^4 \Delta_s(W) \Delta_u^\ast(\tilde d_L) \Big\{
       2 {\rm Re}\{f_{d i}^{L\ast} l_{\tilde d j}^{L} O^{R\ast}_{ij} \}
	           m_i(p_u \cdot p_j)(p_d\cdot s_i^a )
      \nonumber \\&&
       + {\rm Re}\{f_{d i}^{L\ast} l_{\tilde d j}^{L} O^{L\ast}_{ij} \}
 m_j \left[(p_u \cdot p_i)(p_d \cdot s_i^a)-(p_d \cdot p_i)(p_u \cdot s_i^a)  \right]
      \nonumber \\&&
      -{\rm Im}\{f_{d i}^{L\ast} l_{\tilde d j}^{L} O^{L\ast}_{ij} \}
                 m_j[p_u,p_d,p_i,s_i^a]
        \Big\}, 
\label{eq:SigmaPaWd} \\[2mm]
\Sigma_P^a(\tilde u_L \tilde d_L) &=& 
              \frac{g^4}{4} \Delta_t(\tilde u_L)\Delta_u^\ast(\tilde d_L) m_j \Big\{
-{\rm Im}\{f_{u i}^{L\ast} f_{d i}^{L\ast} l_{\tilde u j}^{L} l_{\tilde d j}^{L} \}
	       [p_u,p_d,p_i,s_i^a]
      \nonumber \\&&
+{\rm Re}\{f_{u i}^{L\ast} f_{d i}^{L\ast} l_{\tilde u j}^{L} l_{\tilde d j}^{L} \}
\left[(p_u \cdot p_i)(p_d \cdot s_i^a)-(p_d \cdot p_i)(p_u \cdot s_i^a)  \right]
	       \Big\},
\label{eq:SigmaPaud} 
\end{eqnarray}
with the short-hand notation
\begin{eqnarray}
[p_a,p_b,p_c,p_d] =  \varepsilon_{\mu\nu\rho\sigma}
             \, p_a^\mu \, p_b^\nu \, p_c^\rho \, p_d^\sigma,
\quad {\rm and } \quad \varepsilon_{0123}=1. 
\label{eq:epsilon}
\end{eqnarray}
With our choice of the spin vectors $s_i^a$~(\ref{eq:spinvecNeut})
for neutralino $\tilde\chi_i^0$, 
$\Sigma^{a=3}_P/P$ is the longitudinal polarization of 
neutralino, $\Sigma^{a=1}_P/P$ is the transverse polarization in the 
production plane and $\Sigma^{a=2}_P/P$ is the polarization
normal to the production plane. 
Only if there are non-vanishing CP phases $\phi_{1}$
and/or $\phi_{\mu}$ in the chargino and neutralino sector, 
the polarization $\Sigma^{a=2}_P/P$ normal to the 
production plane is non-zero. Thus it is a probe for
CP violation in the production of a neutralino-chargino pair.
To obtain $\Sigma^{a}_P$ for the charge conjugated process,
$\bar u  d \to \tilde\chi_i^0 \, \tilde\chi_j^-$,
one has to change the signs of Eqs.~(\ref{eq:SigmaPaWW})-(\ref{eq:SigmaPaud}).

\newpage

The coefficients $\Sigma^b_P$, which describe the polarization of the 
chargino $\tilde\chi^+_j$, decompose into
\begin{eqnarray}
\Sigma_P^b &=&
  \Sigma_P^b(W W)+\Sigma_P^b(\tilde u_L \tilde u_L)+\Sigma_P^b(\tilde d_L \tilde d_L)+
 \nonumber\\[2mm]&&
 \Sigma_P^b(W \tilde u_L)+\Sigma_P^b(W \tilde d_L)+\Sigma_P^b(\tilde u_L \tilde d_L),
\label{eq:SigmaChar}
\end{eqnarray}
with
\begin{eqnarray}
\Sigma_P^b(W W) &=&  \frac{g^4}{2}|\Delta_s(W)|^2 \Big\{
        |O^R_{ij}|^2 m_j (p_d \cdot p_i)(p_u \cdot s_j^b )
       -|O^L_{ij}|^2 m_j (p_u \cdot p_i )(p_d \cdot s_j^b)
                \nonumber \\&& 
      +{\rm Re}\{O^L_{ij} O^{R\ast}_{ij} \} m_i \left[  
      (p_d \cdot p_j)(p_u \cdot s_j^b )- (p_u \cdot p_j )(p_d \cdot s_j^b)
      \right]
               \nonumber \\&&
      +{\rm Im}\{O^L_{ij} O^{R\ast}_{ij} \} m_i[p_u,p_d,p_j,s_j^b]
\Big\},
\label{eq:SigmaPbWW}\\[2mm]
\Sigma_P^b(\tilde u_L \tilde u_L) &=& 
              -\frac{g^4}{4}|\Delta_t(\tilde u_L)|^2 |f_{u i}^L|^2
	      |l_{\tilde u j}^{L}|^2 m_j(p_u \cdot p_i)(p_d\cdot s_j^b ),
\label{eq:SigmaPbuu} \\ [2mm]
\Sigma_P^b(\tilde d_L \tilde d_L) &=& 
              \frac{g^4}{4}|\Delta_u(\tilde d_L)|^2 |f_{d i}^L|^2
	      |l_{\tilde d j}^{L}|^2 m_j(p_d \cdot p_i)(p_u\cdot s_j^b ),
\label{eq:SigmaPbdd} \\ [2mm]
\Sigma_P^b(W \tilde u_L) &=& 
      -\frac{\sqrt 2}{4}g^4 \Delta_s(W) \Delta_t^\ast(\tilde u_L) \Big\{
       2 {\rm Re}\{f_{u i}^L l_{\tilde u j}^{L\ast} O^{L\ast}_{ij} \}
	           m_j(p_u \cdot p_i)(p_d\cdot s_j^b )
      \nonumber \\&&
       + {\rm Re}\{f_{u i}^L l_{\tilde u j}^{L\ast} O^{R\ast}_{ij} \}
 m_i \left[(p_u \cdot p_j)(p_d \cdot s_j^b)-(p_d \cdot p_j)(p_u \cdot s_j^b)  \right]
      \nonumber \\&&
      -{\rm Im}\{f_{u i}^L l_{\tilde u j}^{L\ast} O^{R\ast}_{ij} \}
                 m_i[p_u,p_d,p_j,s_j^b]
        \Big\}, 
\label{eq:SigmaPbWu} \\[2mm]
\Sigma_P^b(W \tilde d_L) &=& 
      -\frac{\sqrt 2}{4}g^4 \Delta_s(W) \Delta_u^\ast(\tilde d_L) \Big\{
       2 {\rm Re}\{f_{d i}^{L\ast} l_{\tilde d j}^{L} O^{R\ast}_{ij} \}
	           m_j(p_d \cdot p_i)(p_u \cdot s_j^b )
      \nonumber \\&&
       + {\rm Re}\{f_{d i}^{L\ast} l_{\tilde d j}^{L} O^{L\ast}_{ij} \}
 m_i \left[(p_d \cdot p_j)(p_u \cdot s_j^b)-(p_u \cdot p_j)(p_d \cdot s_j^b)  \right]
      \nonumber \\&&
      -{\rm Im}\{f_{d i}^{L\ast} l_{\tilde d j}^{L} O^{L\ast}_{ij} \}
                 m_i[p_u,p_d,p_j,s_j^b]
        \Big\}, 
\label{eq:SigmaPbWd} \\[2mm]
\Sigma_P^b(\tilde u_L \tilde d_L) &=& 
              -\frac{g^4}{4} \Delta_t(\tilde u_L)\Delta_u^\ast(\tilde d_L) m_i \Big\{
-{\rm Im}\{f_{u i}^{L\ast} f_{d i}^{L\ast} l_{\tilde u j}^{L} l_{\tilde d j}^{L} \}
	       [p_u,p_d,p_j,s_j^b]
      \nonumber \\&&
+{\rm Re}\{f_{u i}^{L\ast} f_{d i}^{L\ast} l_{\tilde u j}^{L} l_{\tilde d j}^{L} \}
\left[(p_d \cdot p_j)(p_u \cdot s_j^b)-(p_u \cdot p_j)(p_d \cdot s_j^b)  \right]
	       \Big\}.
\label{eq:SigmaPbud}
\end{eqnarray}
With our choice of the chargino spin vectors $s_j^b$~(\ref{eq:spinvecChar}),
$\Sigma^{b=3}_P/P$ is the longitudinal polarization of 
chargino~$\tilde\chi^+_j$,
$\Sigma^{b=1}_P/P$ is the transverse polarization in the 
production plane and $\Sigma^{b=2}_P/P$ is the polarization
normal to the production plane. 
To obtain $\Sigma^{b}_P$ for the charge conjugated process,
$\bar u  d \to \tilde\chi_i^0 \, \tilde\chi_j^-$,
one has to change the signs of Eqs.~(\ref{eq:SigmaPbWW})-(\ref{eq:SigmaPbud}).

\newpage

The coefficients $\Sigma^{ab}_P$, which contain the spin vectors
$s^a_i$ of the neutralino $\tilde\chi^0_i$
and $s^b_j$ of the chargino $\tilde\chi^+_j$, are the spin-spin
correlation terms. They  decompose into
\begin{eqnarray}
\Sigma_P^{ab} &=&
  \Sigma_P^{ab}(W W)+\Sigma_P^{ab}(\tilde u_L \tilde u_L)+\Sigma_P^{ab}(\tilde d_L \tilde d_L)+
 \nonumber\\[2mm]&&
 \Sigma_P^{ab}(W \tilde u_L)+\Sigma_P^{ab}(W \tilde d_L)+\Sigma_P^{ab}(\tilde u_L \tilde d_L),
\label{eq:SigmaPab}
\end{eqnarray}
with
\begin{eqnarray}
\Sigma_P^{ab}(W W) &=&  -\frac{g^4}{2}|\Delta_s(W)|^2 \Big[
        |O^L_{ij}|^2 m_i m_j (p_u \cdot s_i^a)(p_d \cdot s_j^b)
       +|O^R_{ij}|^2 m_i m_j (p_d \cdot s_i^a)(p_u \cdot s_j^b)
                \nonumber \\&& 
      + {\rm Re}\{O^L_{ij} O^{R\ast}_{ij} \} 
      g^{ab}
        +{\rm Im}\{O^L_{ij} O^{R\ast}_{ij} \}
      f^{ab}
\Big],
\label{eq:SigmaPabWW}\\[2mm]
\Sigma_P^{ab}(\tilde u_L \tilde u_L) &=& 
              -\frac{g^4}{4}|\Delta_t(\tilde u_L)|^2 |f_{u i}^L|^2
	      |l_{\tilde u j}^{L}|^2 m_i m_j (p_u \cdot s_i^a)(p_d\cdot s_j^b ),
\label{eq:SigmaPabuu} \\ [2mm]
\Sigma_P^{ab}(\tilde d_L \tilde d_L) &=& 
              -\frac{g^4}{4}|\Delta_u(\tilde d_L)|^2 |f_{d i}^L|^2
	      |l_{\tilde d j}^{L}|^2 m_i m_j (p_d \cdot s_i^a)(p_u\cdot s_j^b ),
\label{eq:SigmaPabdd} \\ [2mm]
\Sigma_P^{ab}(W \tilde u_L) &=& 
      -\frac{\sqrt 2}{4}g^4 \Delta_s(W) \Delta_t^\ast(\tilde u_L) \Big[
       2 {\rm Re}\{f_{u i}^L l_{\tilde u j}^{L\ast} O^{L\ast}_{ij} \}
	           m_i m_j(p_u \cdot s_i^a)(p_d\cdot s_j^b )
      \nonumber \\&&
       + {\rm Re}\{f_{u i}^L l_{\tilde u j}^{L\ast} O^{R\ast}_{ij} \}
             g^{ab}
      +{\rm Im}\{f_{u i}^L l_{\tilde u j}^{L\ast} O^{R\ast}_{ij} \}
       f^{ab}
        \Big], 
\label{eq:SigmaPabWu} \\[2mm]
\Sigma_P^{ab}(W \tilde d_L) &=& 
      \frac{\sqrt 2}{4}g^4 \Delta_s(W) \Delta_u^\ast(\tilde d_L) \Big[
       2 {\rm Re}\{f_{d i}^{L\ast} l_{\tilde d j}^{L} O^{R\ast}_{ij} \}
	           m_i m_j(p_d \cdot s_i^a)(p_u \cdot s_j^b )
      \nonumber \\&&
       + {\rm Re}\{f_{d i}^{L\ast} l_{\tilde d j}^{L} O^{L\ast}_{ij} \}
       g^{ab}
      -{\rm Im}\{f_{d i}^{L\ast} l_{\tilde d j}^{L} O^{L\ast}_{ij} \}
                 f^{ab}
        \Big], 
\label{eq:SigmaPabWd} \\[2mm]
\Sigma_P^{ab}(\tilde u_L \tilde d_L) &=& 
              \frac{g^4}{4} \Delta_t(\tilde u_L)\Delta_u^\ast(\tilde d_L)  \Big[
{\rm Re}\{f_{u i}^{L\ast} f_{d i}^{L\ast} l_{\tilde u j}^{L} l_{\tilde d j}^{L} \}
       g^{ab}
-{\rm Im}\{f_{u i}^{L\ast} f_{d i}^{L\ast} l_{\tilde u j}^{L} l_{\tilde d j}^{L} \}
   f^{ab}
	       \Big],
\label{eq:SigmaPabud}
\end{eqnarray}
%
%
and the short-hand notations for the products
\begin{eqnarray}
g^{ab} &=&
 \phantom{+} (p_u \cdot p_d) \left[  (p_j \cdot s_i^a)(p_i \cdot s_j^b)
                        - (p_i \cdot p_j )(s_i^a \cdot s_j^b)\right]
               \nonumber \\&&
  +(p_u \cdot s_i^a) \left[ (p_i \cdot p_j)(p_d \cdot s_j^b)
                        -  (p_d \cdot p_j )(p_i \cdot s_j^b)\right]
               \nonumber \\&&
  +(p_u \cdot p_i) \left[ (p_d \cdot p_j)(s_i^a \cdot s_j^b)
                        - (p_j \cdot s_i^a )(p_d \cdot s_j^b)\right]
               \nonumber \\&&
  +(p_u \cdot s_j^b) \left[ (p_i \cdot p_j)(p_d \cdot s_i^a)
                        - (p_d \cdot p_i )(p_j \cdot s_i^a)\right]
               \nonumber \\&&
  +(p_u \cdot p_j) \left[ (p_d \cdot p_i)(s_i^a \cdot s_j^b)
                        -(p_d \cdot s_i^a)(p_i \cdot s_j^b)\right],
\label{eq:f1ab} \\[2mm]
f^{ab} &=&
   \phantom{+} (p_u \cdot s_i^a)[p_d,p_i,p_j,s_j^b]
             + (p_u \cdot p_i  )[p_d,p_j,s_i^a,s_j^b ]
   \nonumber \\&&
            +(p_d \cdot s_j^b)[p_u,p_i,p_j,s_i^a]
            +(p_d \cdot p_j  )[p_u,p_i,s_i^a,s_j^b]
\label{eq:fab} \\[2mm]
            &=&
   \phantom{+} (p_d \cdot s_i^a)[p_u,p_i,p_j,s_j^b]
             + (p_d \cdot p_i)[p_u,p_j,s_i^a,s_j^b]
    \nonumber \\&&
	    +  (p_u \cdot s_j^b)[p_d,p_i,p_j,s_i^a]
            +  (p_u \cdot p_j)[p_d,p_i,s_i^a,s_j^b].
\label{eq:f3ab} 
\end{eqnarray}
The terms for $\Sigma_P^{ab}$ are the same for the charge conjugated process,
$\bar u  d \to \tilde\chi_i^0 \, \tilde\chi_j^-$.


\subsection{Decay matrices}
     \label{Decay matrices}

Similar to the production matrix, the matrix for neutralino decay 
$ \tilde\chi^0_i \to \tilde\ell_R^- \, \ell_1^+$, Eq.~(\ref{eq:decayNeut}),
can also be expanded in terms of the Pauli matrices
\begin{eqnarray}
\rho_{D_1}(\tilde\chi^0_i)_{\lambda^\prime_i\lambda_i}&=&
        \delta_{\lambda^\prime_i\lambda_i}~D_1+
	\sigma^a_{\lambda^\prime_i\lambda_i}~\Sigma^a_{D_1}. 
\label{eq:rhoD1}
\end{eqnarray}
For the chargino decay
$\tilde\chi^+_j \to \tilde\nu_\ell  \,\ell_3^+$, Eq.~(\ref{eq:decayChar}),
we write
\begin{eqnarray}
\rho_{D_3}(\tilde\chi^\pm_j)_{\lambda^\prime_j\lambda_j} &=&
       \delta_{\lambda^\prime_j\lambda_j}~D_3+
	\sigma^b_{\lambda^\prime_j\lambda_j}~\Sigma^b_{D_3}.
\label{eq:rhoD3}
\end{eqnarray}
The expansion coefficients are given by~\cite{Kittel:2004rp}
\begin{eqnarray}
D_1 &=& \frac{g^2}{2} |f^{R}_{\ell i}|^2 (m_i^2 -m_{\tilde\ell}^2 ),
\label{eq:D1} \\ [2mm]
\Sigma^a_{D_1} &=& \,^{\;\,+}_{(-)} g^2 |f^{R}_{\ell i}|^2 
m_i (s^a_i \cdot p_{\ell_1}),
\label{eq:SigmaaD1} \\ [2mm]
D_3 &= & \frac{g^2}{2} |V_{j1}|^2 
	      (m_j^2 -m_{\tilde\nu_{\ell}}^2 ),
\label{eq:D3} \\ [2mm]
\Sigma^b_{D_3} &=&\;  \,^{\;\,-}_{(+)} g^2 |V_{j1}|^2 
	m_j (s^b_j \cdot p_{\ell_3}),
\label{eq:SigmabD3}
\end{eqnarray}
where the sign in parenthesis in Eq.~(\ref{eq:SigmaaD1})
holds for the charge conjugated process 
$ \tilde\chi^0_i \to \tilde\ell_R^+ \, \ell_1^-$,
and in Eq.~(\ref{eq:SigmabD3})
for the conjugated process
$\tilde\chi^-_j \to \tilde\nu_\ell \, \ell_3^-$.

The decay factor for the subsequent slepton decay
$\tilde\ell_R^\mp\to \tilde\chi^0_1 \,\ell_2^\mp$,
is 
\begin{eqnarray}
D_2(\tilde\ell) &=& g^2 |f^{R}_{\ell 1}|^2 
	( m_{\tilde\ell}^2-m_{\tilde\chi_1^0}^2 ),
\label{eq:D2}
\end{eqnarray}
with all couplings as defined in Appendix~\ref{Lagrangians and couplings}.

\subsection{Squared amplitude of production and decay}
     \label{Squared amplitude of production and decay}

Having defined all density matrices for production and decay,
the amplitude squared of the combined process of neutralino-chargino 
production, Eqs.~(\ref{eq:prod}), and their two-body decay chains, 
Eqs.~(\ref{eq:decayNeut})-(\ref{eq:decayChar}),
is written in the spin-density matrix formalism as~\cite{Haber:1994pe} 
\begin{eqnarray}
	|T|^2=|\Delta(\tilde\chi^0_i)|^2~|\Delta(\tilde\chi^\pm_j)|^2~
              |\Delta(\tilde\ell_R)|^2
 \times \nonumber \\ [2mm]
	\sum_{\lambda_i,\lambda^\prime_i,\lambda_j,\lambda^\prime_j}~
	\rho_{P  }(\tilde\chi^0_i\tilde\chi^\pm_j)^
	     {\lambda_i\lambda^\prime_i\lambda_j\lambda^\prime_j}~
	\rho_{D_1}(\tilde\chi^0_i)_{\lambda^\prime_i\lambda_i}~
        D_2(\tilde\ell_R)~
         \rho_{D_3}(\tilde\chi^\pm_j)_{\lambda^\prime_j\lambda_j }.
 \label{eq:matrixelement}
\end{eqnarray}
The amplitude squared is composed of 
the propagators 
\begin{eqnarray}
     \Delta(k) &=& \frac{i}{s_k -m_k^2 +im_k\Gamma_k},
\label{eq:propagators}
\end{eqnarray}
with mass $m_k$ and width $\Gamma_k$ of the  particles.
Inserting the density matrices
$\rho_P(\tilde\chi^0_i\tilde\chi^\pm_j)$,  Eq.~(\ref{eq:rhoP}), 
$\rho_{D_1}(\tilde\chi^0_i)$,  Eq.~(\ref{eq:rhoD1}), and
$\rho_{D_3}(\tilde\chi^\pm_j)$,  Eq.~(\ref{eq:rhoD3}),
into the formula for the amplitude squared, Eq.~(\ref{eq:matrixelement}),
we obtain
\begin{eqnarray}
    |T|^2=4~|\Delta(\tilde\chi^0_i)|^2~|\Delta(\tilde\chi^\pm_j)|^2~
              |\Delta(\tilde\ell_R)|^2
	       \times \nonumber \\ [2mm]
\left[
  P~D_1~D_3
+ \Sigma^a_P~\Sigma^a_{D_1}~D_3
+ \Sigma^b_P~\Sigma^b_{D_3}~D_1
+ \Sigma^{ab}_P~~\Sigma^a_{D_1}~\Sigma^b_{D_3}
\right]
D_2(\tilde\ell_R),
\label{eq:matrixelement2}
\end{eqnarray}
with an implicit sum over $a=1,2,3$ and $b=1,2,3$.
The amplitude squared $|T|^2$ is now decomposed into an
unpolarized part (first summand), the spin correlations of the neutralino 
(second summand), those of the chargino (third summand), and the spin-spin 
correlations of the neutralino and chargino (fourth summand).

\medskip

We give the amplitude squared  $|T|^2$, Eq.~(\ref{eq:matrixelement2}),  
for the process $\bar u  d \to \tilde\chi_i^0  \,\tilde\chi_j^+$,
followed by $ \tilde\chi^0_i \to \tilde\ell_R^- \, \ell_1^+$,
and $\tilde\chi^+_j \to \tilde\nu_\ell  \ell_3^+$.
To obtain $|T|^2$  for the charge conjugated process,
$\bar u  d \to \tilde\chi_i^0  \,\tilde\chi_j^-$,
followed by $ \tilde\chi^0_i \to \tilde\ell_R^- \, \ell_1^+$,
and $\tilde\chi^-_j \to \tilde\nu_\ell  \,\ell_3^-$,
one has to reverse the sign of the
neutralino spin correlations (second summand),
and the spin-spin correlations (fourth summand).
The unpolarized part (first summand), and the spin correlations 
of the chargino (third summand) stay the same.
For the neutralino decay into a positively charged selectron, 
$ \tilde\chi^0_i \to \tilde\ell_R^+ \, \ell_1^-$,
there is an additional sign change for the
neutralino spin correlations (second summand),
and the spin-spin correlations (fourth summand),
due to the sign change of $\Sigma^a_{D_1}$, see Eq.~(\ref{eq:SigmaaD1}).

\medskip

For the propagators in Eq.~(\ref{eq:matrixelement2}),
we use the narrow width approximation
\begin{eqnarray}
\int|\Delta(k)|^2 ~ d s_k &=& 
\frac{\pi}{m_k\Gamma_k},
\label{eq:narrowwidth}
\end{eqnarray}
which is justified for
$\Gamma_k/m_k\ll1$,
which holds in our case with
$\Gamma_k\lsim {\mathcal O}(1~{\rm GeV}) $.
Note, however, that the naive
${\mathcal O}(\Gamma/m)$-expectation of the error can easily receive
large off-shell corrections of an order of magnitude and more,
in particular at threshold, or due to interferences
with other resonant or non-resonant processes~\cite{Hagiwara:2005wg}.

\medskip

The hard partonic cross section in the laboratory frame is then 
obtained by integrating the squared amplitude, Eq.~(\ref{eq:matrixelement2}),
over the  Lorentz invariant phase space element,
Eq.~(\ref{eq:phasespace}),
\begin{equation}
	d\hat\sigma = \frac{1}{2 \hat s}|T|^2 d{\rm Lips}.
\label{eq:sigmapartonic}
\end{equation}
Finally the hadronic cross section $\sigma $ in the laboratory frame 
is obtained by integrating the partonic cross section 
$\hat\sigma $,  Eq.~(\ref{eq:sigmapartonic}), over the
parton distribution functions
\begin{equation}
     d\sigma = \frac{2\kappa}{3}
          \int dx_1 dx_2 f_u(x_1, \mu^2) f_d(x_2, \mu^2) \hat\sigma,
\label{eq:sigmahadronic}
\end{equation}
which depend on the factorization scale $\mu$,
and the momentum fractions $x_1$, $x_2$ of the  quarks
of the proton center-of mass energy $\sqrt{s}$
in the laboratory frame, such that 
$\hat s = x_1 x_2 s$.
There is a combinatorial color factor of~$1/3$, since the PDFs typically 
sum the quark color ($3$x$3$), but here only quarks with pairing color-anticolor 
($3$ pairs) contribute. The factor of~$2$ takes into account the two possible
assignments that the  quark (antiquark) originates form the left (right) 
incoming proton beam, and vice versa. The factor $\kappa$ takes efficiently 
into account the dominant QCD radiative corrections for the production cross 
section, with typical value $\kappa\approx  1.3$~\cite{CMS}. 
For simplicity however, we set $\kappa= 1$ in our numerical calculations.

\bigskip


By using the completeness relations for the neutralino spin vectors
\begin{eqnarray}
s_i^{a,\,\mu} s_i^{a,\,\nu} = -g^{\mu\nu}
      + \frac{p_i^\mu p_i^\nu}{m_i^2},
\label{eq:completenessneut}
\end{eqnarray}
\noindent
the terms with products of the neutralino spin vectors
 in Eq.~(\ref{eq:matrixelement2}) can be written
\begin{eqnarray}
\Sigma^a_P~\Sigma^a_{D_1} \equiv \Sigma_{PD_1} &=&
  \Sigma_{PD_1}(W W)+
\Sigma_{PD_1}(\tilde u_L \tilde u_L)+
\Sigma_{PD_1}(\tilde d_L \tilde d_L)+
 \nonumber\\[2mm]&&
 \Sigma_{PD_1}(W \tilde u_L)
+\Sigma_{PD_1}(W \tilde d_L)
+\Sigma_{PD_1}(\tilde u_L \tilde d_L),
\label{eq:dSigmaNeut}
\end{eqnarray}
with
\begin{eqnarray}
\Sigma_{PD_1}(W W) &=&  \frac{g^6}{2} |f^{R}_{\ell i}|^2|\Delta_s(W)|^2 \Big\{ 
  |O^L_{ij}|^2(p_d \cdot p_j )\left[  (p_u \cdot p_i) (p_i \cdot p_{\ell_1})- 
                                         m_i^2(p_u \cdot p_{\ell_1}) \right]\nonumber\\[1mm]&&
 -|O^R_{ij}|^2 (p_u \cdot p_j)\left[ (p_d \cdot p_i) (p_i \cdot p_{\ell_1}) -
                                          m_i^2 ( p_d \cdot p_{\ell_1}) \right] \nonumber\\ [1mm]&&
      +{\rm Re}\{O^L_{ij} O^{R\ast}_{ij} \} m_i m_j \left[
                          (p_u \cdot p_i)(p_d \cdot p_{\ell_1})
                         -(p_d \cdot p_i)(p_u \cdot p_{\ell_1})   
                                                     \right]
               \nonumber \\[1mm]&&
      +{\rm Im}\{O^L_{ij} O^{R\ast}_{ij} \}m_i  m_j [p_u,p_d,p_i,p_{\ell_1}] 
\Big\},
\label{eq:SigmaPDaWW} \\ [2mm]
\Sigma_{PD_1}(\tilde u_L \tilde u_L) &=& 
              \frac{g^6}{4}|\Delta_t(\tilde u_L)|^2 |f^{R}_{\ell i}|^2|f_{u i}^L|^2
	      |l_{\tilde u j}^{L}|^2 (p_d \cdot p_j)
              \left[ (p_u \cdot p_i) (p_i \cdot p_{\ell_1})
                    -m_i^2(p_u \cdot p_{\ell_1})\right],
 \nonumber \\ 
\label{eq:SigmaPDauu} \\
\Sigma_{PD_1}(\tilde d_L \tilde d_L) &=& 
              -\frac{g^6}{4}|\Delta_u(\tilde d_L)|^2 |f^{R}_{\ell i}|^2|f_{d i}^L|^2
	      |l_{\tilde d j}^{L}|^2 (p_u \cdot p_j)
            \left[(p_d \cdot p_i) (p_i \cdot p_{\ell_1})
                  -m_i^2(p_d \cdot p_{\ell_1})\right],
 \nonumber \\
\label{eq:SigmaPDadd} \\
\Sigma_{PD_1}(W \tilde u_L) &=& 
   \frac{\sqrt 2}{4}g^6 \Delta_s(W) \Delta_t^\ast(\tilde u_L)|f^{R}_{\ell i}|^2 
     \Big\{ \nonumber \\ [1mm]&&
       2 {\rm Re}\{f_{u i}^L l_{\tilde u j}^{L\ast} O^{L\ast}_{ij} \} (p_d \cdot p_j)
       \left[(p_u \cdot p_i) (p_i \cdot p_{\ell_1}) - m_i^2 (p_u \cdot p_{\ell_1})\right]
      \nonumber \\[1mm]&&
       + {\rm Re}\{f_{u i}^L l_{\tilde u j}^{L\ast} O^{R\ast}_{ij} \} m_i m_j
       \left[(p_u \cdot p_i)(p_d \cdot p_{\ell_1})-(p_d \cdot p_i)(p_u \cdot p_{\ell_1})\right]
      \nonumber \\[1mm]&&
      +{\rm Im}\{f_{u i}^L l_{\tilde u j}^{L\ast} O^{R\ast}_{ij} \}
                m_i m_j[p_u,p_d,p_i,p_{\ell_1}]
        \Big\}, 
\label{eq:SigmaPDaWu} \\[2mm]
\Sigma_{PD_1}(W \tilde d_L) &=& 
      \frac{\sqrt 2}{4}g^6 \Delta_s(W) \Delta_u^\ast(\tilde d_L) |f^{R}_{\ell i}|^2
        \Big\{\nonumber \\[1mm]&&
       2 {\rm Re}\{f_{d i}^{L\ast} l_{\tilde d j}^{L} O^{R\ast}_{ij} \}(p_u \cdot p_j)
       \left[(p_d \cdot p_i) (p_i \cdot p_{\ell_1}) -m_i^2(p_d \cdot p_{\ell_1})\right]
      \nonumber \\[1mm]&&
       + {\rm Re}\{f_{d i}^{L\ast} l_{\tilde d j}^{L} O^{L\ast}_{ij} \} m_i m_j
      \left[(p_d \cdot p_i)(p_u \cdot p_{\ell_1})-(p_u \cdot p_i)(p_d \cdot p_{\ell_1})\right]
  \nonumber \\[1mm]&&
     +{\rm Im}\{f_{d i}^{L\ast} l_{\tilde d j}^{L} O^{L\ast}_{ij} \}
                m_i m_j[p_u,p_d,p_i,p_{\ell_1}]
        \Big\}, 
\label{eq:SigmaPDaWd} \\[2mm]
\Sigma_{PD_1}(\tilde u_L \tilde d_L) &=& 
              \frac{g^6}{4} \Delta_t(\tilde u_L)\Delta_u^\ast(\tilde d_L) |f^{R}_{\ell i}|^2 
             m_i m_j\Big\{
{\rm Im}\{f_{u i}^{L\ast} f_{d i}^{L\ast} l_{\tilde u j}^{L} l_{\tilde d j}^{L} \}
	       [p_u,p_d,p_i,p_{\ell_1}]
      \nonumber \\&&
+{\rm Re}\{f_{u i}^{L\ast} f_{d i}^{L\ast} l_{\tilde u j}^{L} l_{\tilde d j}^{L} \}
 \left[(p_d \cdot p_i)(p_u \cdot p_{\ell_1})-(p_u \cdot p_i)(p_d \cdot p_{\ell_1})\right]
	       \Big\}.
\label{eq:SigmaPDaud} 
\end{eqnarray} 
To obtain $\Sigma_{PD_1}$ for the charge conjugated process,
$\bar u  d \to \tilde\chi_i^0  \,\tilde\chi_j^-$,
and $\tilde\chi^-_j \to \tilde\nu_\ell \, \ell_3^-$,
one has to change the signs of Eqs.~(\ref{eq:SigmaPDaWW})-(\ref{eq:SigmaPDaud}),
due to the sign change of  $\Sigma^a_P$, see Eq.~(\ref{eq:SigmaNeut}).
For the neutralino decay into a positively charged selectron, 
$ \tilde\chi^0_i \to \tilde\ell_R^+ \, \ell_1^-$,
there is an additional sign change in Eqs.~(\ref{eq:SigmaPDaWW})-(\ref{eq:SigmaPDaud}),
due to the sign change of $\Sigma^a_{D_1}$, see Eq.~(\ref{eq:SigmaaD1}).

\newpage

Analogously, by using the completeness relations for the chargino spin
vectors
\begin{eqnarray}
s_j^{b,\,\mu} s_j^{b,\,\nu} = -g^{\mu\nu}
      + \frac{p_j^\mu p_j^\nu}{m_j^2},
\label{eq:completenesschar}
\end{eqnarray}
the terms with products of the chargino spin vectors 
in Eq.~(\ref{eq:matrixelement2}) can be written 
\begin{eqnarray}
\Sigma^b_P~\Sigma^b_{D_3} \equiv \Sigma_{PD_3}  &=&
\Sigma_{PD_3}(W W)+
\Sigma_{PD_3}(\tilde u_L \tilde u_L)+
\Sigma_{PD_3}(\tilde d_L \tilde d_L)+
 \nonumber\\[2mm]&&
\Sigma_{PD_3}(W \tilde u_L)+
\Sigma_{PD_3}(W \tilde d_L)+
\Sigma_{PD_3}(\tilde u_L \tilde d_L),
\label{eq:dSigmaChar}
\end{eqnarray}
with
\begin{eqnarray}
\Sigma_{PD_3}(W W) &=&  -\frac{g^6}{2}|\Delta_s(W)|^2 |V_{j1}|^2 
         \Big\{ 
      |O^R_{ij}|^2 (p_d \cdot p_i)
 \left[(p_u \cdot p_j) (p_j \cdot p_{\ell_3})-m_j^2(p_u \cdot p_{\ell_3})\right]
        \nonumber \\[1mm]&&
       -|O^L_{ij}|^2  (p_u \cdot p_i )
 \left[(p_d \cdot p_j) (p_j \cdot p_{\ell_3})-m_j^2(p_d \cdot p_{\ell_3})\right]
                \nonumber \\[1mm]&& 
      +{\rm Re}\{O^L_{ij} O^{R\ast}_{ij} \} m_i m_j
   \left[(p_u \cdot p_j )(p_d \cdot p_{\ell_3})-(p_d \cdot p_j)(p_u \cdot p_{\ell_3})\right]
               \nonumber \\[1mm]&& 
      -{\rm Im}\{O^L_{ij} O^{R\ast}_{ij} \} m_i m_j[p_u,p_d,p_j,p_{\ell_3} ]
\Big\},
\label{eq:SigmaPDbWW}\\[2mm]
\Sigma_{PD_3}(\tilde u_L \tilde u_L) &=& 
              \frac{g^6}{4}|\Delta_t(\tilde u_L)|^2 |V_{j1}|^2|f_{u i}^L|^2
	      |l_{\tilde u j}^{L}|^2 (p_u \cdot p_i)
 \left[(p_d \cdot p_j) (p_j \cdot p_{\ell_3})-m_j^2(p_d \cdot p_{\ell_3})\right],
\nonumber \\&&\label{eq:SigmaPDbuu} \\ 
\Sigma_{PD_3}(\tilde d_L \tilde d_L) &=& 
             - \frac{g^6}{4}|\Delta_u(\tilde d_L)|^2  |V_{j1}|^2|f_{d i}^L|^2
	      |l_{\tilde d j}^{L}|^2 (p_d \cdot p_i)
\left[(p_u \cdot p_j) (p_j \cdot p_{\ell_3})-m_j^2(p_u \cdot p_{\ell_3})\right],
\nonumber \\&&\label{eq:SigmaPDbdd} \\ 
\Sigma_{PD_3}(W \tilde u_L) &=& 
      \frac{\sqrt 2}{4}g^6 \Delta_s(W) \Delta_t^\ast(\tilde u_L) |V_{j1}|^2\Big\{\nonumber \\&&
       2 {\rm Re}\{f_{u i}^L l_{\tilde u j}^{L\ast} O^{L\ast}_{ij} \}(p_u \cdot p_i)
  \left[(p_d \cdot p_j) (p_j \cdot p_{\ell_3})-m_j^2(p_d \cdot p_{\ell_3})\right]
      \nonumber \\[1mm]&&
       + {\rm Re}\{f_{u i}^L l_{\tilde u j}^{L\ast} O^{R\ast}_{ij} \}m_i m_j
   \left[(p_d \cdot p_j)(p_u \cdot p_{\ell_3})-(p_u \cdot p_j)(p_d \cdot p_{\ell_3})\right]
      \nonumber \\[1mm]&&
      +{\rm Im}\{f_{u i}^L l_{\tilde u j}^{L\ast} O^{R\ast}_{ij} \}
                m_i m_j[p_u,p_d,p_j,p_{\ell_3} ]
        \Big\}, 
\label{eq:SigmaPDbWu} \\[2mm]
\Sigma_{PD_3}(W \tilde d_L) &=& 
      \frac{\sqrt 2}{4}g^6 \Delta_s(W) \Delta_u^\ast(\tilde d_L) |V_{j1}|^2\Big\{\nonumber \\&&
       2 {\rm Re}\{f_{d i}^{L\ast} l_{\tilde d j}^{L} O^{R\ast}_{ij} \}(p_d \cdot p_i)
  \left[(p_u \cdot p_j) (p_j \cdot p_{\ell_3})-m_j^2(p_u \cdot p_{\ell_3})\right]
      \nonumber \\[1mm]&&
       + {\rm Re}\{f_{d i}^{L\ast} l_{\tilde d j}^{L} O^{L\ast}_{ij} \} m_i m_j
  \left[(p_u \cdot p_j)(p_d \cdot p_{\ell_3})-(p_d \cdot p_j)(p_u \cdot p_{\ell_3})\right]
      \nonumber \\[1mm]&&
      +{\rm Im}\{f_{d i}^{L\ast} l_{\tilde d j}^{L} O^{L\ast}_{ij} \}
                 m_im_j[p_u,p_d,p_j,p_{\ell_3} ]
        \Big\}, 
\label{eq:SigmaPDbWd} \\[2mm]
\Sigma_{PD_3}(\tilde u_L \tilde d_L) &=& 
              \frac{g^6}{4} \Delta_t(\tilde u_L)\Delta_u^\ast(\tilde d_L)|V_{j1}|^2 m_i m_j 
 \Big\{
 {\rm Im}\{f_{u i}^{L\ast} f_{d i}^{L\ast} l_{\tilde u j}^{L} l_{\tilde d j}^{L} \}
	       [p_u,p_d,p_j,p_{\ell_3} ]
      \nonumber \\[1mm]&&
 +{\rm Re}\{f_{u i}^{L\ast} f_{d i}^{L\ast} l_{\tilde u j}^{L} l_{\tilde d j}^{L} \}
  \left[(p_u \cdot p_j)(p_d \cdot p_{\ell_3})-(p_d \cdot p_j)(p_u \cdot p_{\ell_3})\right]
	       \Big\}.
\label{eq:SigmaPDbud}
\end{eqnarray}
The terms $\Sigma_{PD_3}$ are the same for the charge conjugated process,
$\bar u  d \to \tilde\chi_i^0 \, \tilde\chi_j^-$,
and $\tilde\chi^-_j \to \tilde\nu_\ell \, \ell_3^-$,
due to the sign change of both 
$\Sigma^b_P$, see Eq.~(\ref{eq:SigmaChar}),
and
 $\Sigma^b_{D_3}$, see Eq.~(\ref{eq:SigmabD3}).

Finally, by using the completeness relations for both 
the chargino and neutralino spin vectors, 
 Eq.~(\ref{eq:completenessneut}) and Eq.~(\ref{eq:completenesschar}),
respectively, 
the terms with products of the chargino and neutralino spin vectors 
in Eq.~(\ref{eq:matrixelement2}) can be written 
\begin{eqnarray}
 \Sigma^{ab}_P~~\Sigma^a_{D_1}~\Sigma^b_{D_3} &\equiv& \Sigma_{P D_1 D_3} =
 \Sigma_{PD_1D_3}(W W)+
\Sigma_{PD_1D_3}(\tilde u_L \tilde u_L)+
\Sigma_{PD_1D_3}(\tilde d_L \tilde d_L)+
 \nonumber\\[2mm]&&
\Sigma_{PD_1D_3}(W \tilde u_L)+
\Sigma_{PD_1D_3}(W \tilde d_L)+
\Sigma_{PD_1D_3}(\tilde u_L \tilde d_L), 
\label{eq:SigmaPD1D3ab}
\end{eqnarray}
with
\begin{eqnarray}
\Sigma_{PD_1D_3}(W W) &=&  \frac{g^8}{2}|\Delta_s(W)|^2|f^{R}_{\ell i}|^2|V_{j1}|^2 \Big[
        |O^L_{ij}|^2 g_1
       +|O^R_{ij}|^2 g_2
                \nonumber \\[1mm]&& 
      + {\rm Re}\{O^L_{ij} O^{R\ast}_{ij} \} m_i m_j g_3
        +{\rm Im}\{O^L_{ij} O^{R\ast}_{ij} \} m_i m_j f
\Big],
\label{eq:SigmaPD1D3abWW}\\[2mm]
\Sigma_{PD_1D_3}(\tilde u_L \tilde u_L) &=& 
              \frac{g^8}{4}|\Delta_t(\tilde u_L)|^2 |f^{R}_{\ell i}|^2|V_{j1}|^2|f_{u i}^L|^2
	      |l_{\tilde u j}^{L}|^2 g_1,
\label{eq:SigmaPD1D3abuu} \\ [2mm]
\Sigma_{PD_1D_3}(\tilde d_L \tilde d_L) &=& 
              \frac{g^8}{4}|\Delta_u(\tilde d_L)|^2 |f^{R}_{\ell i}|^2|V_{j1}|^2|f_{d i}^L|^2
	      |l_{\tilde d j}^{L}|^2 g_2,
\label{eq:SigmaPD1D3abdd} \\ [2mm]
\Sigma_{PD_1D_3}(W \tilde u_L) &=& 
      \frac{\sqrt 2}{4}g^8 \Delta_s(W) \Delta_t^\ast(\tilde u_L)|f^{R}_{\ell i}|^2|V_{j1}|^2 \Big[
       2 {\rm Re}\{f_{u i}^L l_{\tilde u j}^{L\ast} O^{L\ast}_{ij} \}
	           g_1 
      \nonumber \\&&
       + {\rm Re}\{f_{u i}^L l_{\tilde u j}^{L\ast} O^{R\ast}_{ij} \} m_i m_j g_3
        +{\rm Im}\{f_{u i}^L l_{\tilde u j}^{L\ast} O^{R\ast}_{ij} \} m_i m_j f
        \Big], \qquad
\label{eq:SigmaPD1D3abWu} \\[2mm]
\Sigma_{PD_1D_3}(W \tilde d_L) &=& 
      -\frac{\sqrt 2}{4}g^8 \Delta_s(W) \Delta_u^\ast(\tilde d_L)|f^{R}_{\ell i}|^2|V_{j1}|^2 \Big[
       2 {\rm Re}\{f_{d i}^{L\ast} l_{\tilde d j}^{L} O^{R\ast}_{ij} \}
	           g_2 
      \nonumber \\&&
      + {\rm Re}\{f_{d i}^{L\ast} l_{\tilde d j}^{L} O^{L\ast}_{ij} \} m_i m_j g_3
       -{\rm Im}\{f_{d i}^{L\ast} l_{\tilde d j}^{L} O^{L\ast}_{ij} \} m_i m_j f
        \Big], 
\label{eq:SigmaPD1D3abWd} \\[2mm]
\Sigma_{PD_1D_3}(\tilde u_L \tilde d_L) &=& 
              -\frac{g^8}{4} \Delta_t(\tilde u_L)\Delta_u^\ast(\tilde d_L) 
               |f^{R}_{\ell i}|^2|V_{j1}|^2  m_i m_j \Big[
  \nonumber \\[1mm]&&
 {\rm Re}\{f_{u i}^{L\ast} f_{d i}^{L\ast} l_{\tilde u j}^{L} l_{\tilde d j}^{L} \} g_3
-{\rm Im}\{f_{u i}^{L\ast} f_{d i}^{L\ast} l_{\tilde u j}^{L} l_{\tilde d j}^{L} \} f
	       \Big], \qquad \quad
\label{eq:SigmaPD1D3abud}
\end{eqnarray}
and the short-hand notations for the kinematic functions
\begin{eqnarray}
g_1 &=&
\Big[(p_u \cdot p_i) (p_i \cdot p_{\ell_1})-m_i^2(p_u \cdot p_{\ell_1})\Big]
\Big[(p_d \cdot p_j) (p_j \cdot p_{\ell_3})-m_j^2(p_d \cdot p_{\ell_3})\Big],\qquad \quad
\label{eq:g1}\\[2mm]
g_2 &=&
\Big[(p_d \cdot p_i) (p_i \cdot p_{\ell_1})-m_i^2(p_d \cdot p_{\ell_1})\Big]
\Big[(p_u \cdot p_j) (p_j \cdot p_{\ell_3})-m_j^2(p_u \cdot p_{\ell_3})\Big],\qquad \quad
\label{eq:g2}
\end{eqnarray}
\begin{eqnarray}
g_3 &=&
 \phantom{+} (p_u \cdot p_d) \left[  (p_j \cdot p_{\ell_1})(p_i \cdot p_{\ell_3})
                        - (p_i \cdot p_j )(p_{\ell_1} \cdot p_{\ell_3})\right]
               \nonumber \\&&
  +(p_u \cdot p_{\ell_1}) \left[ (p_i \cdot p_j)(p_d \cdot p_{\ell_3})
                        -  (p_d \cdot p_j )(p_i \cdot p_{\ell_3})\right]
               \nonumber \\&&
  +(p_u \cdot p_i) \left[ (p_d \cdot p_j)(p_{\ell_1} \cdot p_{\ell_3})
                        - (p_j \cdot p_{\ell_1} )(p_d \cdot p_{\ell_3})\right]
               \nonumber \\&&
  +(p_u \cdot p_{\ell_3}) \left[ (p_i \cdot p_j)(p_d \cdot p_{\ell_1})
                        - (p_d \cdot p_i )(p_j \cdot p_{\ell_1})\right]
               \nonumber \\&&
  +(p_u \cdot p_j) \left[ (p_d \cdot p_i)(p_{\ell_1} \cdot p_{\ell_3})
                        -(p_d \cdot p_{\ell_1})(p_i \cdot p_{\ell_3})\right],
\label{eq:f1} \\[2mm]
f &=&
   \phantom{+} (p_u \cdot p_{\ell_1})[p_d,p_i,p_j,p_{\ell_3}]
             + (p_u \cdot p_i  )[p_d,p_j,p_{\ell_1},p_{\ell_3} ]
   \nonumber \\&&
            +(p_d \cdot p_{\ell_3})[p_u,p_i,p_j,p_{\ell_1}]
            +(p_d \cdot p_j  )[p_u,p_i,p_{\ell_1},p_{\ell_3}]
\label{eq:f2} \\[2mm]
%
%
      &=&
   \phantom{+} (p_d \cdot p_{\ell_1})[p_u,p_i,p_j,p_{\ell_3}]
             + (p_d \cdot p_i)[p_u,p_j,p_{\ell_1},p_{\ell_3}]
    \nonumber \\&&
	    +  (p_u \cdot p_{\ell_3})[p_d,p_i,p_j,p_{\ell_1}]
            +  (p_u \cdot p_j)[p_d,p_i,p_{\ell_1},p_{\ell_3}].
\label{eq:f3} 
\end{eqnarray}
To obtain the terms $\Sigma_{PD_1D_3}$ for the charge conjugated process,
$\bar u  d \to \tilde\chi_i^0 \, \tilde\chi_j^-$,
and $\tilde\chi^-_j \to \tilde\nu_\ell  \,\ell_3^-$,
one has to change the signs of Eqs.~(\ref{eq:SigmaPD1D3abWW})-(\ref{eq:SigmaPD1D3abud}),
due to the sign change of  $\Sigma^b_{D_3}$, see Eq.~(\ref{eq:SigmabD3}).
For the neutralino decay into a positively charged selectron, 
$ \tilde\chi^0_i \to \tilde\ell_R^+ \, \ell_1^-$,
there is an additional sign change in Eqs.~(\ref{eq:SigmaPD1D3abWW})-(\ref{eq:SigmaPD1D3abud}),
due to the sign change of $\Sigma^a_{D_1}$, see Eq.~(\ref{eq:SigmaaD1}).


\section{Charge conjugation and asymmetries
  \label{CPasymmetries}}

In this Appendix we discuss the dependence of our asymmetries on the
charges of the near and far leptons, $\ell_1$ and $\ell_2$
respectively, from the neutralino decay, and on the charge of the
lepton $\ell_3$ from the chargino decay (which also determines the
chargino charge). Note first that changing the chargino charge also
requires a charge conjugation of the initial state, \textit{i.e.}  $u
\bar d \to \tilde\chi_i^0 \,\tilde\chi_j^+$, becomes $\bar u d \to
\tilde\chi_i^0 \,\tilde\chi_j^-$. Since the total charge of the three
leptons is easy to measure, $u \bar d \to \tilde\chi_i^0
\,\tilde\chi_j^+$ production can be distinguished from its charge
conjugated production, $\bar u d \to \tilde\chi_i^0
\,\tilde\chi_j^-$. Due to the Majorana nature of the neutralino, the
near lepton produced in its decay chain has equal probability to have
positive or negative charge, independent of the charge of the initial
state.

\medskip

The chargino spin correlation asymmetry depends on these charges as follows:
\begin{eqnarray}
 \mathcal{A} [p_u,p_{\bar d},p_{\tilde \chi_j^+} ,p_{\ell_3^+}] &=&
 \mathcal{A} [p_{\bar u},p_d,p_{\tilde \chi_j^-} ,p_{\ell_3^-}] ,
\end{eqnarray}
see the remarks regarding the sign change in the relevant contribution to the
amplitude squared for the charge conjugated processes in Appendix~\ref{Squared
  amplitude of production and decay}, after Eq.~(\ref{eq:SigmaPDbud}).
The asymmetry of the neutralino-chargino spin-spin correlations satisfies
\begin{eqnarray}
 \mathcal{A}[f(p_u, p_{\bar d}, p_{\tilde \chi_i^0}, p_{\tilde \chi_j^+},
     p_{\ell_1^+}, p_{\ell_3^+})]  &=&
 \mathcal{A}[f(p_{\bar u}, p_d, p_{\tilde \chi_i^0}, p_{\tilde \chi_j^-},
   p_{\ell_1^+}, p_{\ell_3^-})]\nonumber \\ &=& 
-\mathcal{A}[f(p_{\bar u}, p_d, p_{\tilde \chi_i^0}, p_{\tilde \chi_j^-},
  p_{\ell_1^-}, p_{\ell_3^-})], 
\end{eqnarray}
see the remarks after Eq.~(\ref{eq:f3}), and the T-odd product $f$ as given in
Eq.~(\ref{eq:xf2}). For the neutralino spin correlation asymmetry we have
\begin{eqnarray}
     \mathcal{A} [p_u,p_{\bar d},p_{\tilde \chi_i^0} ,p_{\ell_1^+}] 
&=& -\mathcal{A} [p_u,p_{\bar d},p_{\tilde \chi_i^0} ,p_{\ell_1^-}] \nonumber \\
&=& -\mathcal{A} [p_{\bar u},p_d,p_{\tilde \chi_i^0} ,p_{\ell_1^+}] \nonumber \\
&=& +\mathcal{A} [p_{\bar u},p_d,p_{\tilde \chi_i^0} ,p_{\ell_1^-}] ,
\label{eq:signNspin}
\end{eqnarray}
see the remarks  after Eq.~(\ref{eq:SigmaPDaud}). 
Although these relations concern the asymmetries which contain the inaccessible quark 
momenta and neutralino and/or chargino momenta, these relations are essential to 
understand the properties under charge conjugation of the accessible asymmetries, where
these momenta have been replaced and approximated, as discussed in 
Section~\ref{construct}.

\medskip

As discussed in Section~\ref{neutmom}, the neutralino momentum cannot be 
reconstructed and thus is replaced by the momentum of the far lepton $\ell_2$. As shown 
in Section~\ref{neutmom}, the corresponding asymmetry depends on the charge of the near
($\ell_1$) and far ($\ell_2$) leptons as follows
\begin{eqnarray}
 \mathcal{A} [p_u,p_{\bar d},p_{\ell_1^+} ,p_{\ell_2^-}]   &=&
-\mathcal{A} [p_u,p_{\bar d},p_{\ell_1^-} ,p_{\ell_2^+}] \nonumber\\&=&
+\mathcal{A} [p_u,p_{\bar d},p_{\ell_2^+} ,p_{\ell_1^-}] \nonumber\\&=&
-\mathcal{A} [p_u,p_{\bar d},p_{\ell_2^-} ,p_{\ell_1^+}].
\label{eq:chargeN} 
\end{eqnarray}
The sign change in the first step originates from Eq.~(\ref{eq:SigmaaD1}), 
and the sign change in the second and third steps is simply due to the interchange 
of the momenta of $\ell_1$ and $\ell_2$ in the antisymmetric epsilon product. 
Thus we need not determine from which vertex, near or far, the leptons $\ell_1$ and 
$\ell_2$ originate. Instead in the epsilon product one would just group the leptons 
according to their charge, and use the corresponding asymmetry
\begin{eqnarray}
 \mathcal{A} [p_u, p_{\bar d}, p_{\ell^+}, p_{\ell^-}]&=&
-\mathcal{A} [p_u, p_{\bar d}, p_{\ell^-}, p_{\ell^+}].
\end{eqnarray}
The triple product asymmetry fulfills similar relations to Eq.~(\ref{eq:chargeN}),
concerning the interchange of the near and far leptons, and of their charges,
\begin{eqnarray}
 \mathcal{A}(\mathbf{p}_{\ell_1^+}, \mathbf{p}_{\ell_2^-}, \mathbf{p}_{\ell_3})   &=& 
-\mathcal{A}(\mathbf{p}_{\ell_1^-}, \mathbf{p}_{\ell_2^+}, \mathbf{p}_{\ell_3})
  \nonumber \\&=&
+\mathcal{A}(\mathbf{p}_{\ell_2^+}, \mathbf{p}_{\ell_1^-}, \mathbf{p}_{\ell_3}) 
   \nonumber\\&=&
-\mathcal{A}(\mathbf{p}_{\ell_2^-}, \mathbf{p}_{\ell_1^+}, \mathbf{p}_{\ell_3}),
\end{eqnarray}
which each hold for the chargino charge fixed $\ell_3= \ell_3^+$ or $\ell_3^-$.
Thus also for the triple product asymmetry, only the charges of the two leptons from 
the neutralino decay have to be tagged, and for measurements one would use the triple 
product asymmetry
\begin{eqnarray}
 \mathcal{A}(\mathbf{p}_{\ell^+}, \mathbf{p}_{\ell^-}, \mathbf{p}_{\ell_3})   &=& 
-\mathcal{A}(\mathbf{p}_{\ell^-}, \mathbf{p}_{\ell^+}, \mathbf{p}_{\ell_3}). 
\end{eqnarray}
Concerning now the charged conjugated production, 
the neutralino spin correlation asymmetry obeys
\begin{eqnarray}
  \mathcal{A} [p_u,   p_{\bar d}, p_{\ell^+}, p_{\ell^-}]  &=&
- \mathcal{A} [p_{\bar u}, p_{d}, p_{\ell^+}, p_{\ell^-}].
\end{eqnarray}
Note however, that no such simple relations for charge conjugation of the initial state
hold neither for the asymmetries where the quark momenta are approximated, see 
Eqs.~(\ref{eq:quarkapproxu}), (\ref{eq:quarkapproxd}), nor for the triple product 
asymmetry,
\begin{eqnarray}
 \mathcal{A}[p_u^{\rm aprx}, p_{\bar d}^{\rm aprx}, p_{\ell^+}, p_{\ell^-}]&\neq& 
-\mathcal{A}[p_{\bar u}^{\rm aprx}, p_{d}^{\rm aprx}, p_{\ell^+}, p_{\ell^-}], \\
 \mathcal{A}(\mathbf{p}_{\ell^+}, \mathbf{p}_{\ell^-}, \mathbf{p}_{\ell_3^+})&\neq& 
-\mathcal{A}(\mathbf{p}_{\ell^+}, \mathbf{p}_{\ell^-}, \mathbf{p}_{\ell_3^-}).
\end{eqnarray}
The initial partonic systems $\bar u d$ and $ u \bar d$ have different boost
distributions due to the PDFs, thus the efficiencies to guess the initial quark 
directions are different. Recall that the triple product asymmetry is not  
Lorentz invariant.
Thus also here the different boost distributions of the initial partonic systems,  
$\bar u d$ as opposed to $ u \bar d$, will give different values for the triple 
product asymmetry for $pp \to\tilde\chi_i^0 \,\tilde\chi_j^+$ production, and its charge 
conjugated process $pp \to\tilde\chi_i^0 \,\tilde\chi_j^-$, in general. 
Moreover, as pointed out in Section~\ref{construct}, the asymmetries
$\mathcal{A}[p_u^{\rm aprx}, p_{\bar d}^{\rm aprx}, p_{\ell^+}, p_{\ell^-}]$,
as well as
$\mathcal{A}(\mathbf{p}_{\ell^+}, \mathbf{p}_{\ell^-}, \mathbf{p}_{\ell_3^+})$,
receive contributions both from the neutralino spin correlations and from the 
spin-spin correlations. Since these two contributions react differently under charge 
conjugation, also the corresponding asymmetries have a different magnitude. 

\end{appendix}
\newpage



\end{document}